\documentclass[preprint]{elsarticle}

\usepackage[left=3cm,right=3cm,top=5cm,bottom=5cm,a4paper]{geometry}
\usepackage{amsmath}
\usepackage{graphicx}
\usepackage{dcolumn}
\usepackage{bm}
\usepackage{subfigure}
\usepackage{xcolor}
\usepackage[utf8]{inputenc}
\usepackage[T1]{fontenc}
\usepackage{mathptmx}
\usepackage{setspace}
\usepackage{tabularx}
\usepackage{amsmath}
\usepackage{graphicx}
\usepackage{dcolumn}
\usepackage{bm}
\usepackage{subfigure}
\usepackage{xcolor}
\usepackage[utf8]{inputenc}
\usepackage[T1]{fontenc}
\usepackage{mathptmx}
\usepackage{amssymb}
\usepackage[normalem]{ulem}
\journal{Physics Letters A}
\begin{document}
\begin{frontmatter}
\title{Comprehensive Gyrokinetic Study of Eigenstate Transitions in Fast Ion-Driven Electrostatic Drift Instabilities}
\author[label1]{B.J. Kang\corref{cor1}}
\affiliation[label1]{organization={National Institute for Fusion Science, Toki, Gifu 509-5292, Japan}}
\cortext[cor1]{corresponding author}
\ead{byungjun.kang@nifs.ac.jp}
\author[label1]{H. Sugama}
\author[label2]{T.-H. Watanabe}
\affiliation[label2]{organization={Department of Physics, Nagoya University, Nagoya 464-8602, Japan}}
\author[label1]{M. Nunami}
%%%%%%%%%%%%%%%%%%%%%%%%%%%%%%%%%%%%%%%%%%%%%%%%%%%%%%%%%%%%%%%%%%%%%%%%%%%%%%%%%%%%%%%%%%%%%%%%%%%%%%%%%%%%%%%%%%%%%%%%%%%%%%%%%%%%%%%%%%%%%%%%%%%%%%%%%%%%%%%%%%%%%%%%%%%%
%%%%%%%%%%%%%%%%%%%%%%%%%%%%%%%%%%%%%%%%%%%%%%%%%%%%%%%%%%%%%%%%%%%%%%%%%%%%%%%%%%%%%%%%%%%%%%%%%%%%%%%%%%%%%%%%%%%%%%%%%%%%%%%%%%%%%%%%%%%%%%%%%%%%%%%%%%%%%%%%%%%%%%%%%%%%
%%%%%%%%%%%%%%%%%%%%%%%%%%%%%%%%%%%%%%%%%%%%%%%%%%%%%%%%%%%%%%%%%%%%%%%%%%%%%%%%%%%%%%%%%%%%%%%%%%%%%%%%%%%%%%%%%%%%%%%%%%%%%%%%%%%%%%%%%%%%%%%%%%%%%%%%%%%%%%%%%%%%%%%%%%%%
\begin{abstract}
This study comprehensively investigates fast ion-driven drift instability, extending the theory in [B. J. Kang and T. S. Hahm, Phys. Plasmas 26, 042501 (2019)]. The eigenmode equation, including the resonant contribution of passing fast ions, is derived and solved using the shooting method. Passing fast ions significantly affect the instability in weak negative shear or moderate positive shear plasmas. Eigenstate transitions to non-ground states occur more readily in weak magnetic shear, high safety factor, and long wavelength perturbations. Linear gyrokinetic simulations using the GKV code verify the theory, showing good agreement with shooting method results. The estimated quasilinear transport indicates that the net energy flux can be inward, without contradicting the second law of thermodynamics. These findings have important implications for heating efficiency and plasma confinement in the heating process, such as Ion Cyclotron Resonance Heating (ICRH) in future fusion devices.
\end{abstract}
%%%%%%%%%%%%%%%%%%%%%%%%%%%%%%%%%%%%%%%%%%%%%%%%%%%%%%%%%%%%%%%%%%%%%%%%%%%%%%%%%%%%%%%%%%%%%%%%%%%%%%%%%%%%%%%%%%%%%%%%%%%%%%%%%%%%%%%%%%%%%%%%%%%%%%%%%%%%%%%%%%%%%%%%%%%%
%%%%%%%%%%%%%%%%%%%%%%%%%%%%%%%%%%%%%%%%%%%%%%%%%%%%%%%%%%%%%%%%%%%%%%%%%%%%%%%%%%%%%%%%%%%%%%%%%%%%%%%%%%%%%%%%%%%%%%%%%%%%%%%%%%%%%%%%%%%%%%%%%%%%%%%%%%%%%%%%%%%%%%%%%%%%
%%%%%%%%%%%%%%%%%%%%%%%%%%%%%%%%%%%%%%%%%%%%%%%%%%%%%%%%%%%%%%%%%%%%%%%%%%%%%%%%%%%%%%%%%%%%%%%%%%%%%%%%%%%%%%%%%%%%%%%%%%%%%%%%%%%%%%%%%%%%%%%%%%%%%%%%%%%%%%%%%%%%%%%%%%%%
\end{frontmatter}
%%%%%%%%%%%%%%%%%%%%%%%%%%%%%%%%%%%%%%%%%%%%%%%%%%%%%%%%%%%%%%%%%%%%%%%%%%%%%%%%%%%%%%%%%%%%%%%%%%%%%%%%%%%%%%%%%%%%%%%%%%%%%%%%%%%%%%%%%%%%%%%%%%%%%%%%%%%%%%%%%%%%%%%%%%%%
%%%%%%%%%%%%%%%%%%%%%%%%%%%%%%%%%%%%%%%%%%%%%%%%%%%%%%%%%%%%%%%%%%%%%%%%%%%%%%%%%%%%%%%%%%%%%%%%%%%%%%%%%%%%%%%%%%%%%%%%%%%%%%%%%%%%%%%%%%%%%%%%%%%%%%%%%%%%%%%%%%%%%%%%%%%%
%%%%%%%%%%%%%%%%%%%%%%%%%%%%%%%%%%%%%%%%%%%%%%%%%%%%%%%%%%%%%%%%%%%%%%%%%%%%%%%%%%%%%%%%%%%%%%%%%%%%%%%%%%%%%%%%%%%%%%%%%%%%%%%%%%%%%%%%%%%%%%%%%%%%%%%%%%%%%%%%%%%%%%%%%%%%
\section{INTRODUCTION}\label{Sec: Introduction}
Predicting turbulence and transport in magnetized fusion plasmas in the presence of fast ions remains one of the most important research topics in the magnetic fusion energy community because many fast ions are generated either through external heating processes or fusion reactions. Nearly half a century ago, the possibility that fusion product $\alpha$-particles could resonantly interact with and excite Alfv\'enic instabilities was recognized \cite{RoesnbluthPRL1975, HasegawaPF1976}. Since then, the behavior of electromagnetic Alfv\'enic instabilities driven by $\alpha$-particles in burning plasmas has been extensively studied, as reviewed in \cite{ChenRMP2016}. 
On the other hand, studies of small-scale electrostatic fluctuations in burning plasmas have mainly focused on how high-energy fusion products are transported due to the well-understood microinstabilities, such as the Ion Temperature Gradient (ITG) mode \cite{AngioniPoP2008, ZhangPRL2008}, and the Collisionless Trapped Electron Mode (CTEM) \cite{YangPoP2018}.  In addition, fast ions produced by external heating methods like Neutral Beam Injection (NBI) and Ion Cyclotron Resonance Heating (ICRH) are generally considered to have stabilizing effects on ITG modes through various mechanisms as discussed in the review by Citrin and Mantica \cite{CitrinPPCF2023}. For instance, with ICRH, electromagnetic waves are launched into the plasma at a frequency resonant with the ion cyclotron motion at specific locations, allowing a small population of minority ions to efficiently absorb the power and be heated up to MeV-range energies \cite{KazakovNF2015}. The fast ion tails created by the ICRH can generate steep fast ion temperature gradients. In such conditions, fast ions can suppress ITG instability through electrostatic wave-particle resonance \cite{SienaNF2018, SienaPoP2019}.\\
\indent  Reversed magnetic shear (or negative magnetic shear) plasmas are known as one of the leading operation scenarios for a steady-state tokamak reactor characterized by good confinement, high stability, and a high fraction of self-sustaining bootstrap current \cite{KesselPRL1994, TurnbullPRL1995}. The stabilizing role of reversed shear plasmas in CTEM due to the precession reversal of trapped electrons \cite{BeerPoP1997, LiPPCF2002} and enhanced confinement with Internal Transport Barriers (ITBs) \cite{LevintonPRL1995, StraitPRL1995, NazikianPRL2005} has been widely accepted. Additionally, the reversed magnetic shear further enhances the stabilizing effect of the electrostatic wave-particle resonance on ITG mode \cite{SienaPoP2019}. However, recent studies suggest that fast ions can destabilize a certain electrostatic instability in reversed shear plasmas. A new instability driven by the resonance between toroidal precession of trapped fast ions and drift waves propagating in the electron diamagnetic direction for steep fast ion temperature gradients in reversed shear plasmas was theoretically predicted \cite{KangPoP2019}, and various properties of this instability have been explored through both linear and nonlinear gyrokinetic simulations \cite{KangPoP2020, KangPLA2021}. While the electrostatic wave-particle resonance between fast ions and ITG mode propagating in the ion diamagnetic direction and its stabilizing effects are well-established, this new instability demonstrates that fast ions can destabilize electron drift waves, particularly in reversed shear plasmas with steep fast ion temperature gradients.\\
\indent Despite these insights, several questions about the fast ion-driven electron drift instability remain unresolved. Gyrokinetic simulations have observed unstable modes even in moderate positive magnetic shear \cite{KangPoP2020}, despite theoretical predictions that the instability primarily arises in reversed magnetic shear plasmas, where a significant fraction of trapped fast ions reverse their precession direction. Additionally, parity transitions in the parallel mode structure have also been observed, which cannot be explained by previous theoretical models. Theoretical models in \cite{KangPoP2019, KangPoP2020} used bounce-averaged gyrokinetic equations to describe trapped fast ion dynamics, assuming that the non-adiabatic response of passing fast ions was negligible. However, the odd parity component of the mode cannot be explained by these models because the bounce-averaged potential vanishes for odd functions along the field line, eliminating the source of the instability. \\
\indent Drift waves are ubiquitous in nonuniform magnetized plasmas and are widely regarded as one of the dominant mechanisms driving turbulent transport. Although it has been theoretically understood for a long time \cite{ChenPF1980, HortonRMP1999} that toroidal drift waves, such as ITG mode and Trapped Electron Mode (TEM), can have multiple eigenstates, most studies have focused on the ground state branch because it is usually the most significant branch. However, it was reported that the most unstable branch can change from a ground state to a non-ground state under strong temperature gradient plasma parameters, as shown by local gyrokinetic simulations \cite{WangNF2012}. Furthermore, it was found that high-order electrostatic drift modes become increasingly important under strong gradient parameters, both linearly \cite{XiePoP2015, XiePoP2016} and nonlinearly \cite{XiePRL2017}. These studies suggest that the parity transitions observed in \cite{KangPoP2020} could be linked to eigenstate transitions, given that the fast ion-driven drift instability tends to become more unstable in the strong gradient regime. Moreover, the gyrokinetic simulation results from \cite{KangPoP2020} revealed that the instabilities in positive shear plasmas are significantly influenced by passing fast ions, which were previously neglected in the theoretical models. This indicates the need to extend earlier models to account for the non-adiabatic response of passing fast ions and better understand the conditions leading to eigenstate transitions.\\
\indent In this work, we perform linear stability analyses of the fast ion-driven drift instability, focusing on the role of fast ions in eigenstate transitions. We extend the theoretical models in \cite{KangPoP2019, KangPoP2020} by including the non-adiabatic response of passing fast ions. The eigenmode equation in the ballooning coordinate \cite{ConnorPRS1979} is solved using the shooting method, and our findings are further validated through gyrokinetic simulations using the GKV code \cite{WatanabeNF2006}. This study aims to provide a more comprehensive understanding of the fast ion-driven drift instability by considering the contributions of passing fast ions and various eigenstate transitions. \\
\indent The main results of this work are as follows: \\
\indent (1)  We extend the analytic theory of the fast ion-driven drift instability, explicitly including the resonant contribution from passing fast ions. The eigenmode equation in the ballooning coordinate is derived and solved using the shooting method. \\
\indent (2) The inclusion of passing fast ions reveals that their resonant contribution significantly affects the instability, particularly in weak negative shear and moderate positive shear plasmas. Eigenstate transitions to non-ground states occur widely under steep fast ion temperature gradients, and they are more prevalent in weak magnetic shear, high safety factor, and long wavelength perturbations. \\
\indent (3) In regimes with weak magnetic shear, high safety factor, and long wavelength perturbations, the oscillatory property of the toroidicity-induced potential dominates, making various non-ground states the most unstable. \\
\indent (4) Linear gyrokinetic simulations using the GKV code are performed to validate the theoretical model. The simulation results generally show good agreement with the prediction from the shooting method.\\
\indent (5) We estimate the quasilinear transport caused by the fast ion-driven drift instability. Our analysis suggests that under specific conditions, the net energy flux can be inward, consistent with the second law of thermodynamics. \\
\indent The rest of this paper is organized in the following order. In Sec. \ref{Sec: Theoretical model}, we introduce our theoretical model, including passing fast ion dynamics, and derive the eigenmode equation using gyrokinetic equations in the ballooning coordinate. In Sec. \ref{Sec: Linear stability analysis}, we solve the eigenmode equation using the shooting method and investigate the various properties of the instability. We also perform gyrokinetic simulations and verify the theoretical predictions using the GKV code in the flux-tube toroidal geometry. Turbulent fluxes driven by the fast ion-driven drift instability are calculated using the quasilinear model in Sec. \ref{Sec: Quasilinear transport}. Finally, the summary and discussions of this work are presented in Sec. \ref{Sec: Conclusion}.
\clearpage
%%%%%%%%%%%%%%%%%%%%%%%%%%%%%%%%%%%%%%%%%%%%%%%%%%%%%%%%%%%%%%%%%%%%%%%%%%%%%%%%%%%%%%%%%%%%%%%%%%%%%%%%%%%%%%%%%%%%%%%%%%%%%%%%%%%%%%%%%%%%%%%%%%%%%%%%%%%%%%%%%%%%%%%%%%%%
%%%%%%%%%%%%%%%%%%%%%%%%%%%%%%%%%%%%%%%%%%%%%%%%%%%%%%%%%%%%%%%%%%%%%%%%%%%%%%%%%%%%%%%%%%%%%%%%%%%%%%%%%%%%%%%%%%%%%%%%%%%%%%%%%%%%%%%%%%%%%%%%%%%%%%%%%%%%%%%%%%%%%%%%%%%%
%%%%%%%%%%%%%%%%%%%%%%%%%%%%%%%%%%%%%%%%%%%%%%%%%%%%%%%%%%%%%%%%%%%%%%%%%%%%%%%%%%%%%%%%%%%%%%%%%%%%%%%%%%%%%%%%%%%%%%%%%%%%%%%%%%%%%%%%%%%%%%%%%%%%%%%%%%%%%%%%%%%%%%%%%%%%
\section{THEORETICAL MODEL}\label{Sec: Theoretical model}
\indent The linear theory of the fast ion-driven electrostatic drift instability, neglecting the non-adiabatic response of passing fast ions, was examined \cite{KangPoP2019, KangPoP2020}, based on the gyrokinetic equations in general toroidal geometry \cite{FriemanPF1982, HahmPF1988}. In this work, we use an electrostatic gyrokinetic model to investigate the fast ion-driven drift instability because this study focuses on electrostatic modes, continuing previous theoretical investigations. This choice simplifies the treatment of eigenstate transitions and allows us to isolate the specific effects of fast ions on the drift instability.
In this section, we briefly review and extend these theoretical models by considering the resonant contribution from passing fast ions. \\
\indent Each ion species is described by the linearized gyrokinetic equation in the electrostatic limit. Assuming the equilibrium distribution function is Maxwellian, the collisionless linearized gyrokinetic equation in a toroidal geometry is written as:
\begin{equation} \label{gyrokinetic_0}
\begin{split}
&\left[i\omega_{\bold{k_\perp}} + \frac{\mathrm{v}_\perp^2/2 + \mathrm{v}_\parallel^2}{R_0\Omega_{c\sigma}}\left( \sin\theta ik_r + \cos\theta ik_\theta \right) - \mathrm{v}_\parallel \nabla_\parallel \right]\delta g_{\sigma\bold{k_\perp}} \\
&= i\left( \omega_{\bold{k_\perp}} - \omega_{*\sigma}^T \right)f_{M\sigma} \frac{Z_\sigma e}{T_{\sigma 0}}J_0\left(k_{\perp}\rho_{\perp \sigma} \right)\delta \phi_{\bold{k_\perp}}  =0,
\end{split}
\end{equation}
where $\sigma$ denotes an arbitrary plasma species, $\omega_{\bold{k_\perp}}$ is the mode frequency for the perpendicular wavenumber vector $\bold{k_\perp}$, $R_0$ is the major radius of a tokamak, $\Omega_{c\sigma}=\frac{Z_\sigma e B_0}{m_\sigma c}$ is the cyclotron frequency, $Z_\sigma$ is the charge number, $e$ is the elementary charge, $B_0$ is the amplitude of the equilibrium magnetic field, $m_\sigma$ is the mass, and $c$ is the speed of light. Additional parameters include the poloidal angle $\theta$, the radial wave number $k_r$, and the poloidal wave number $k_\theta$. The term $\delta g_{\sigma\bold{k_\perp}} = \delta f_{\sigma\bold{k_\perp}} + f_{M\sigma}\frac{Z_\sigma e \delta\phi_{\bold{k_\perp}}}{T_{\sigma 0}} $ represents the non-adiabatic part of the perturbed distribution function, where the Maxwellian distribution function is defined as $f_{M\sigma}=\frac{n_{\sigma0}}{(2\pi)^{3/2}\mathrm{v}_{t\sigma}^{3}}e^{-(\mathrm{v}_\parallel^2+\mathrm{v}_\perp^2)/2\mathrm{v}_{t\sigma}^2}$. Here, $n_{\sigma 0}$ is the equilibrium density, $\mathrm{v}_{t\sigma}=\sqrt{T_{\sigma 0}/m_\sigma}$, $T_{ \sigma 0}$ is the equilibrium temperature,  and $\delta\phi_{\bold{k_\perp}}$ is the perturbed electrostatic potential. $\omega_{*\sigma}^T=\omega_{*\sigma}\left[1+\eta_\sigma\left(\frac{\mathrm{v}_\parallel^2+\mathrm{v}_\perp^2}{2\mathrm{v}_{t\sigma}^2}-\frac{3}{2}\right)\right]$ incorporates the effects of gradients of equilibrium density and temperature, where $\omega_{*\sigma}=-\frac{k_\theta \rho_{t\sigma} v_{t\sigma}}{L_{n\sigma}}$, $\rho_{t\sigma} = \frac{\mathrm{v}_{t\sigma}}{\Omega_{c\sigma}}$, $\eta_\sigma = L_{n\sigma}/L_{T\sigma}$, $L_{n\sigma}^{-1} = -\frac{1}{n_{\sigma 0}} \frac{dn_{\sigma 0}}{dr}$ and $L_{T\sigma}^{-1} = -\frac{1}{T_{\sigma 0}} \frac{dT_{\sigma 0}}{dr}$. $J_0$ is the zeroth order Bessel function of the first kind, $k_\perp$ is the perpendicular wave number, and $\rho_{\perp \sigma} = \frac{\mathrm{v}_{\perp}}{\Omega_{c\sigma}}$. Using this framework, the perturbed density can be calculated as:
\begin{equation}\label{eqn:density}
\delta n_{\sigma\bold{k_\perp}} = -n_{\sigma 0} \frac{Z_\sigma e\delta\phi_\bold{k_\perp}}{T_{\sigma 0}} + \int d^3\mathrm{v} J_0\left(k_{\perp}\rho_{\perp \sigma} \right)\delta g_{\sigma\bold{k_\perp}}.
\end{equation}
Note that Eq. (\ref{gyrokinetic_0}) is a gyrokinetic equation in guiding center coordinate, where $J_0(k_\perp \rho_{\perp\sigma})$ in Eq. (\ref{gyrokinetic_0}) represents the Finite Larmor Radius (FLR) effect resulting from the push-forward transformation (particle coordinate $\rightarrow$ guiding center coordinate). Conversely, Eq. (\ref{eqn:density}) is derived in particle coordinate, where $J_0(k_\perp \rho_{\perp\sigma})$ in Eq.  (\ref{eqn:density}) comes from the pull-back transformation (guiding center coordinate $\rightarrow$ particle coordinate). Additionally, $\mathrm{v}_\parallel$ in Eq. (\ref{gyrokinetic_0}) is not constant along a particle's orbit due to the magnetic mirror force in a non-uniform magnetic field. To consider this effect, the gyrokinetic equation is often expressed in energy $E$ and magnetic moment $\mu$, constants of the motion. However, we employ a transformed gyrokinetic equation expressed in terms of $(\bold{\mathrm{v}_\perp},\mathrm{v}_\parallel)$, where we assume $\mathrm{v}_\parallel$ is constant to simplify the analysis and focus on the dynamics of passing particles. This approach allows for a more analytically tractable exploration of the parallel dynamics. For trapped particle dynamics, we use the bounce-averaged gyrokinetic equation \cite{GangPFB1990, FongPoP1999}, where the orbit-averaged motion describes the effects of a non-uniform magnetic field. \\
\indent To examine the mode structure along the magnetic field line, the eigenmode analysis is conducted in the ballooning coordinate. In a circular tokamak plasma, the perturbed electrostatic potential, $\delta\phi$, can be expressed as
\begin{equation}\label{phi_rtheta}
\delta\phi(r,\theta) = \sum_{n=-\infty}^{\infty}\sum_{j=-\infty}^{\infty} \delta\phi_{n,j}(r) e^{i(m_0+j)\theta-in\zeta},
\end{equation}
where $r$ is the minor radius, $n$ is the toroidal mode number, $\zeta$ is the toroidal angle, and $m_0$ is the poloidal mode number at the reference radius $r_0$. For high mode numbers $(n,m)$, quasi-translational invariance can be assumed \cite{ChenPF1980}, indicating that neighboring poloidal harmonics, $\delta \phi_{n,j}$, have a similar shape. This assumption leads to
\begin{equation}
\delta \phi_{n,j}(Z) = \delta\phi_0 \left( \frac{r-r_0}{\Delta r_n}-j\right),
\end{equation}
where $\Delta r_n = 1/(k_\theta \hat{s})$ defines the radial distance between modes, $\hat{s}$ is the magnetic shear measured at $r=r_0$, and $\delta\phi_0(Z)$ describes the shape function of poloidal harmonics near $r=r_0$. The Fourier transform of $\delta\phi_0(Z)$ is defined as
\begin{equation}
\delta\phi_0(Z) = \frac{1}{\sqrt{2\pi}} \int_{-\infty}^{\infty} e^{iZ\hat{\theta}}\delta\hat{\phi}_0(\hat{\theta})d\hat{\theta},
\end{equation} 
where $\hat{\theta}$ is the extended poloidal angle, also known as the ballooning angle, and spans from $-\infty$ to $\infty$. Applying the same formalism to the non-adiabatic part of the perturbed distribution function $\delta g_\sigma$ yields the gyrokinetic equation in ballooning representation:
\begin{equation} \label{gyrokinetic_ballooning}
\begin{split}
&\left(i\omega -i \omega_{d \mathrm{v} \sigma } - \mathrm{v}_\parallel \nabla_\parallel \right)\delta\hat{g}_{\sigma}(\hat{\theta},\bold{\mathrm{v}}) \\
&= i(\omega-\omega_{*\sigma}^T) f_{M\sigma}\frac{Z_\sigma e}{T_{\sigma 0 }}J_0\left(k_{\perp}\rho_{\perp \sigma} \right)\delta\hat{\phi}(\hat{\theta}).
\end{split}
\end{equation}
\noindent Here, $\omega_{d\sigma\mathrm{v} }=\omega_{d\sigma}\big(\frac{2\mathrm{v}_\parallel^2+\mathrm{v}_\perp^2}{2\mathrm{v}_{t\sigma}^2} \big)$ is the magnetic drift frequency and
$\omega_{d\sigma}=\omega_{*\sigma}\epsilon_{n\sigma} \left( \cos\hat{\theta} + \hat{s}\hat{\theta}\sin\hat{\theta} \right)$ is the velocity-independent part of the magnetic drift frequency, where $\epsilon_{n\sigma} = L_{n\sigma} / {R_0}$ is the normalized density gradient length. For notational simplicity, subscript $\bold{k_\perp}$ is omitted from here.
Note that $\nabla_\parallel$ in Eq. (\ref{gyrokinetic_ballooning}) acts as an operator on $\delta\hat{g}_\sigma$. Consequently, the parallel structure of the modes may be described by an integral Fredholm equation of the second kind \cite{RomanelliPFB1989}. However, such a formulation tends to be analytically intractable. Instead, we describe the parallel dynamics using a semi-local approximation, meaning $\nabla_\parallel$ is interchangeably used with $ik_\parallel = i\hat{k}_{\hat{\theta}}/qR$ and is treated as a number, not as an operator. This approximation aligns with the Wentzel-Krameres-Brillouin (WKB) description of the mode structure. Here, $\hat{k}_{\hat{\theta}}$ represents the dimensionless wave number in the ballooning angle $\hat{\theta}$ and should not be confused with the poloidal wave number $k_\theta$. From Eq. (\ref{gyrokinetic_ballooning}), we express the non-adiabatic part of the perturbed distribution function as
\begin{equation} \label{ghat}
\delta\hat{g}_{\sigma}  = \frac{\omega - \omega_{*\sigma}^T}{\omega - \omega_{d  \sigma\mathrm{v} } - k_\parallel \mathrm{v}_\parallel} f_{M\sigma} \frac{Z_\sigma e}{T_{\sigma 0}}J_0\left(k_{\perp}\rho_{\perp \sigma} \right)\delta\hat{\phi}
\end{equation}\\
\indent To capture the behavior of fast ions in present-day tokamaks, it is important to consider Finite Orbit Width (FOW) effects due to the large drift orbit size of fast ions. In this study, FOW effects are incorporated into the gyrokinetic model through the radial drift component $\omega_{d\sigma \mathrm{v}}$, and similar considerations are also made in other works, such as \cite{SienaNF2018, SienaPoP2019} through with different notations. These effects explicitly affect the resonant contribution of fast ions. This approach has been validated by previous studies, such as \cite{SugamaPoP2006, SugamaJPP2006}, which demonstrated that the same gyrokinetic equations successfully describe FOW effects in the context of Landau damping on Geodesic Acoustic Mode (GAM) oscillation in both tokamaks and helical systems. The precession drift motion of fast ions, which requires accounting for FOW in a sheared magnetic field, is also included in our formulation and will be shown later.  \\
%%%%%%%%%%%%%%%%%%%%%%%%%%%%%%%%%%%%%%%%%%%%%%%%%%%%%%%%%%%%%%%%%%%%%%%%%%%%%%%%%%%%%%%%%%%%%%%%%%%%%%%%%%%%%%%%%%%%%%%%%%%%%%%%%%%%%%%%%%%%%%%%%%%%%%%%%%%%%%%%%%%%%%%%%%%%
%%%%%%%%%%%%%%%%%%%%%%%%%%%%%%%%%%%%%%%%%%%%%%%%%%%%%%%%%%%%%%%%%%%%%%%%%%%%%%%%%%%%%%%%%%%%%%%%%%%%%%%%%%%%%%%%%%%%%%%%%%%%%%%%%%%%%%%%%%%%%%%%%%%%%%%%%%%%%%%%%%%%%%%%%%%%
%%%%%%%%%%%%%%%%%%%%%%%%%%%%%%%%%%%%%%%%%%%%%%%%%%%%%%%%%%%%%%%%%%%%%%%%%%%%%%%%%%%%%%%%%%%%%%%%%%%%%%%%%%%%%%%%%%%%%%%%%%%%%%%%%%%%%%%%%%%%%%%%%%%%%%%%%%%%%%%%%%%%%%%%%%%%
\subsection{FAST ION DYNAMICS}\label{Subsec: Fast ion dynamics}
\indent To include the contribution of passing fast ions, we express the density response of fast ions as follows:
\begin{equation} \label{deltan_f}
\frac{\delta n_{f}}{n_{f0}} = \left( C_T \alpha_T\hat{D}_{f}^T + \alpha_P \hat{D}_{f}^P \right) \frac{Z_f e\delta \hat{\phi}}{T_{f0}},
\end{equation}
where $\hat{D}_{f}^T$ and $\hat{D}_{f}^P$ denote the perturbed density response functions of trapped fast ions and passing fast ions, respectively. The constants $\alpha_T$ and $\alpha_P = 1-\alpha_T$ are the relative fractions of trapped and passing fast ions, respectively.  In a large aspect ratio tokamak, $\alpha_T \simeq \sqrt{2\epsilon}$ where $\epsilon = r/R_0$. The appearance of a constant $C_T$ in Eq. (\ref{deltan_f}) results from the tearing parity mode not contributing to the trapped fast ion density response. In this paper, we adopt the definition of parity transformation of fields and distribution function in general toroidal geometry of \cite{SugamaPPCF2011}, along with the classification of instabilities presented in \cite{IshizawaJPP2015}. The ballooning parity mode does not change the sign of the perturbed distribution function and electrostatic potential upon parity transformation. In contrast, the tearing parity mode reverses the sign of the perturbed distribution function and electrostatic potential upon parity transformation. Therefore, $C_T = 1$ for ballooning parity mode and $C_T=0$ for tearing parity mode.
%%%%%%%%%%%%%%%%%%%%%%%%%%%%%%%%%%%%%%%%%%%%%%%%%%%%%%%%%%%%%%%%%%%%%%%%%%%%%%%%%%%%%%%%%%%%%%%%%%%%%%%%%%%%%%%%%%%%%%%%%%%%%%%%%%%%%%%%%%%%%%%%%%%%%%%%%%%%%%%%%%%%%%%%%%%%
%%%%%%%%%%%%%%%%%%%%%%%%%%%%%%%%%%%%%%%%%%%%%%%%%%%%%%%%%%%%%%%%%%%%%%%%%%%%%%%%%%%%%%%%%%%%%%%%%%%%%%%%%%%%%%%%%%%%%%%%%%%%%%%%%%%%%%%%%%%%%%%%%%%%%%%%%%%%%%%%%%%%%%%%%%%%
%%%%%%%%%%%%%%%%%%%%%%%%%%%%%%%%%%%%%%%%%%%%%%%%%%%%%%%%%%%%%%%%%%%%%%%%%%%%%%%%%%%%%%%%%%%%%%%%%%%%%%%%%%%%%%%%%%%%%%%%%%%%%%%%%%%%%%%%%%%%%%%%%%%%%%%%%%%%%%%%%%%%%%%%%%%%
\subsubsection{Trapped fast ion} \label{Subsubsec: Trapped fast ion}
\begin{figure}[h]
\centering
\includegraphics[width=0.35\textwidth ]{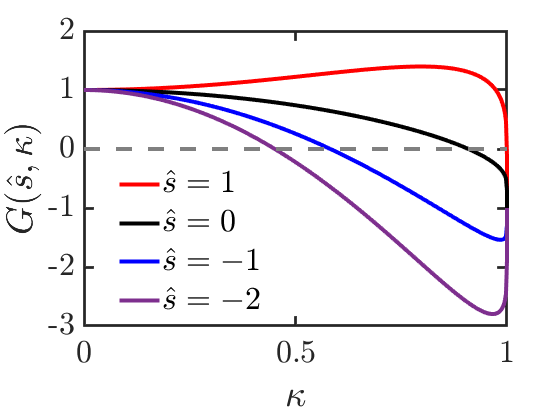}
\caption{Pitch angle dependence of the function $G(\hat{s},\kappa)$ for trapped particles from \cite{KadomtsevSPJ1967}. The $\hat{s}=-2$ case is obtained from our calculation. } 
\label{fig:Gplot}
\end{figure}

This section reviews the dynamics of trapped fast ions and their contribution to the instability. Toroidal precession is the orbit-averaged toroidal drift motion of guiding centers in axisymmetric geometry, with the characteristic frequency defined as \cite{KadomtsevSPJ1967}
\begin{equation}\label{eqn:TP_frequency}
\left\langle \omega_{d\sigma} \right\rangle_b = \omega_{*\sigma} \frac{L_{n\sigma}E_\sigma}{R_0 T_{\sigma 0}} G(\hat{s},\kappa)
\end{equation}
where $E_\sigma=m_\sigma \mathrm{v}^2/2$ and $\kappa = \sqrt{\frac{ 1-\mu_\sigma B_{min}/E_\sigma}{2b_B} }$ is the pitch angle parameter, $\mu_\sigma=m_\sigma \mathrm{v}_\perp^2/2B$ is the magnetic moment, $b_B=(B_{max}-B_{min})/2B_{max}$, $B_{max}$ and $B_{min}$ are the maximum and minimum of $|\bold{B}|$ that a particle experiences during its unperturbed orbit excursion, respectively. $G(\hat{s},\kappa)$ is a dimensionless quantity characterizing the direction and magnitude of the toroidal precession drift. In figure \ref{fig:Gplot}, we can see that precession reversal happens for barely trapped particles ($\kappa \lesssim 1)$ in negative shear plasmas since $\left\langle \omega_{d\sigma} \right\rangle_b \propto G$. Therefore, many fast ions reverse their toroidal precession direction and can resonate with electron drift waves in reversed shear plasmas. 
\\
\noindent This resonance occurs when the toroidal precession motion of trapped particles aligns with the phase velocity of the electron drift wave and can lead to instability.  \\
\indent For $\omega\ll\omega_{bf}$ where $\omega_{bf}$ is a bounce frequency of fast ion, the trapped fast ion dynamics can be described by the bounce-averaged gyrokinetic equation \cite{GangPFB1990, FongPoP1999}. The resonant contribution of trapped fast ion can be expressed as a response function $\hat{D}_{f}^T$. The calculation of $\hat{D}_{f}^T$ is detailed in \cite{KangPoP2019}, and its transformation in ballooning coordinates is outlined in \cite{KangPoP2020}. However, an error in the expression for $\hat{D}_{f}^T$ in Eq. (12) of \cite{KangPoP2020} needs to be corrected: $\sqrt{\hat{E}_{f0}e^{-\hat{E}_{f0}}}$ should be revised to $\sqrt{\hat{E}_{f0}}e^{-\hat{E}_{f0}}$. Consequently, $\hat{D}_{f}^T$ is correctly expressed as
\begin{equation}
\begin{split}
\hat{D}_{f}^T(\Omega,\hat{\theta}) = &-1 -2i\sqrt{\frac{2}{\pi}} \hat{E}_{f0}^{3/2}e^{-\hat{E}_{f0}}\left( 2 + \frac{4}{3} \cos\hat{\theta} \right) \\
&\times \left\{ 1 - \frac{\Omega_{*f}}{\Omega}\left[ 1+ \eta_f \left( \hat{E}_{f0} - \frac{3}{2}\right) \right] \right\}J^2_0\left( \sqrt{2b_f\hat{E}_{f0}}\right),
\end{split}
\end{equation}
where $\Omega=\omega/\omega_{*e}$, $\Omega_{*f} = \omega_{*f}/\omega_{*e}$, $b_f = k_\perp^2 \rho_{\perp f}^2$, and $\hat{E}_{f0}=\frac{\omega_r R_0}{\omega_{*f}L_{nf}G_{avg}}$. $G_{avg}$ is further defined in Eq. (10) and (11) of \cite{KangPoP2020}.
%%%%%%%%%%%%%%%%%%%%%%%%%%%%%%%%%%%%%%%%%%%%%%%%%%%%%%%%%%%%%%%%%%%%%%%%%%%%%%%%%%%%%%%%%%%%%%%%%%%%%%%%%%%%%%%%%%%%%%%%%%%%%%%%%%%%%%%%%%%%%%%%%%%%%%%%%%%%%%%%%%%%%%%%%%%%
%%%%%%%%%%%%%%%%%%%%%%%%%%%%%%%%%%%%%%%%%%%%%%%%%%%%%%%%%%%%%%%%%%%%%%%%%%%%%%%%%%%%%%%%%%%%%%%%%%%%%%%%%%%%%%%%%%%%%%%%%%%%%%%%%%%%%%%%%%%%%%%%%%%%%%%%%%%%%%%%%%%%%%%%%%%%
%%%%%%%%%%%%%%%%%%%%%%%%%%%%%%%%%%%%%%%%%%%%%%%%%%%%%%%%%%%%%%%%%%%%%%%%%%%%%%%%%%%%%%%%%%%%%%%%%%%%%%%%%%%%%%%%%%%%%%%%%%%%%%%%%%%%%%%%%%%%%%%%%%%%%%%%%%%%%%%%%%%%%%%%%%%%
\subsubsection{Passing fast ions}\label{Passing fast ion}
\indent The density response function of passing fast ions can be expressed as:
\begin{equation} \label{DfP}
\hat{D}_{f}^P(\Omega,\hat{\theta}) = -1 + \frac{1}{n_{f0}} \int d^3 \mathrm{v} \frac{\omega - \omega_{*f}^T}{\omega - \omega_{df\mathrm{v}} - k_\parallel \mathrm{v}_\parallel} J_0^2\left(k_{\perp}\rho_{\perp \sigma} \right)f_{Mf}.
\end{equation}
The resonant denominator in Eq. (\ref{DfP}), expressed as $\omega - \omega_{df}\left(\mathrm{v}_\parallel^2/\mathrm{v}_{tf}^2 + \mathrm{v}_\perp^2/2\mathrm{v}_{tf}^2 \right)- k_\parallel \mathrm{v}_\parallel$, can be reformulated by completing the square as:
\begin{equation}\label{eqn:resonance_condition}
\frac{\omega}{\omega_{df}} + \frac{k_\parallel^2 \mathrm{v}_{tf}^2}{4\omega_{df}^2} = \left( \frac{k_\parallel \mathrm{v}_{tf}}{2\omega_{df}} + \frac{\mathrm{v}_{\parallel}}{\mathrm{v}_{tf}} \right)^2 + \frac{\mathrm{v}_\perp^2}{2\mathrm{v}_{tf}^2}.
\end{equation}
The left-hand side of Eq. (\ref{eqn:resonance_condition}) can be negative, while the right-hand side remains positive for all values of $\mathrm{v}$. This indicates that the resonance condition can not be satisfied if $\omega_{df}<0$ $(\because \omega>0)$. Therefore, no particles can resonate if $\omega_{df}<0$ and $\omega>-\frac{k_\parallel^2 \mathrm{v}_{tf}^2}{4\omega_{df}}$. Here, we neglect the imaginary part of $\omega$ to obtain a useful physical insight, which can be justified with $\omega_{r} \gg \omega_i$. Given that the sign of $\omega_{df}$ frequently alternates with changes in $\hat{\theta}$, this will influence the eigenmode structure along the field line. Additionally, this condition suggests that the negative magnetic shear may be more favorable to this type of resonance, a characteristic shared with trapped fast ions, because $\omega_{df}$ can reverse its sign more easily.  \\
\indent Generally, it is not easy to calculate the contribution of the resonant denominator $\omega-\omega_{df\mathrm{v}}-k_\parallel \mathrm{v}_\parallel$. A simplified approach involves retaining one and expanding the other, complicating the evaluation of relative contributions. Instead, we follow the procedure of \cite{BeerPoP1996} to calculate the resonant contribution of the passing particles. For $\Im(\omega/\omega_{*e})>0$, the resonant denominator can be written as
\begin{equation} \label{Resdenominator}
\frac{1}{\omega - \omega_{df\mathrm{v}} - k_\parallel \mathrm{v}_\parallel} = - \frac{i}{\omega_{*e}}\int_0^{\infty} d\tau e^{i\tau (\omega - \omega_{df\mathrm{v}} - k_\parallel \mathrm{v}_\parallel)/\omega_{*e}}.
\end{equation}
This allows the calculation of  $\hat{D}_f^P$ as detailed in the following expression:
\begin{equation} \label{DfP_detail}
\begin{split}
&\hat{D}_{f}^P(\Omega, \hat{\theta}) = -1 \\
& - \frac{i}{\sqrt{2\pi}}\int_0^{\infty}d\tau \int_0^{\infty} d\hat{\mathrm{v}}_{\perp f} \hat{\mathrm{v}}_{\perp f} \int_{-\infty}^{\infty} d\hat{\mathrm{v}}_{\parallel f} \left\{ \Omega -\Omega_{*f} \left[ 1 + \eta_f \left( \frac{\hat{\mathrm{v}}_{\parallel f}^2+\hat{\mathrm{v}}_{\perp f}^2}{2}-\frac{3}{2} \right) \right] \right\} \times \\
&e^{i\tau [\Omega - z_{\parallel f} \hat{\mathrm{v}}_{\parallel f}- \Omega_{df}(\hat{\mathrm{v}}_{\parallel f}^2+\hat{\mathrm{v}}_{\perp f}^2/2)]}e^{-(\hat{\mathrm{v}}_{\parallel f}^2+\hat{\mathrm{v}}_{\perp f}^2)/2}J_0^2\left(\sqrt{b_f}\hat{\mathrm{v}}_{\perp f}\right),
\end{split}
\end{equation}
where $\hat{\mathrm{v}}_{\parallel f}  = \mathrm{v}_\parallel / \mathrm{v}_{tf}$,  $\hat{\mathrm{v}}_{\perp f} = \mathrm{v}_\perp / \mathrm{v}_{tf}$, $z_{\parallel f} =| k_\parallel| \mathrm{v}_{tf}/\omega_{*e}$ and $\Omega_{df}=\omega_{df}/\omega_{*e}$. The integrals over $\hat{\mathrm{v}}_{\perp f}$ in Eq. (\ref{DfP_detail}) are calculated as follows:
\begin{equation}\label{vperp_integral1}
\int_0^{\infty} d\hat{\mathrm{v}}_{\perp f} \hat{\mathrm{v}}_{\perp f} e^{-(1+i\Omega_{df}\tau)\hat{\mathrm{v}}_{\perp f}^2/2} J_0^2\left(\sqrt{b_f}\hat{\mathrm{v}}_{\perp f}\right) = \frac{e^{-b_f/(1+i\Omega_{df}\tau)}}{1+i\Omega_{df}\tau}I_0\left(\frac{b_f}{1+i\Omega_{df}\tau} \right),
\end{equation}
and
\begin{equation}\label{vperp_integral2}
\begin{split}
&\int_0^{\infty} d\hat{\mathrm{v}}_{\perp f} \hat{\mathrm{v}}_{\perp f}^3 e^{-(1+i\Omega_{df}\tau)\hat{\mathrm{v}}_{\perp f}^2/2} J_0^2\left(\sqrt{b_f}\hat{\mathrm{v}}_{\perp f}\right) \\
&= 2 \frac{e^{-b_f/(1+i\Omega_{df}\tau)}}{\left( 1+i\Omega_{df}\tau\right) ^2}I_0\left(\frac{b_f}{1+i\Omega_{df}\tau} \right) \left[ 1- \frac{b_f}{1+i\Omega_{df}\tau} + \frac{b_f}{1+i\Omega_{df}\tau} \frac{I_1\left(\frac{b_f}{1+i\Omega_{df}\tau} \right)}{I_0\left(\frac{b_f}{1+i\Omega_{df}\tau} \right)} \right],
\end{split}
\end{equation}
where $I_0$ and $I_1$ are the modified Bessel functions of the first kind of zeroth order and first order, respectively. The integrals over $\hat{\mathrm{v}}_{\parallel f}$ are calculated as follows:
\begin{equation}\label{vparallel_integral1}
\int_{-\infty}^{\infty} d\hat{\mathrm{v}}_{\parallel f}  e^{-(1+2i\Omega_{df}\tau)\hat{\mathrm{v}}_{\parallel f}^2/2}e^{-iz_{\parallel f} \tau  \hat{\mathrm{v}}_{\parallel f} } = \sqrt{2\pi} \frac{e^{-\tau^2 z_{\parallel f}^2/2(1+2\Omega_{df}\tau)}}{\sqrt{1+2i\Omega_{df}\tau}}, 
\end{equation}
and
\begin{equation}\label{vparallel_integral2}
\int_{-\infty}^{\infty} d\hat{\mathrm{v}}_{\parallel f} \hat{\mathrm{v}}_{\parallel f}^2 e^{-(1+2i\Omega_{df}\tau)\hat{\mathrm{v}}_{f\parallel}^2/2}e^{-iz_{\parallel f} \tau \hat{\mathrm{v}}_{\parallel f} } = \sqrt{2\pi}  \frac{e^{-\tau^2 z_{\parallel f}^2/2(1+2\Omega_{df}\tau)}}{(1+2i\Omega_{df}\tau)^{5/2}}( 1+ 2i\Omega_{df}\tau - \tau^2 z_{\parallel f}^2 ).
\end{equation}
Combining these results, the density response function of passing fast ions $\hat{D}_f^P$ is expressed as
\begin{equation} \label{DfP_decomposition}
\hat{D}_{f}^P = \hat{D}_{f1}^P +\Omega_{*f}\hat{D}_{f2}^P + \eta_f \Omega_{*f}\hat{D}_{f3}^P,
\end{equation}
where $\hat{D}_{f}^P$ is decomposed into three distinct parts, each representing different physical physical phenomena:
\begin{equation} \label{DfP1}
\begin{split}
&\hat{D}_{f1}^P = -1 - i \int_0^{\infty}d\tau e^{i\tau\Omega}e^{-\tau^2z_{\parallel f}^2/2(1+2i\Omega_{df}\tau)}e^{-b_f/(1+i\Omega_{df}\tau)} I_0 \left( \frac{b_f}{1+i\Omega_{df}\tau} \right) \times \\
&\left[ \frac{\Omega}{(1+i\Omega_{df}\tau)\sqrt{1+2i\Omega_{df}\tau}}\right],
\end{split}
\end{equation}
\begin{equation} \label{DfP2}
\begin{split}
&\hat{D}_{f2}^P =  i \int_0^{\infty}d\tau e^{i\tau\Omega}e^{-\tau^2z_{f\parallel }^2/2(1+2i\Omega_{df}\tau)}e^{-b_f/(1+i\Omega_{df}\tau)} I_0 \left( \frac{b_f}{1+i\Omega_{df}\tau} \right) \times \\
&\left[ \frac{1}{(1+i\Omega_{df}\tau)\sqrt{1+2i\Omega_{df}\tau}}\right],
\end{split}
\end{equation}
and
\begin{equation} \label{DfP3}
\begin{split}
&\hat{D}_{f3}^P = - i \int_0^{\infty}d\tau e^{i\tau\Omega}e^{-\tau^2z_{ \parallel f}^2/2(1+2i\Omega_{df}\tau)}e^{-b_f/(1+i\Omega_{df}\tau)} I_0 \left( \frac{b_f}{1+i\Omega_{df}\tau} \right) \times \\
&\left\{ \frac{3/2}{(1+i\Omega_{df}\tau)\sqrt{1+2i\Omega_{df}\tau}}\right. \\
&- \left[ \frac{1-\frac{b_f}{1+i\Omega_{df}\tau} + \frac{b_f}{1+i\Omega_{df}\tau} I_1\left( \frac{b_f}{1+i\Omega_{df}\tau}\right)/I_0\left( \frac{b_f}{1+i\Omega_{df}\tau}\right) }{(1+i\Omega_{df}\tau)^2\sqrt{1+2i\Omega_{df}\tau}}  \right] \\
& \left. - \left[ \frac{1+2i\Omega_{df}\tau-\tau^2z_{\parallel f}^2}{2(1+i\Omega_{df}\tau)(1+2i\Omega_{df}\tau)^{5/2}} \right] \right\}.
\end{split}
\end{equation}
$\hat{D}_{f1}^P$ corresponds to the adiabatic response and Landau damping. $\hat{D}_{f2}^P$ and $\hat{D}_{f3}^P$ are associated with the equilibrium density gradient and temperature gradient drive, respectively. With $\Omega_{*f}<0$, $\hat{D}_{f3}^P$ is particularly crucial for the stability analysis, given its potential to yield a positive imaginary value of $\Omega$ under the influence of $\eta_f$.\\
\indent  %Since $\hat{D}_f^P$ includes complicated integrals, it is difficult to get a physical interpretation from it. 
A useful insight can be obtained by simplifying this problem by assuming the slab limit ($\Omega_{df}\rightarrow 0$). Eqs. (\ref{DfP1}) - (\ref{DfP3}) can be considerably simplified with the usual plasma dispersion function Z \cite{Fried1961}. Then, we write 
\begin{equation} \label{DfP1_slablimit}
\hat{D}_{f1}^P  = -1 - e^{-b_f}I_0(b_f)\frac{\Omega}{\sqrt{2}z_{\parallel f}} Z \left( \frac{\Omega}{\sqrt{2}z_{\parallel f}} \right),
\end{equation}
\begin{equation} \label{DfP2_slablimit}
\hat{D}_{f2}^P = \frac{e^{-b_f}I_0(b_f)}{\sqrt{2}z_{\parallel f}} Z \left( \frac{\Omega}{\sqrt{2}z_{\parallel f}} \right),
\end{equation}
and
\begin{equation} \label{DfP3_slablimit}
\hat{D}_{f3}^P  = -e^{-b_f}I_0(b_f) \left\{ \frac{1}{\sqrt{2}z_{\parallel f}} Z \left( \frac{\Omega}{\sqrt{2}z_{\parallel f}} \right)\left[b_f\left( 1 - \frac{I_1(b_f)}{I_0(b_f)}\right) +   \frac{z_{\parallel f}^2-\Omega^2}{2z_{\parallel f}^2} \right] -\frac{\Omega}{2z_{\parallel f}^2} \right\}.
\end{equation}
The linear growth rate of the instability is closely related to the imaginary parts of these functions. $\Im(\hat{D}_{f3}^P)$ must be negative to make the mode unstable given that both $\Im(\hat{D}_{f1}^P)$ and $\Im(\hat{D}_{f2}^P)\Omega_{*f}$ are negative. For simplicity, we have neglected the imaginary part of $\Omega$ ($\because |\Re(\Omega)|\gg |\Im(\Omega)|$). Then, the necessity for the instability can be expressed as:
\begin{equation} \label{Df3condition_slablimit}
\frac{\Omega^2}{z_{\parallel f}^2}<1+2b_f\left( 1-\frac{I_1(b_f)}{I_0(b_f)} \right).
\end{equation}
This condition essentially takes the form $\omega^2 < (\text{const.}) k_\parallel^2 \mathrm{v}_{tf}^2$, where the constant is positive and expected to be in the order of unity. This implies that the thermal speed of fast ions should be faster than the parallel propagation speed of the mode, a condition well satisfied for high-temperature fast ions. Using the inequality (\ref{Df3condition_slablimit}), the instability condition $\Im(\hat{D}_f^P)>0$ is expressed in terms of $\eta_f$ as
\begin{equation} \label{ins_condition1}
\eta_f > \frac{1-\frac{\Omega }{\Omega_{*f}}}{ b_f\left( 1- \frac{I_1(b_f)}{I_0(b_f)}\right) + \frac{1}{2}\left(1-\frac{\Omega^2}{z_{\parallel f}^2}\right)} 
\end{equation}
Therefore, the mode becomes unstable when $\eta_f$ exceeds a critical value, a property similar to that of trapped fast ions. The theoretical model in \cite{KangPoP2019} also claims that the instability occurs when $\eta_f$ exceeds a critical value, i.e., when the fast ion temperature profile is more peaked than the density profile.\\ 
%%%%%%%%%%%%%%%%%%%%%%%%%%%%%%%%%%%%%%%%%%%%%%%%%%%%%%%%%%%%%%%%%%%%%%%%%%%%%%%%%%%%%%%%%%%%%%%%%%%%%%%%%%%%%%%%%%%%%%%%%%%%%%%%%%%%%%%%%%%%%%%%%%%%%%%%%%%%%%%%%%%%%%%%%%%%
%%%%%%%%%%%%%%%%%%%%%%%%%%%%%%%%%%%%%%%%%%%%%%%%%%%%%%%%%%%%%%%%%%%%%%%%%%%%%%%%%%%%%%%%%%%%%%%%%%%%%%%%%%%%%%%%%%%%%%%%%%%%%%%%%%%%%%%%%%%%%%%%%%%%%%%%%%%%%%%%%%%%%%%%%%%%
%%%%%%%%%%%%%%%%%%%%%%%%%%%%%%%%%%%%%%%%%%%%%%%%%%%%%%%%%%%%%%%%%%%%%%%%%%%%%%%%%%%%%%%%%%%%%%%%%%%%%%%%%%%%%%%%%%%%%%%%%%%%%%%%%%%%%%%%%%%%%%%%%%%%%%%%%%%%%%%%%%%%%%%%%%%%
\subsection{BACKGROUND MAIN ION AND ELECTRON DYNAMICS}
\indent For the background main ion, we adopt a fluid ion approximation and consider a long perpendicular wavelength regime. Expanding Eq. (\ref{ghat}) to the first order in $\frac{\omega_{di}}{\omega}$ and to the second order in $\frac{k_\parallel \mathrm{v}_\parallel}{\omega}$, and integrating it in velocity space yields
\begin{equation}\label{deltan_i}
\frac{ \delta n_{i}  } { n_{i0}} = \hat{D}_{i} \frac{e\delta \hat{\phi}}{T_{i0}}.
\end{equation}
Here, $\hat{D}_i$ is calculated as
\begin{equation}\label{Di}
\hat{D}_{i} (\Omega,\hat{\theta}) = \frac{T_{i0} L_{ne}}{\Omega T_{e0} L_{ni}} + \left[ 1 + (1+\eta_i)\frac{ T_{i0}L_{ne} }{\Omega T_{e0}L_{ni} } \right] \left( -b_i + 2 \frac{\Omega_{di}}{\Omega} + \frac{z_{\parallel i}^2}{\Omega^2} \right),
\end{equation}
where $b_i = k_\perp^2 \rho_{ti}^2$, $\Omega_{di} = \omega_{di}/\omega_{*e}$ and $z_{ \parallel i} = |k_\parallel| \mathrm{v}_{ti} / \omega_{*e}$. The detailed calculation for $\hat{D}_{i}$ was shown in Eqs. (21) - (27) of  \cite{KangPoP2019}, so we don't repeat it here.\\
\indent To examine the electron response, we assume an adiabatic response for passing electrons since $\omega \sim \omega_{*e} \ll k_\parallel \mathrm{v}_{te}$. The trapped electron response is assumed to be negligible since we focus on the plasma parameter space where conventional microinstabilities like ITG mode and TEM remain stable. The electron density response is then calculated as:
\begin{equation}\label{deltan_e}
\frac{\delta n_{e}}{n_{e0}} = \frac{e\delta\hat{\phi}}{T_{e0}}.
\end{equation}
%%%%%%%%%%%%%%%%%%%%%%%%%%%%%%%%%%%%%%%%%%%%%%%%%%%%%%%%%%%%%%%%%%%%%%%%%%%%%%%%%%%%%%%%%%%%%%%%%%%%%%%%%%%%%%%%%%%%%%%%%%%%%%%%%%%%%%%%%%%%%%%%%%%%%%%%%%%%%%%%%%%%%%%%%%%%
%%%%%%%%%%%%%%%%%%%%%%%%%%%%%%%%%%%%%%%%%%%%%%%%%%%%%%%%%%%%%%%%%%%%%%%%%%%%%%%%%%%%%%%%%%%%%%%%%%%%%%%%%%%%%%%%%%%%%%%%%%%%%%%%%%%%%%%%%%%%%%%%%%%%%%%%%%%%%%%%%%%%%%%%%%%%
%%%%%%%%%%%%%%%%%%%%%%%%%%%%%%%%%%%%%%%%%%%%%%%%%%%%%%%%%%%%%%%%%%%%%%%%%%%%%%%%%%%%%%%%%%%%%%%%%%%%%%%%%%%%%%%%%%%%%%%%%%%%%%%%%%%%%%%%%%%%%%%%%%%%%%%%%%%%%%%%%%%%%%%%%%%%
\subsection{EIGENMODE EQUATION IN THE BALLOONING COORDINATE} \label{Subsec: Eigenmode equation}
\indent By combining Eqs. (\ref{deltan_f}), (\ref{deltan_i}) and (\ref{deltan_e}), the quasi-neutrality condition $\delta n_e = \delta n_i + Z_f \delta n_f$ can be written as
\begin{equation}\label{EMequation0}
\left[ 1-\frac{n_{i0}}{n_{e0}}\frac{T_{e0}}{T_{i0}} \hat{D}_{i} - Z_f\frac{n_{f0}}{n_{e0}} \frac{T_{e0}}{T_{f0}} \left( C_T \alpha_T \hat{D}_{f}^T + \alpha_P \hat{D}_{f}^P \right) \right]\hat{\phi} = 0,
\end{equation}
where $\hat{\phi}=\frac{e\delta\hat{\phi}}{T_{e0}}$. Eq. (\ref{EMequation0}) can be transformed into a differential equation by rewriting $k_\parallel^2$ in $\hat{D}_i$ as $-\frac{1}{q^2R_0^2}\frac{\partial^2 }{\partial \hat{\theta}^2}$. It can be formulated as follows:
\begin{equation}\label{EMequation}
\left[ \frac{1}{\Omega^2 p_s^2} \frac{\partial^2}{\partial\hat{\theta}^2} + Q(\Omega,\hat{\theta}) - W(\Omega,\hat{\theta}) \right] \hat{\phi}(\hat{\theta}) = 0.
\end{equation}
Here, $Q(\Omega,\hat{\theta})$ corresponds to a potential function from the background main ion response and can be expressed as:
\begin{equation}\label{Q}
\begin{split}
&Q(\Omega,\hat{\theta}) = \\
&\left( \frac{n_{e0}}{n_{i0} } - \frac{L_{ne}}{\Omega L_{ni}}\right) \left[ 1 + (1+\eta_i)\frac{T_{i0}L_{ne}}{\Omega T_{e0} L_{ni}}\right]^{-1}  + b_{\theta} (1+\hat{s}^2\hat{\theta}^2) + \frac{2\epsilon_{ne}}{\Omega}(\cos{\hat{\theta}} + \hat{s}\hat{\theta}\sin{\hat{\theta}}),
\end{split}
\end{equation}
where $p_{s}=qb_{\theta}^{1/2}/\epsilon_{ne}$, $\epsilon_{ne} = L_{ne}/R_0$, $b_{\theta} = k_\theta^2 \rho_s^2$, $\rho_s = \frac{c_s}{\Omega_{ci}}$ and $c_s = \sqrt{T_{e0}/m_i}$. $Q$ can be decomposed into two parts:
\begin{equation}
Q(\Omega,\hat{\theta}) = Q_{slab}(\Omega,\hat{\theta})  + Q_{tor}(\Omega,\hat{\theta}),
\end{equation}
where 
\begin{equation}
Q_{slab}(\Omega,\hat{\theta}) =\left( \frac{n_{e0}}{n_{i0} } - \frac{L_{ne}}{\Omega L_{ni}}\right) \left[ 1 + (1+\eta_i)\frac{T_{i0}L_{ne}}{\Omega T_{e0} L_{ni}}\right]^{-1}  + b_{\theta} (1+\hat{s}^2\hat{\theta}^2)
\end{equation}
and
\begin{equation}
Q_{tor}(\Omega,\hat{\theta}) =\frac{2\epsilon_{ne}}{\Omega}(\cos{\hat{\theta}} + \hat{s}\hat{\theta}\sin{\hat{\theta}}).
\end{equation}
$Q_{slab}$ represents the slab-like potential, and it exists even if toroidal mode coupling is absent. $Q_{tor}$ corresponds to the toroidicity-induced potential, representing the toroidal mode coupling effect. \\
$W$ represents the contribution from the fast ion response and is written as:
\begin{equation}\label{W}
W(\Omega,\hat{\theta}) =\lambda_{f} \left[ C_T\alpha_T D_{f}^T(\Omega, \hat{\theta}) + \alpha_P D_{f}^P(\Omega,\hat{\theta}) \right],
\end{equation}
where $\lambda_{f} = Z_f^2 \frac{n_{f0}}{n_{i0}}\frac{T_{e0}}{T_{f0}} \left[ 1 + (1+\eta_i)\frac{T_{i0}L_{ne}}{\Omega T_{e0} L_{ni}}\right]^{-1}$. Note that $-Q+W$ corresponds to the effective potential in Eq. (\ref{EMequation}). Given that $\lambda_{f} \ll 1$, $W$ becomes  a minor perturbation to $Q$, i.e., $|W/Q|\ll 1$. This allows us to apply the WKB approximation for $k_\parallel$ within $W$, such that $k_\parallel^2 = Q/q^2R_0^2$. Given that the effective potential is symmetric about $\hat{\theta}=0$, the eigenfunctions will necessarily be symmetric or anti-symmetric. As discussed in Sec. \ref{Subsec: Fast ion dynamics}, these eigenfunctions are classified as ballooning parity mode (symmetric) and tearing parity mode (anti-symmetric), respectively \cite{IshizawaJPP2015}. The eigenmode equation Eq. (\ref{EMequation}) is solved using the shooting method to calculate the eigenvalue $\Omega$ and eigenfunction $\hat{\phi}$ for both parity modes. Note that the general solution of Eq. (\ref{EMequation}) is a linear sum of the ballooning parity and tearing parity modes. \\
\indent Typically, quantum numbers are closely related to the eigenvalues and eigenfunctions of differential equations like Eq. (\ref{EMequation}). However, specifying the quantum number using the shooting method is difficult because it only provides proper eigenvalue and eigenfunction pairs. In principle, the WKB eigenvalue condition is determined by the phase integral methods \cite{WhiteJCP1979} in complex ballooning angle space. Between the two appropriate zeros of the effective potential function, the quantization condition can be expressed as:
\begin{equation}\label{DSrelation}
\Omega p_s \int_{-\hat{\theta}_t}^{\hat{\theta}_t} d\hat{\theta}\sqrt{Q(\Omega, \hat{\theta})-W(\Omega,\hat{\theta})} = \left(l+\frac{1}{2}\right) \pi,
\end{equation}
where $l=0,1,2, \ldots$ is a quantum number, and $\hat{\theta}_t$ is a turning point where $Q-W=0$. The integration in Eq. (\ref{DSrelation}) is performed along the anti-Stokes line, a path in the complex ballooning angle space where $\int d\hat{\theta}\sqrt{Q-W}$ remains real. According to the Phase Integral Methods, a solution with particular parity for $\hat{\theta}\rightarrow -\hat{\theta}$ can be chosen when the differential equation is symmetric about $\hat{\theta}=0$. An odd quantum number corresponds to an odd solution, and an even quantum number corresponds to an even solution. Unfortunately, determining the quantum number for this problem is challenging for several reasons. Firstly, $W$ is not well-defined in complex ballooning angle space due to Eq. (\ref{Resdenominator}). Additionally, the oscillatory property of $Q_{tor}$ makes it difficult to specify a proper pair of turning points, which will be discussed in Sec. \ref{Sec: Linear stability analysis}. The presence of unstable modes that do not follow Eq. (\ref{DSrelation}), such as the Bloch state (resulting from periodic behavior of potential) and modes with non-zero Bloch angles, complicates the determination of the quantum number. Therefore, we focus on providing a qualitative description of the eigenstate transition rather than calculating the exact quantum number. To avoid confusion, we clearly define the following terms. The ground state refers to the eigenstate with the lowest value of $\Omega_r$ among all the eigenstates. On the other hand,  a non-ground state corresponds to any eigenstate with a value of $\Omega_r$ greater than that of the ground state. Furthermore, we define the meaning of the order of an eigenstate such that a higher-order state corresponds to an eigenstate with a higher value of $\Omega_r$.
\clearpage
%%%%%%%%%%%%%%%%%%%%%%%%%%%%%%%%%%%%%%%%%%%%%%%%%%%%%%%%%%%%%%%%%%%%%%%%%%%%%%%%%%%%%%%%%%%%%%%%%%%%%%%%%%%%%%%%%%%%%%%%%%%%%%%%%%%%%%%%%%%%%%%%%%%%%%%%%%%%%%%%%%%%%%%%%%%%
%%%%%%%%%%%%%%%%%%%%%%%%%%%%%%%%%%%%%%%%%%%%%%%%%%%%%%%%%%%%%%%%%%%%%%%%%%%%%%%%%%%%%%%%%%%%%%%%%%%%%%%%%%%%%%%%%%%%%%%%%%%%%%%%%%%%%%%%%%%%%%%%%%%%%%%%%%%%%%%%%%%%%%%%%%%%
%%%%%%%%%%%%%%%%%%%%%%%%%%%%%%%%%%%%%%%%%%%%%%%%%%%%%%%%%%%%%%%%%%%%%%%%%%%%%%%%%%%%%%%%%%%%%%%%%%%%%%%%%%%%%%%%%%%%%%%%%%%%%%%%%%%%%%%%%%%%%%%%%%%%%%%%%%%%%%%%%%%%%%%%%%%%
\section{LINEAR STABILITY ANALYSIS}\label{Sec: Linear stability analysis}
In this section, we solve the eigenmode equation Eq. (\ref{EMequation}) using the shooting method. The solutions are required to decay at the infinity, i.e., $\hat{\phi}\rightarrow 0$ as $|\hat{\theta}|\rightarrow \infty$. The initial values are set as follows: $\hat{\phi}(0) = 1$, $\hat{\phi}'(0)=0$ for the ballooning pairty modes, and $\hat{\phi}(0) = 0$, $\hat{\phi}'(0)=1$ for the tearing parity modes, where $\hat{\phi}'=\partial \hat{\phi}/\partial \hat{\theta}$. For the ballooning parity mode, $\hat{\phi}(\hat{\theta})$ is further normalized by a real constant to make $\max(|\hat{\phi}_r|,|\hat{\phi}_i|)=1$. For the tearing parity mode, $\hat{\phi}(\hat{\theta})$ is further normalized by the maximum complex value of $\hat{\phi}(\hat{\theta})$. \\
\indent The solution calculated by the shooting method describes well the average behavior of the instability, but the details of a mode at a given $\bold{k}$, such as the linear growth rate, mainly related to $k_\parallel$ might deviate from the exact solution. The difficulty lies in the dependence of $\gamma_{lin}$ on $\Im(W)$, which varies with $k_\parallel$. Recall that $k_\parallel$ in $W$ comes from the differential operator $\nabla_\parallel$ on $\hat{g}_f$. This term is easily handled with the WKB approximation, where $k_\parallel$ is treated as a number. The validity of the local treatment of $k_\parallel$ requires that the scale length of $Q$ is much longer than that of $\hat{\phi}$, which is not always strictly satisfied. Therefore, comparing the analytical prediction with the results of gyrokinetic codes would be desirable.  For this purpose, we perform gyrokinetic simulations using the GKV code \cite{WatanabeNF2006}. GKV is a Vlasov simulation code based on $\delta f$ gyrokinetic equations in a local flux-tube domain where the background densities, temperatures, and their gradients are fixed. It employs Clebsh-type coordinates $(x,y,z)$ \cite{Haeseleer1991}, which are referred to as the radial, field-line-label, and field-aligned coordinates, respectively, and can include various physics such as kinetic electrons, electromagnetic effects, collision, and multi-species ions. In the electrostatic limit, GKV solves the following equation: 
\begin{equation}\label{GKVequation}
\begin{split}
&\frac{\partial \delta f_\sigma}{\partial t} + \mathrm{v}_\parallel \bold{b}\cdot \nabla\delta f_{\sigma} + \frac{c}{B_0} \{ \Phi_\sigma, \delta f_\sigma\} + \bold{\mathrm{v}}_{d\sigma } \cdot \nabla \delta f_\sigma - \mu_\sigma (\bold{b}\cdot \nabla \Omega_{c \sigma} ) \frac{\partial \delta f_\sigma }{\partial \mathrm{v}_\parallel} \\
&= (\bold{\mathrm{v}}_{*\sigma } - \bold{\mathrm{v}}_{d\sigma } - \mathrm{v}_\parallel \bold{b} ) \cdot \frac{e\nabla \Phi_\sigma }{T_{\sigma}} f_{M\sigma } + C(\delta f_\sigma ),
\end{split}
\end{equation}
where $\bold{b}$ is the unit vector parallel to the magnetic field, $\left\{ \Phi_\sigma , \delta f_\sigma  \right\}=\frac{\partial \Phi_\sigma }{\partial x}\frac{\partial \delta f_\sigma }{\partial y} - \frac{\partial \Phi}{\partial y} \frac{\partial \delta f}{\partial x}$, $\bold{\mathrm{v}_{d\sigma }}$ is the magnetic drift velocity, $\bold{\mathrm{v}}_{*\sigma }$ is the diamagnetic velocity due to the equilibrium pressure gradient, $\Omega_{c\sigma }$ is the cyclotron frequency, $B_0$ is magnetic field strength on the magnetic axis, $\Phi_\sigma $ is the electrostatic potential averaged over the gyromotion, $f_{M\sigma }$ is the Maxwellian distribution function, and $C(\delta f_\sigma )$ is the collision term. Further details about Eq. (\ref{GKVequation}) are described in \cite{WatanabeNF2006}.\\
\indent Although many linear properties of this instability were investigated by GKW code \cite{PeetersCPC2009} in \cite{KangPoP2020}, it has been found that the version of GKW code used in \cite{KangPoP2020} was not able to accurately calculate the real frequency $\omega_r$ of the tearing parity mode for the electrostatic instability. The GKW code calculates $\omega_r$ from the phase angle of the field components. For simplicity, assuming the field component is expressed as $\hat{\phi}(z,t)=\hat{\phi}(z)e^{-i\omega t}$ where $\omega = \omega_r + i\gamma_{lin}$. In the electrostatic limit, the GKW code calculates $\omega_r$ as:
\begin{equation}
\omega_{r}(t) = \left[ Arg \left( \int_{-n_\theta\pi}^{n_\theta\pi} dz\hat{\phi}(z,t+\Delta t) \right) - Arg \left( \int_{-n_\theta\pi}^{n_\theta\pi} dz\hat{\phi}(z,t) \right) \right]/\Delta t,
\end{equation}
where $n_\theta$ is an integer associated with the simulation domain of field-aligned coordinate. In this case, the tearing parity component of $\hat{\phi}$ vanishes due to the symmetric integration of $\hat{\phi}$ along the field line. Therefore, $\omega_r$ calculated by GKW code in \cite{KangPoP2020} only included $\omega_r$ of the ballooning parity mode. This can be problematic when the tearing parity mode becomes the most unstable. Note that the GKW code properly calculated the linear growth rate $\gamma_{lin}$ since it is calculated from $|\hat{\phi}|^2$. In contrast, the GKV code calculates the $\omega_r$ from the inner product of the field components as follows:
\begin{equation}
\omega(t) = -\ln \left[ \int_{-n_\theta\pi}^{n_\theta\pi} dz\hat{\phi}^*(z,t)\hat{\phi}(z,t+\Delta t) / \int_{-n_\theta\pi}^{n_\theta\pi} dz \left| \hat{\phi}(z,t) \right|^2 \right] / i\Delta t,
\end{equation}
where the superscript $*$ indicates the complex conjugate. Due to the inner product of two field components, the tearing parity component of $\hat{\phi}$ can survive after the symmetric integration along the field line. Thus, the GKV code can properly calculate $\omega_r$, making it more suitable for investigating the fast ion-driven drift instability. \\
\indent From Eq. (\ref{EMequation}), we can deduce that the eigenmode structure is primarily influenced by the potential $Q$ since $|Q|\gg|W|$. $Q_{slab}$ dominates as $|\hat{\theta}|\rightarrow \infty$ because $Q_{slab}\rightarrow b_\theta \hat{s}^2 \hat{\theta}^2$ and $Q_{tor} \rightarrow \frac{2\epsilon_{ne}}{\Omega_r}\hat{s}\hat{\theta}$. In contrast, $Q_{tor}$ becomes dominant (or comparable) in the inner region. Therefore, $Q_{tor}$ modulates on the anti-well potential structure of $Q_{slab}$. The plasma parameters related to these potentials, such as $\hat{s}$ and $b_\theta$, will play crucial roles in forming the eigenmode structure. The safety factor $q$ can significantly affect the instability because it is closely associated with the parallel dynamics $k_\parallel$. Furthermore, $q$ can contribute to the eigenmode structure because it is related to the depth of the potential well structure. Recall that the derivative operator in Eq. (\ref{EMequation}) is divided by $\Omega^2 p_s^2$ and $p_s\propto q$. It is easier to understand this with the WKB solution of Eq. (\ref{EMequation}), which is expressed as
\begin{equation}\label{WKBsolution}
\hat{\phi}_{WKB} = \exp\left[ \pm i\Omega p_s\int^{\hat{\theta}} d\hat{\theta}  (Q-W)^{1/2}\right].
\end{equation}
Here, we employ the WKB solution for qualitative description because it can accurately describe the general behavior of the solution, except near the turning points. The eikonal of the eigenfunction is related to the phase in the ballooning coordinate $\hat{\theta}$ and proportional to $\Omega q$. As $\Omega_r q$ increases, the mode oscillates more frequently as $\hat{\theta}$ varies, corresponding to the mode structure of the high-order state in a general sense. Therefore, in this section, we focus on how these parameters $\hat{s}$, $q$, and $k_\theta$ influence the eigenvalue and eigenmode structure of the instability. Additionally, we further investigate the effects of other parameters, such as $T_{f0}$, $n_{f0}$, $R_0/L_{Tf}$, and $R_0/L_{nf}$, on the fast ion-driven drift instability.\\
\indent For the GKV simulation, we consider a collisionless plasma, with the equilibrium modeled as circular concentric flux surfaces \cite{LapillonnePoP2009} with $r/a = 0.5$ and $R_0/a=3$. The plasma consists of adiabatic electrons, deuterium bulk ions, and $\alpha$ particles. The simulation uses 30 grid points per poloidal turn along the field line. The velocity space is discretized into 64 points for the parallel velocity and 32 points for the magnetic moment. Simulation parameters are varied appropriately depending on the specific objectives of each simulation. The shooting method uses the same parameter settings as the GKV simulations for consistency. \\
%%%%%%%%%%%%%%%%%%%%%%%%%%%%%%%%%%%%%%%%%%%%%%%%%%%%%%%%%%%%%%%%%%%%%%%%%%%%%%%%%%%%%%%%%%%%%%%%%%%%%%%%%%%%%%%%%%%%%%%%%%%%%%%%%%%%%%%%%%%%%%%%%%%%%%%%%%%%%%%%%%%%%%%%%%%%
%%%%%%%%%%%%%%%%%%%%%%%%%%%%%%%%%%%%%%%%%%%%%%%%%%%%%%%%%%%%%%%%%%%%%%%%%%%%%%%%%%%%%%%%%%%%%%%%%%%%%%%%%%%%%%%%%%%%%%%%%%%%%%%%%%%%%%%%%%%%%%%%%%%%%%%%%%%%%%%%%%%%%%%%%%%%
%%%%%%%%%%%%%%%%%%%%%%%%%%%%%%%%%%%%%%%%%%%%%%%%%%%%%%%%%%%%%%%%%%%%%%%%%%%%%%%%%%%%%%%%%%%%%%%%%%%%%%%%%%%%%%%%%%%%%%%%%%%%%%%%%%%%%%%%%%%%%%%%%%%%%%%%%%%%%%%%%%%%%%%%%%%%
\subsection{Magnetic shear variation}\label{Subsec: Magnetic shear variation}
\begin{figure}[htb!]
\centering
\subfigure[]{ \includegraphics[width=0.35\textwidth ]{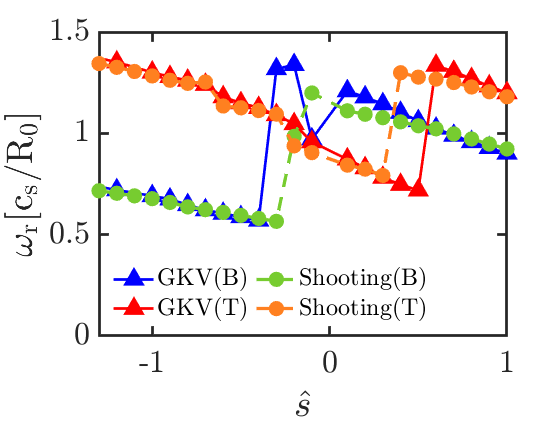}\label{fig:shearscan_ky0.2_omega_r}}
\subfigure[]{ \includegraphics[width=0.35\textwidth ]{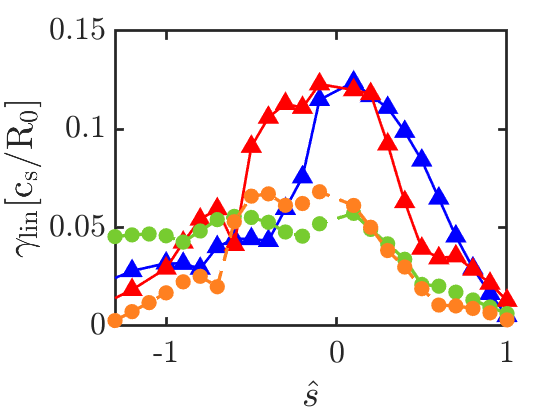} \label{fig:shearscan_ky0.2_omega_i}}
\caption{ (a) Real frequency ($\omega_r$) and (b) the linear growth rate ($\gamma_{lin}$) as functions of $\hat{s}$ of the most unstable ballooning parity mode and tearing parity mode.  }
\label{fig:shearscan_ky0.2_omega}
\end{figure}

\begin{figure}[htb!]
\centering
\subfigure[]{ \includegraphics[width=0.24\textwidth ]{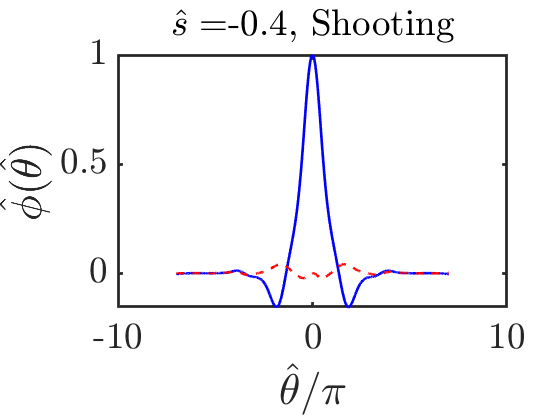}\label{fig:shear-0.4_phiz_B_MU}}
\subfigure[]{ \includegraphics[width=0.24\textwidth ]{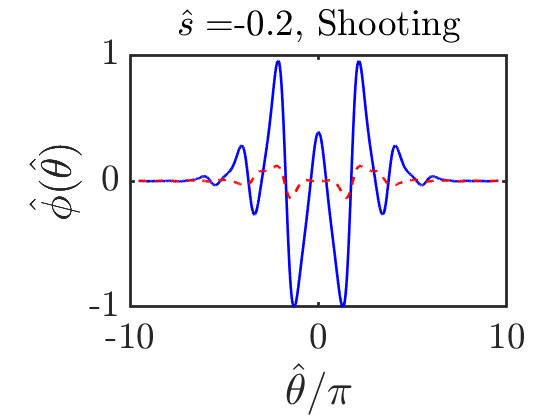}\label{fig:shear-0.2_phiz_B_MU}}
\subfigure[]{ \includegraphics[width=0.24\textwidth ]{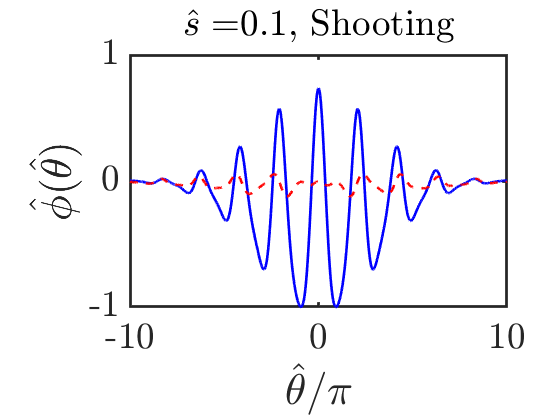}\label{fig:shear0.1_phiz_B_MU}}
\hfill
\subfigure[]{ \includegraphics[width=0.24\textwidth ]{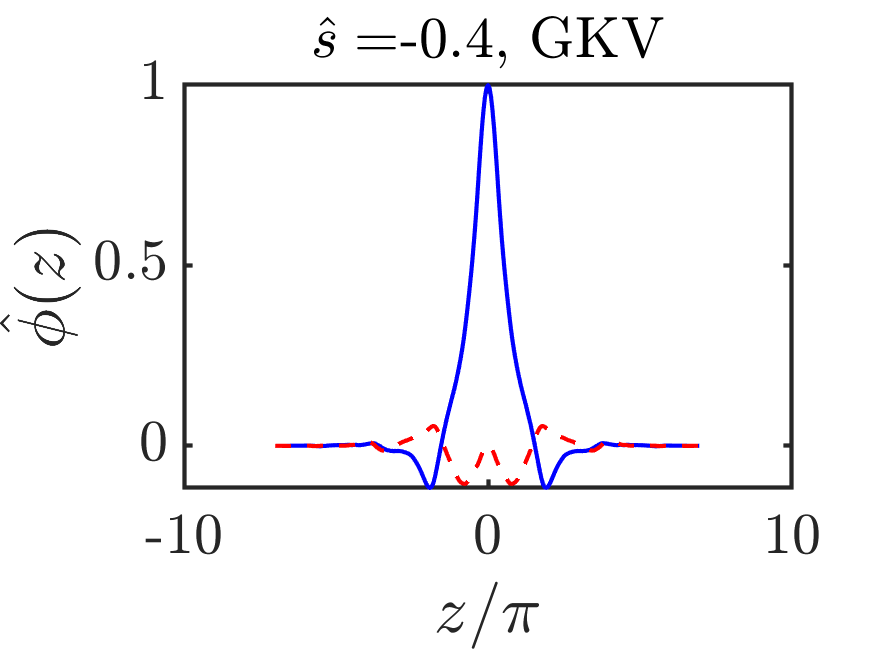}\label{fig:shear-0.4_phi_z1_GKV_even}}
\subfigure[]{ \includegraphics[width=0.24\textwidth ]{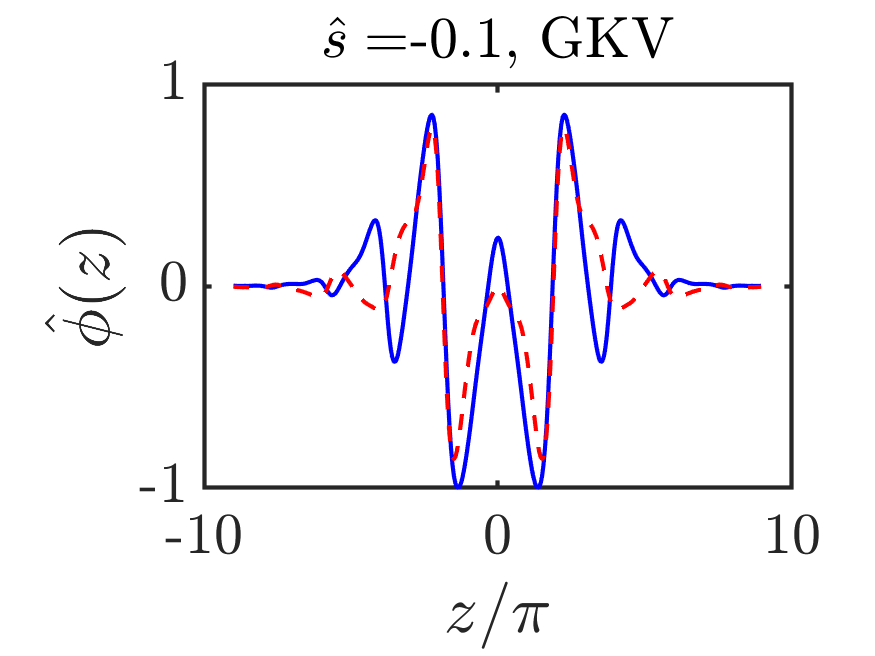}\label{fig:shear-0.1_phi_z1_GKV_even}}
\subfigure[]{ \includegraphics[width=0.24\textwidth ]{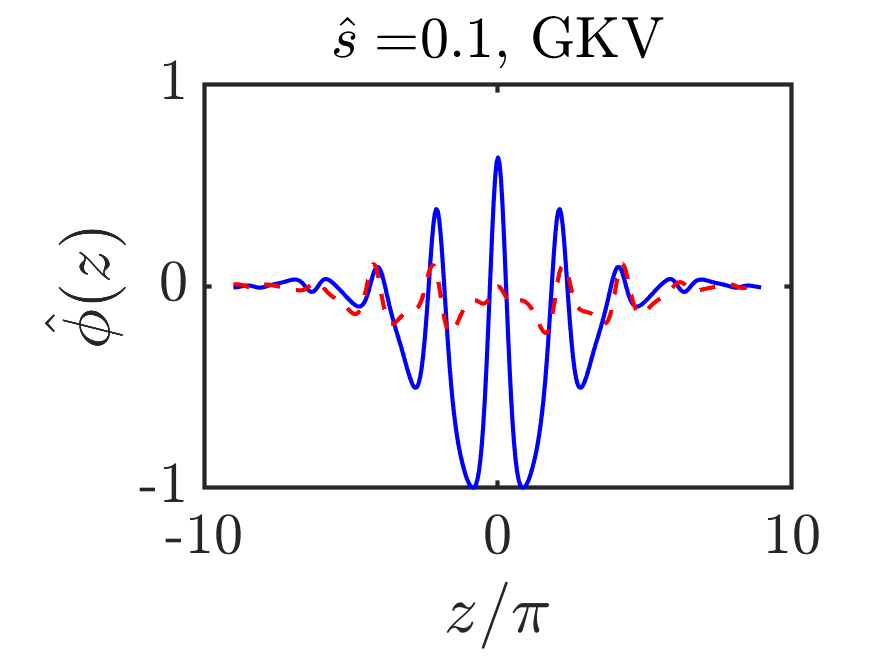}\label{fig:shear0.1_phi_z1_GKV_even}}
\caption{ Eigenmode structures of the most unstable ballooning parity modes calculated by the shooting method and GKV code. The blue solid line and red dotted line correspond to the real and imaginary parts of the eigenmode, respectively.   }
\label{fig:phiz_B_shearscan_ky0.2}
\end{figure}

\begin{figure}[htb!]
\centering
\subfigure[]{ \includegraphics[width=0.24\textwidth ]{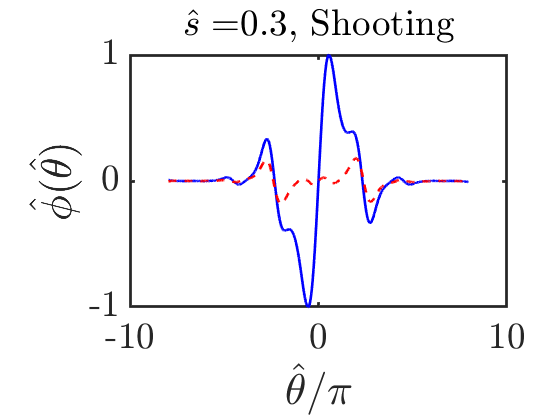}\label{fig:shear0.3_phiz_T_MU}}
\subfigure[]{ \includegraphics[width=0.24\textwidth ]{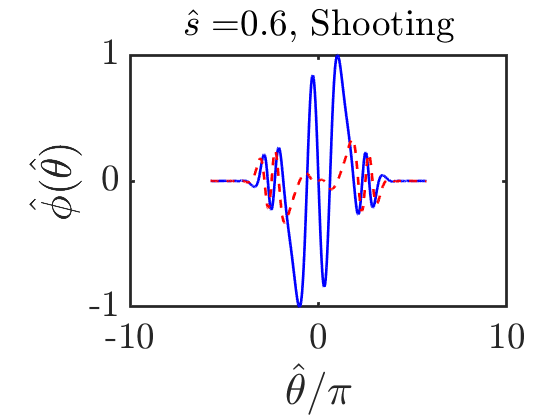}\label{fig:shear0.6_phiz_T_MU}}\\
\subfigure[]{ \includegraphics[width=0.24\textwidth ]{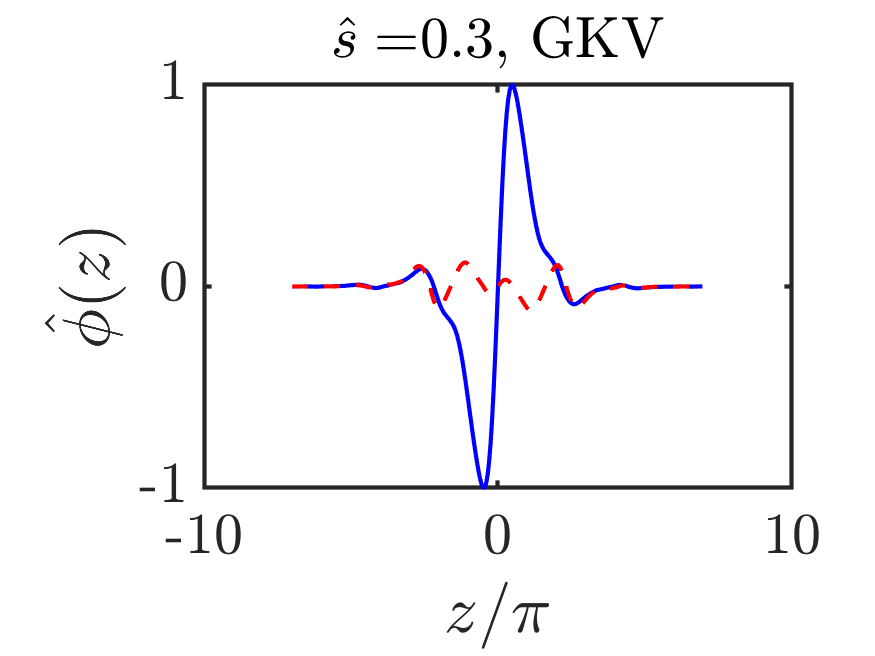}\label{fig:shear0.3_phi_z2_GKV_odd}}
\subfigure[]{ \includegraphics[width=0.24\textwidth ]{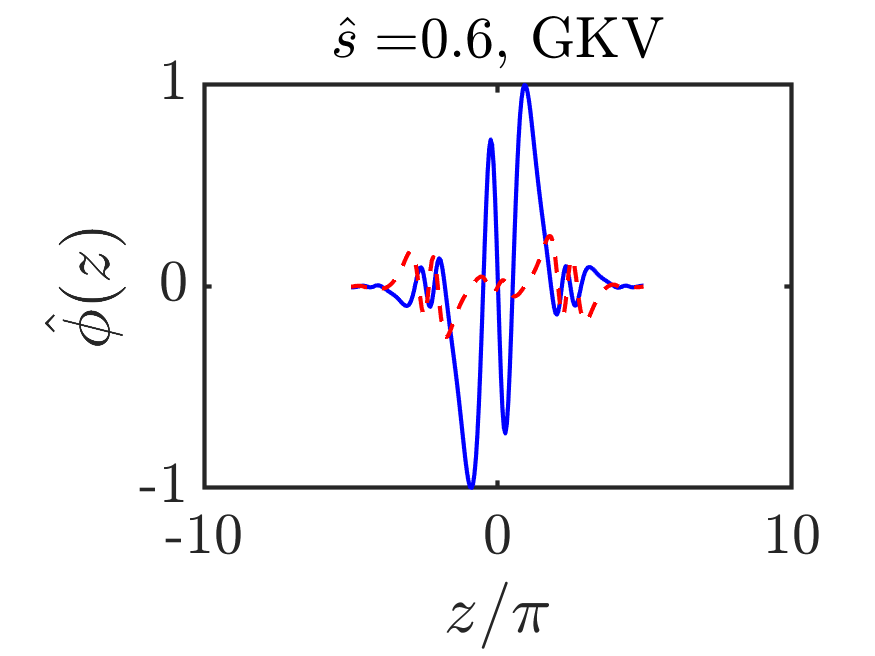}\label{fig:shear0.6_phi_z2_GKV_odd}}
\caption{ Eigenmode structures of the most unstable tearing parity modes calculated by the shooting method and GKV code. }
\label{fig:phiz_T_shearscan_ky0.2}
\end{figure}

\begin{figure}[htb!]
\centering
\subfigure[]{ \includegraphics[width=0.35\textwidth ]{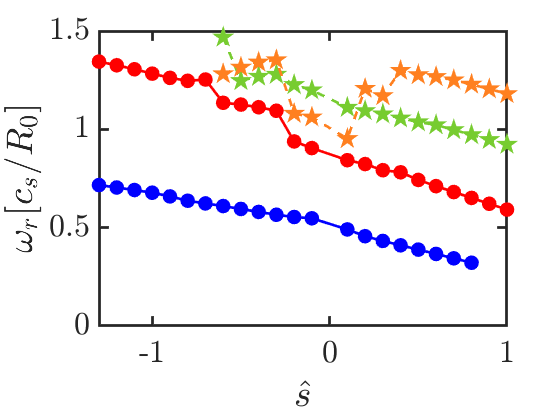}\label{fig:shearscan_ky0.2_omega_r_GH}}
\subfigure[]{ \includegraphics[width=0.35\textwidth ]{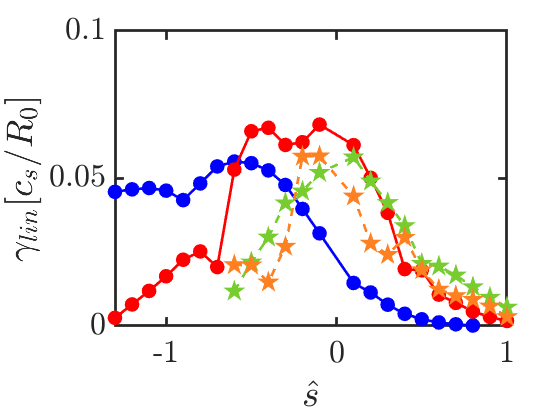} \label{fig:shearscan_ky0.2_omega_i_GH}}
\caption{ (a) Real frequency ($\omega_r$) and (b) linear growth rate ($\gamma_{lin}$) of various eigenstates calculated by the shooting method.  The blue (red) circle corresponds to the lowest-order state of the ballooning parity (tearing parity) mode, and the green (orange) star corresponds to the most unstable mode, except the lowest-order state of the ballooning parity (tearing parity) mode. }
\label{fig:shearscan_ky0.2_omega_GH}
\end{figure}

\begin{figure}[htb!]
\centering
\subfigure[]{ \includegraphics[width=0.24\textwidth ]{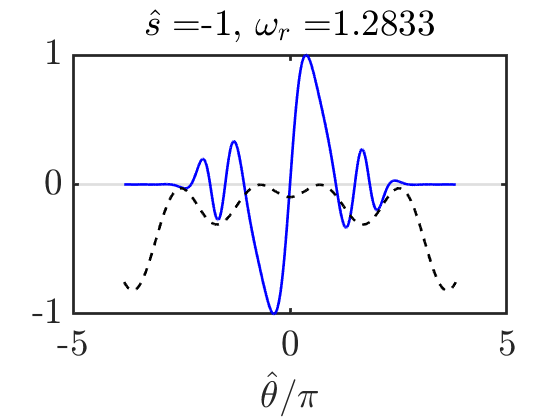}\label{fig:shear-1.0_ky0.2_phiz_and_Q_T_G}}
\subfigure[]{ \includegraphics[width=0.24\textwidth ]{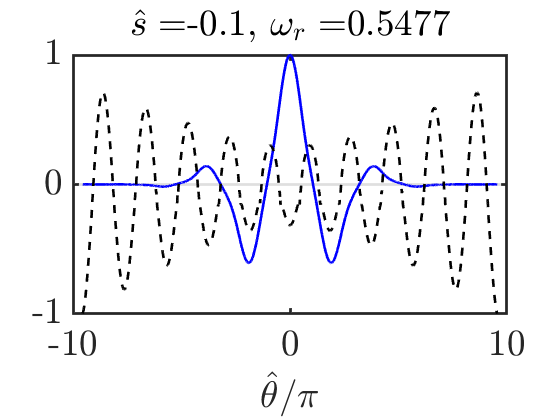}\label{fig:shear-0.1_ky0.2_phiz_and_Q_B_G}}
\subfigure[]{ \includegraphics[width=0.24\textwidth ]{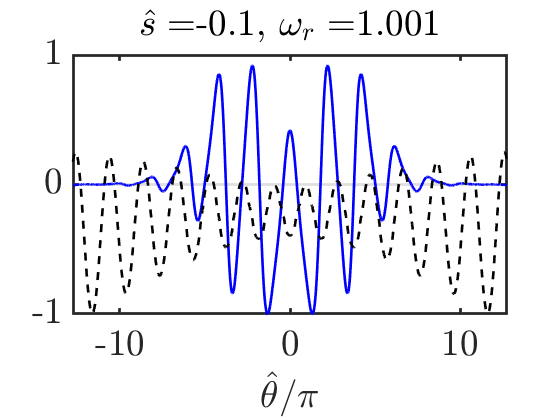}\label{fig:shear-0.1_ky0.2_phiz_and_Q_B_E1}}
\caption{ Real part of the eigenfunction (blue solid line) and the effective potential (black dotted line) of various eigenstates. Here, $\omega_r$ is normalized by $c_s/R_0$, and the effective potential is normalized by its maximum absolute value within the given ballooning angle range. }
\label{fig:strong_shear_phiz_and_Q}
\end{figure}

\indent Figure \ref{fig:shearscan_ky0.2_omega} shows the normalized real frequency ($\omega_r$) and linear growth rate ($\gamma_{lin}$) of ballooning parity (B) and tearing parity (T) modes as functions of the magnetic shear,  calculated by the GKV code and the shooting method for $q=1.4$, $k_\theta \rho_s =0.2$, $T_{i0}/T_{e0}=0.1$, $T_{f0}/T_{e0}=50$, $R_0/L_{Te}=R_0/L_{Ti}=0$, $R_0/L_{ne}=R_0/L_{ni}=R_0/L_{nf}=3$, $R_0/L_{Tf}=30$, $n_{f0}/n_{e0}=0.1$.  
The overall trends in $\omega_r$ and $\gamma_{lin}$ from the GKV simulations and the shooting method are consistent. Both methods show peaks in $\gamma_{lin}$ around the zero shear, but the GKV result shows higher peaks and more pronounced fluctuations than the shooting method. The tearing parity mode tends to be more unstable than the ballooning parity mode in a weak negative shear regime.  These results suggest that the fast ion-driven drift instability is most amplified in weak magnetic shear plasmas, and parity transition to the tearing parity mode tends to occur in weak negative shear regimes. \\
\indent Both methods show a general trend of decreasing $\omega_r$ as $\hat{s}$ increases, with significant shifts in $\omega_r$, and the magnetic shear values at these transitions calculated by both methods are similar. These significant shifts in $\omega_r$ are associated with eigenstate transitions, as shown in figures \ref{fig:phiz_B_shearscan_ky0.2} and \ref{fig:phiz_T_shearscan_ky0.2}. Each parity mode undergoes eigenstate transitions around the magnetic shear values where $\omega_r$ is shifted abruptly. The overall trend of the eigenstate transition calculated by the shooting method and GKV code is consistent. These results indicate that the shooting method can provide a clear picture of the general trends of the instability but may miss some details captured by the GKV simulations. 
\begin{figure}[htb!]
\centering
\subfigure[]{ \includegraphics[width=0.35\textwidth ]{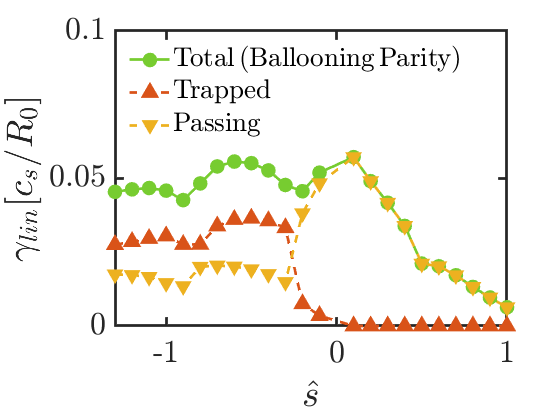}\label{fig:shearscan_ky0.2_omega_i_TRPS}}
\caption{ The contribution of trapped and passing fast ion to the linear growth rate of the most unstable ballooning parity mode calculated by the shooting method.  }
\label{fig:shearscan_ky0.2_omega_TRPS}
\end{figure}
Nevertheless, a more detailed physical interpretation can be achieved through the shooting method. Figure \ref{fig:shearscan_ky0.2_omega_TRPS} shows the distinct contributions of trapped and passing fast ions to the most unstable ballooning parity mode, where each contribution is calculated in the absence of the other by the shooting method. In the negative shear regime, where a sufficient number of trapped fast ions reverse their toroidal precession motion, the contribution from trapped fast ions dominates. Conversely, as the magnetic shear increases, trapped fast ions are less likely to reverse their toroidal precession motion, and the contribution from passing fast ions becomes dominant. This result suggests that the unstable modes observed in the weak negative shear and moderate positive shear regime, which could not explained by \cite{KangPoP2019,KangPoP2020}, are primarily driven by the resonant contribution of the passing fast ions.\\
For more detail, we have further investigated the real frequency and the linear growth rate of various eigenstates using the shooting method. With the GKV code, it is challenging to calculate each eigenstate's frequency and mode structure except the most unstable mode due to the exponential growth of each eigenstate in time. In figure \ref{fig:shearscan_ky0.2_omega_GH}, the blue (red) circle corresponds to the lowest-order state of the ballooning parity (tearing parity) mode, and the green (orange) star corresponds to the most unstable mode, except the lowest-order state of the ballooning parity (tearing parity) mode. Here, we refer to the latter as "the most unstable high-order state" for convenience. 
In the case of ballooning parity mode, the linear growth rate of the lowest-order state peaks around $\hat{s}=-0.6$ and decreases significantly as $\hat{s}$ increases, while the linear growth rate of the most unstable high-order state peaks around zero shear. Consequently, the eigenstate transition of the ballooning parity mode occurs near $\hat{s}=-0.2$, where the linear growth rate of the most unstable high-order state becomes greater than that of the lowest-order state. In the case of the tearing parity mode, the linear growth rates of the lowest-order and the most unstable high-order states reach their maximum near the zero shear. However, the lowest-order state is more unstable than the most unstable high-order state in most magnetic shear regimes. The linear growth rate of the most unstable high-order state becomes slightly larger than the lowest-order state for $\hat{s}\geq 0.4$, triggering an eigenstate transition of the tearing parity mode. These results clearly demonstrate that various eigenstates can coexist, and the eigenstate transition occurs between these eigenstates. \\
\indent Interestingly, the ballooning parity modes show more complex structures in the weak magnetic shear regime than the strong shear regime. Table. \ref{table:comparing_shear} shows the list of coexisting unstable eigenstates calculated by the shooting method, for $\hat{s}=-1.0$, $\hat{s}=-0.6$ and $\hat{s}=-0.1$. The number of coexisting unstable eigenstates increases as $|\hat{s}|$ decreases. 
\begin{table*}[htb!] \label{table:comparing_shear}
\caption{List of unstable eigenstates for $\hat{s}=-1.0$,$\hat{s}=-0.6$ and $\hat{s}=-0.1$. Here, frequencies are normalized by $c_s/R_0$.}     
\centering  
\begin{tabular}{l l l c l l l c l l l}
\hline\hline
$\hat{s}=-1.0$ &&&& $\hat{s}=-0.6$&&&&$\hat{s}=-0.1$\\
\hline\hline
$\omega_r$ &$\gamma_{lin}$& Parity && $\omega_r$& $\gamma_{lin}$& Parity&&$\omega_r$&$\gamma_{lin}$&Parity\\
\hline
\textbf{0.6777}&\textbf{0.0459}&\textbf{Ballooning}&&\textbf{0.6098}&\textbf{0.0557}&\textbf{Ballooning}&&0.5477&0.0314&Ballooning\\
1.2833&0.0169&Tearing&&0.8354&0.0076&Ballooning&&\textbf{0.9048} &\textbf{0.0683} & \textbf{Tearing}\\
&&&&1.1350&0.0530&Tearing&&0.9373&0.0445&Ballooning\\
&&&&1.2128&0.0080&Ballooning&&1.0010 &0.0500 &Ballooning\\
&&&&1.2818&0.0207&Tearing&&1.0618 &0.0576 &Tearing\\
&&&&1.4683&0.0118&Ballooning&&1.1993 &0.0520 &Ballooning\\
&&&&&&&&1.2502 &0.0315 &Tearing\\
&&&&&&&&1.2750 &0.0315 &Ballooning\\
&&&&&&&&1.3834 &0.0378 &Tearing\\
&&&&&&&&1.4400 &0.0378 &Ballooning\\
&&&&&&&&1.4706 &0.0152 &Tearing\\
\hline              
\end{tabular}       
\end{table*}
This behavior of the eigenmode structure is primarily due to the effective potential structure. As discussed in Sec. \ref{Subsec: Eigenmode equation}, the potential $Q$ is divided into slab-like potential $Q_{slab}$ and toroidicity-induced potential $Q_{tor}$. The slab-like potential $Q_{slab}$ dominates as $|\hat{\theta}|\rightarrow \infty$ and the effective potential approximately becomes $-Q\rightarrow -b_\theta\hat{s}^2\hat{\theta}^2$, which is an anti-well structure. Since $Q_{slab}\propto b_\theta \hat{s}^2$ and $Q_{tor} \propto 1/\Omega_r$, the slab-like potential becomes dominant at relatively more interior regions as $b_\theta$, $|\hat{s}|$ and $\Omega_r$ increases. Here we neglect the effect of $W(\Omega,\hat{\theta})$ and imaginary part of the eigenvalue $\Omega_i$ to the mode structure for qualitative desription, because $|Q|\gg|W|$ and $|\Omega_r| \gg |\Omega_i|$.  For high values of $b_\theta$, $|\hat{s}|$ and $\Omega_r$, $Q_{slab}$ dominates at almost every ballooning angle, and the effective potential structure becomes a negative anti-well structure. The wave can freely convect outward in this potential structure, and the slab-like eigenmodes are damped \cite{ChenPF1980}. Therefore, the eigenvalue $\Omega_r$ of unstable eigenmode is limited, and this limitation is more strict for large $b_\theta$ and $|\hat{s}|$, which reduces the number of unstable eigenmodes. For example, figures \ref{fig:shear-1.0_ky0.2_phiz_and_Q_T_G} show that the eigenmodes are localized in local potential wells, and the effective potential structure is approximately negative anti-well structure for $\hat{s}=-1.0$.  
In contrast, for low $b_\theta$ and $|\hat{s}|$, $Q_{tor}$ becomes dominant within a relatively wide range of $\hat{\theta}$. The oscillatory property of $Q_{tor}$ makes the behavior of eigenmode structure more complicated, as shown in figures \ref{fig:shear-0.1_ky0.2_phiz_and_Q_B_G} and \ref{fig:shear-0.1_ky0.2_phiz_and_Q_B_E1}. The effective potentials have oscillating structures, allowing various unstable eigenmodes to coexist. In short, the toroidicity-induced potential is responsible for the complex mode structure at the weak magnetic shear regime, and the eigenstate transition to a high-order state is more likely to occur in weak magnetic shear plasmas. Since the instability becomes more unstable at the weak magnetic shear regime, the effect of the eigenstate transition to a high-order state should be considered in this regime.
%%%%%%%%%%%%%%%%%%%%%%%%%%%%%%%%%%%%%%%%%%%%%%%%%%%%%%%%%%%%%%%%%%%%%%%%%%%%%%%%%%%%%%%%%%%%%%%%%%%%%%%%%%%%%%%%%%%%%%%%%%%%%%%%%%%%%%%%%%%%%%%%%%%%%%%%%%%%%%%%%%%%%%%%%%%%
%%%%%%%%%%%%%%%%%%%%%%%%%%%%%%%%%%%%%%%%%%%%%%%%%%%%%%%%%%%%%%%%%%%%%%%%%%%%%%%%%%%%%%%%%%%%%%%%%%%%%%%%%%%%%%%%%%%%%%%%%%%%%%%%%%%%%%%%%%%%%%%%%%%%%%%%%%%%%%%%%%%%%%%%%%%%
%%%%%%%%%%%%%%%%%%%%%%%%%%%%%%%%%%%%%%%%%%%%%%%%%%%%%%%%%%%%%%%%%%%%%%%%%%%%%%%%%%%%%%%%%%%%%%%%%%%%%%%%%%%%%%%%%%%%%%%%%%%%%%%%%%%%%%%%%%%%%%%%%%%%%%%%%%%%%%%%%%%%%%%%%%%%
\subsection{$q$ variation}\label{Subsec: q variation}
In this section, we analyze how the safety factor $q$ influences the instability and eigenmode structure for two cases, $\hat{s}=-1.0$ and $\hat{s}=-0.1$. The other parameters are given as : $r/a = 0.5$, $R_0/a=3$, $k_\theta \rho_s =0.2$, $T_{i0}/T_{e0}=0.1$, $T_{f0}/T_{e0}=50$, $R_0/{L_{Te}}=R_0/{L_{Ti}}=0$, $R_0/L_{ne}=R_0/L_{ni}=R_0/L_{nf}=3$, $R_0/L_{Tf}=30$, $n_{f0}/n_{e0}=0.1$.
%%%%%%%%%%%%%%%%%%%%%%%%%%%%%%%%%%%%%%%%%%%%%%%%%%%%%%%%%%%%%%%%%%%%%%%%%%%%%%%%%%%%%%%%%%%%%%%%%%%%%%%%%%%%%%%%%%%%%%%%%%%%%%%%%%%%%%%%%%%%%%%%%%%%%%%%%%%%%%%%%%%%%%%%%%%%
%%%%%%%%%%%%%%%%%%%%%%%%%%%%%%%%%%%%%%%%%%%%%%%%%%%%%%%%%%%%%%%%%%%%%%%%%%%%%%%%%%%%%%%%%%%%%%%%%%%%%%%%%%%%%%%%%%%%%%%%%%%%%%%%%%%%%%%%%%%%%%%%%%%%%%%%%%%%%%%%%%%%%%%%%%%%
%%%%%%%%%%%%%%%%%%%%%%%%%%%%%%%%%%%%%%%%%%%%%%%%%%%%%%%%%%%%%%%%%%%%%%%%%%%%%%%%%%%%%%%%%%%%%%%%%%%%%%%%%%%%%%%%%%%%%%%%%%%%%%%%%%%%%%%%%%%%%%%%%%%%%%%%%%%%%%%%%%%%%%%%%%%%
\subsubsection{$\hat{s}=-1.0$}\label{Subsubsec: s=-1.0}
\begin{figure}[htb!]
\centering
\subfigure[]{ \includegraphics[width=0.35\textwidth ]{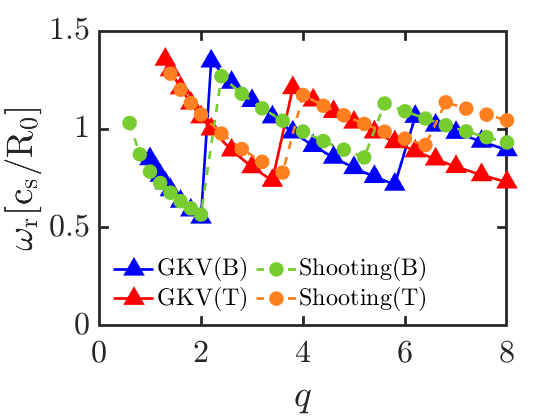}\label{fig:GKV_qscan_omega_r2}}
\subfigure[]{ \includegraphics[width=0.35\textwidth ]{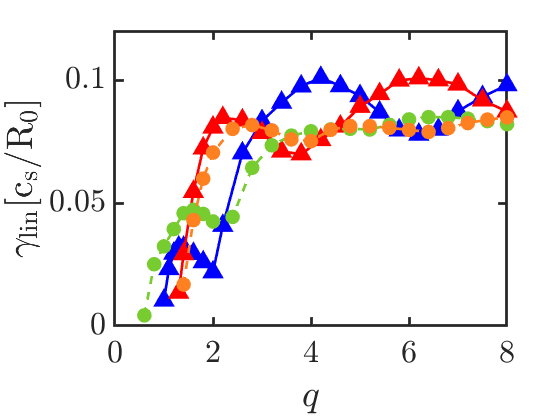}\label{fig:GKV_qscan_omega_i2}}
\caption{ (a) Real frequency ($\omega_r$) and (b) the linear growth rate ($\gamma_{lin}$) for $\hat{s}=-1.0$ as functions of $q$ of the most unstable ballooning parity mode and tearing parity mode. }
\label{fig:GKV_qscan_shear-1.0}
\end{figure}
\begin{figure}[htb!]
\centering
\subfigure[]{ \includegraphics[width=0.24\textwidth ]{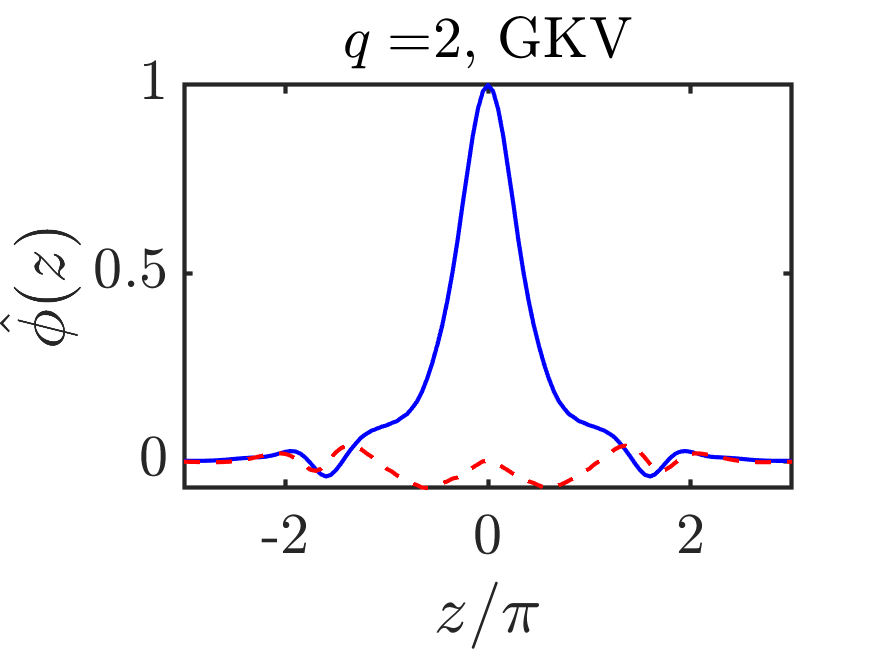}\label{fig:q2.0_phi_z1_GKV_even}}
\subfigure[]{ \includegraphics[width=0.24\textwidth ]{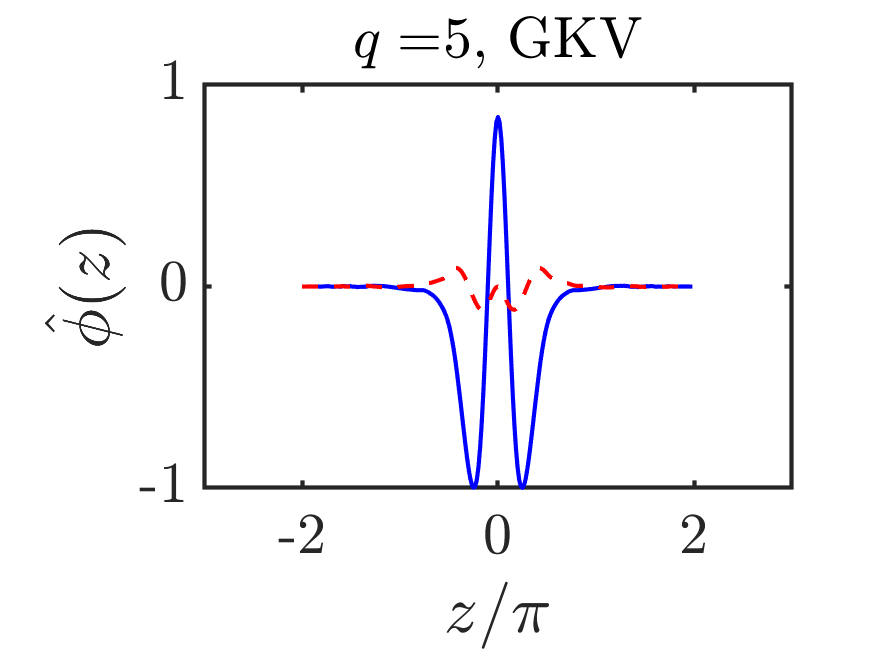}\label{fig:q5.0_phi_z1_GKV_even}}
\subfigure[]{ \includegraphics[width=0.24\textwidth ]{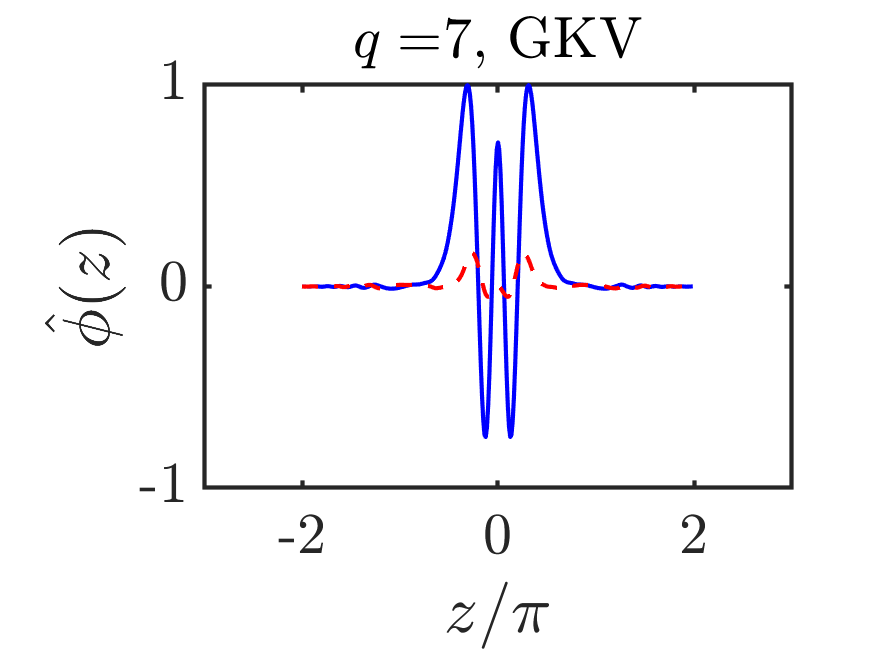}\label{fig:q7.0_phi_z1_GKV_even}}
\caption{ Eigenmode structures of the most unstable ballooning parity mode for (a) $q=2.0$, (b) $q=5.0$, and $q=7.0$, calculated by the GKV code.}
\label{fig:GKV_phiz_B_qscan}
\end{figure}
\begin{figure}[htb!]
\centering
\subfigure[]{ \includegraphics[width=0.24\textwidth ]{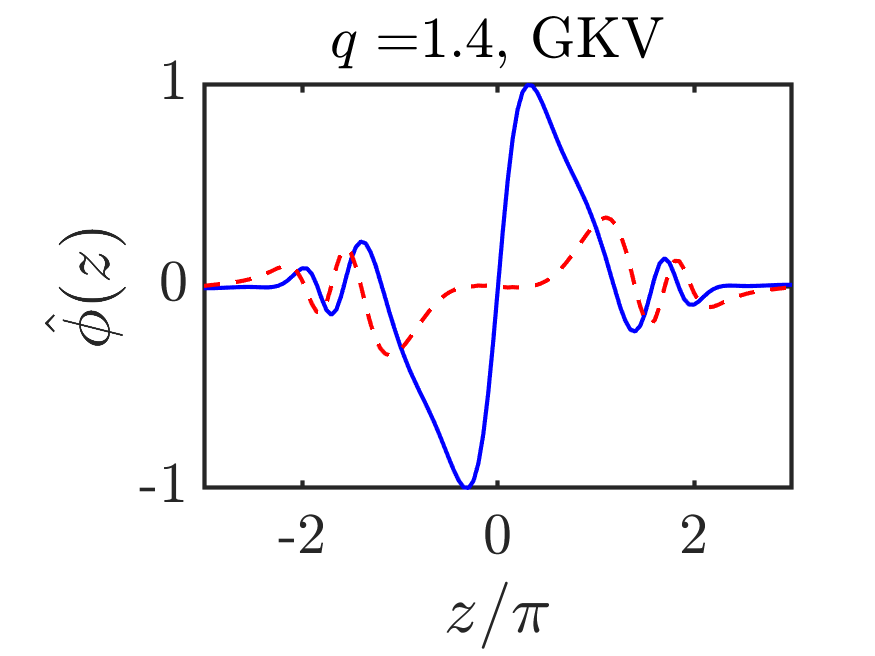}\label{fig:q1.4_phi_z2_GKV_odd}}
\subfigure[]{ \includegraphics[width=0.24\textwidth ]{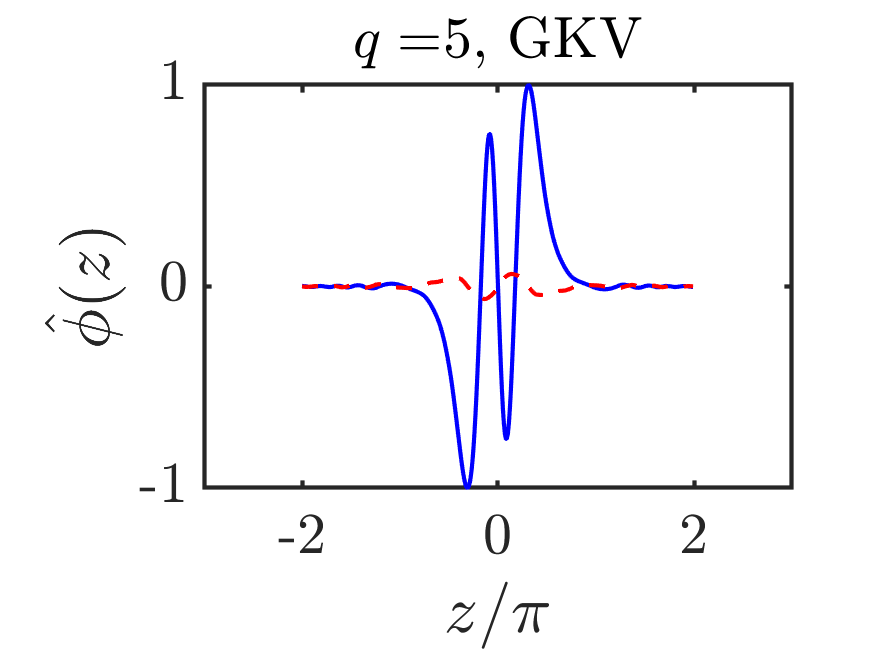}\label{fig:q5.0_phi_z2_GKV_odd}}
\caption{ Eigenmode structures of the most unstable tearing parity mode for (a) $q=1.4$ and (b) $q=5.0$, calculated by the GKV code. }
\label{fig:GKV_phiz_T_qscan}
\end{figure}
\begin{figure}[htb!]
\centering
\subfigure[]{ \includegraphics[width=0.35\textwidth ]{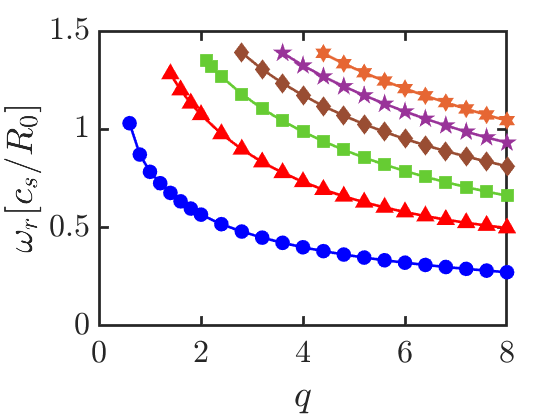}\label{fig:qscan_omega_r}}
\subfigure[]{ \includegraphics[width=0.35\textwidth ]{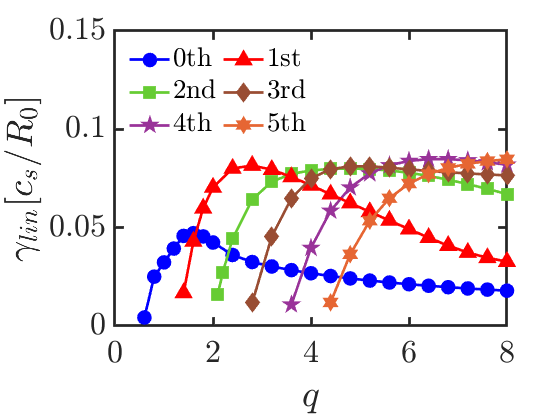} \label{fig:qscan_omega_i}}
\caption{ (a) The real frequency ($\omega_r$) and (b) linear growth rate ($\gamma_{lin}$) of 0th, 1st, 2nd, 3rd, 4th and 5th order state, calculated by the shooting method.}
\label{fig:qscan_shear-1.0_GH}
\end{figure}

Figure \ref{fig:GKV_qscan_shear-1.0} shows the dependence of the real frequency and linear growth rate on the safety factor $q$, calculated by the GKV code and the shooting method. 
The linear growth rate of both parity modes initially increases significantly with $q$ and then reaches a saturation regime with minor fluctuations. The amplitude of minor fluctuations of $\gamma_{lin}$ within the saturation regime calculated by the GKV code is larger than that calculated by the shooting method. Both methods show that $\omega_r$ decreases as $q$ increases, exhibiting some significant shifts. Figures \ref{fig:GKV_phiz_B_qscan} and \ref{fig:GKV_phiz_T_qscan} show that these shifts in $\omega_r$ are related to the eigenmode transition. The parallel mode structure becomes more localized around the center and evolves to higher-order eigenstates as $q$ increases. \\
\begin{figure}[htb!]
\centering
\subfigure[]{ \includegraphics[width=0.24\textwidth ]{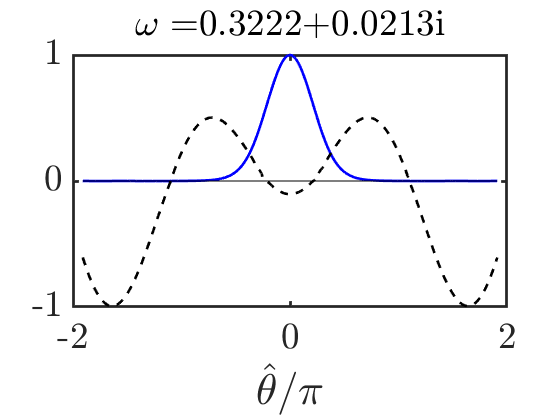}\label{fig:q6_shear-1.0_phiz_and_Q_B_G}}
\subfigure[]{ \includegraphics[width=0.24\textwidth ]{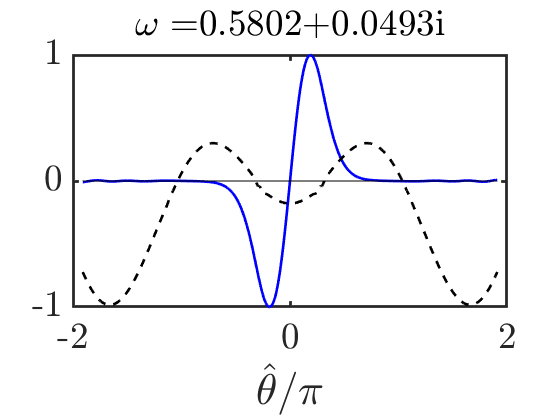}\label{fig:q6_shear-1.0_phiz_and_Q_T_G}}
\subfigure[]{ \includegraphics[width=0.24\textwidth ]{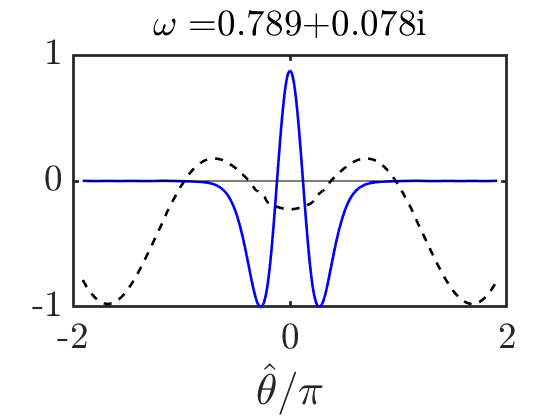}\label{fig:q6_shear-1.0_phiz_and_Q_B_E1}}
\hfill
\subfigure[]{ \includegraphics[width=0.24\textwidth ]{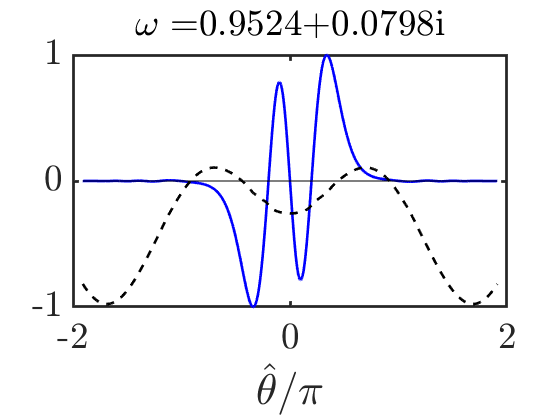}\label{fig:q6_shear-1.0_phiz_and_Q_T_E1}}
\subfigure[]{ \includegraphics[width=0.24\textwidth ]{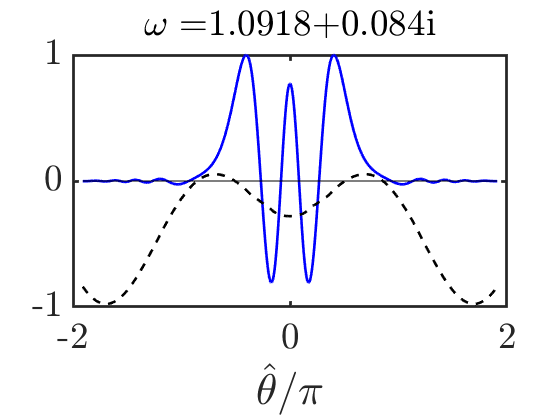}\label{fig:q6_shear-1.0_phiz_and_Q_B_E2}}
\subfigure[]{ \includegraphics[width=0.24\textwidth ]{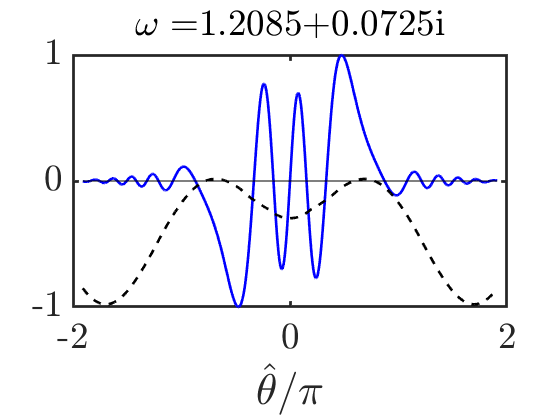}\label{fig:q6_shear-1.0_phiz_and_Q_T_E2}}
\caption{ Real part of the eigenfunction (blue solid line) and the effective potential (black dotted line) of various eigenstates in increasing order of $\omega_r$ for $\hat{s}=-1.0$ and $q=6.0$.  }
\label{fig:q6_shear-1.0_phiz_and_Q}
\end{figure}
\indent For more details, we further investigate the behavior of various eigenstates using the shooting method. In the case of $\hat{s}=-1.0$, the eigenstates can be clearly categorized according to the mode structure and $\omega_r$ because the eigenmodes are well-localized in a single local potential well, as discussed in the previous section. Figure \ref{fig:q6_shear-1.0_phiz_and_Q} shows various eigenstates' eigenmode and effective potential structures for $q=6.0$. Based on this, we categorize the order of eigenstates as 0th (ground state), 1st, 2nd, and so on, in increasing order of $\omega_r$. Even-numbered and odd-numbered eigenstates correspond to ballooning and tearing parity, respectively. Figure \ref{fig:qscan_shear-1.0_GH} shows each eigenstate's real frequency and linear growth rate as functions of $q$. The linear growth rate of each eigenstate follows a similar trend: it increases with $q$ up to a specific point and then slowly decreases. The $q$ value at which $\gamma_{lin}$ is maximized increases with the order of the eigenstate, and the most unstable mode becomes a higher-order eigenstate as $q$ increases. These eigenstate transitions are related to the depth of the potential well structure. For a high $q$ value, the depth of the potential well becomes deeper, and the eikonal of Eq. (36) becomes larger. This makes the mode structure fluctuate more frequently in $\hat{\theta}$, corresponding to the mode structure of a high-order state. Since the magnetic shear $\hat{s}$ and safety factor $q$ is closely related ($\because \hat{s}=\frac{r}{q}\frac{dq}{dr}$), the eigenstate transition to the high-order state with increasing $q$ suggests a complex global behavior of the fast ion-driven drift instability. 
%%%%%%%%%%%%%%%%%%%%%%%%%%%%%%%%%%%%%%%%%%%%%%%%%%%%%%%%%%%%%%%%%%%%%%%%%%%%%%%%%%%%%%%%%%%%%%%%%%%%%%%%%%%%%%%%%%%%%%%%%%%%%%%%%%%%%%%%%%%%%%%%%%%%%%%%%%%%%%%%%%%%%%%%%%%%
%%%%%%%%%%%%%%%%%%%%%%%%%%%%%%%%%%%%%%%%%%%%%%%%%%%%%%%%%%%%%%%%%%%%%%%%%%%%%%%%%%%%%%%%%%%%%%%%%%%%%%%%%%%%%%%%%%%%%%%%%%%%%%%%%%%%%%%%%%%%%%%%%%%%%%%%%%%%%%%%%%%%%%%%%%%%
%%%%%%%%%%%%%%%%%%%%%%%%%%%%%%%%%%%%%%%%%%%%%%%%%%%%%%%%%%%%%%%%%%%%%%%%%%%%%%%%%%%%%%%%%%%%%%%%%%%%%%%%%%%%%%%%%%%%%%%%%%%%%%%%%%%%%%%%%%%%%%%%%%%%%%%%%%%%%%%%%%%%%%%%%%%%
\subsubsection{$\hat{s}=-0.1$}\label{Subsubsec: s=-0.1}
\begin{figure}[htb!]
\centering
\subfigure[]{ \includegraphics[width=0.3\textwidth ]{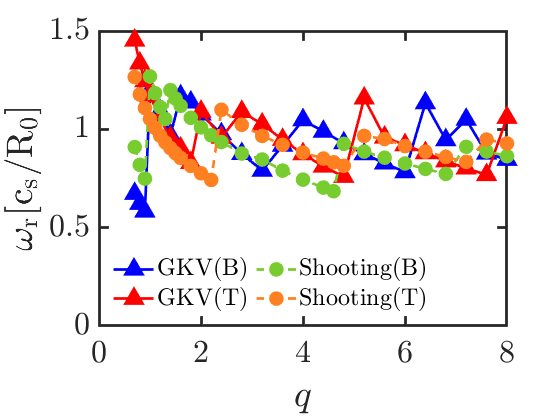}\label{fig:GKV_qscan_shear-0.1_omega_r2}}
\subfigure[]{ \includegraphics[width=0.3\textwidth ]{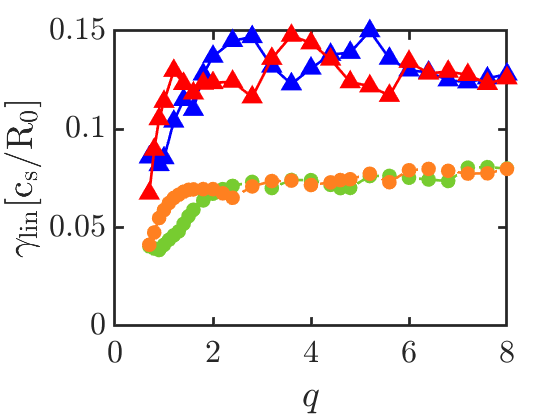}\label{fig:GKV_qscan_shear-0.1_omega_i2}}
\caption{ (a) Real frequency ($\omega_r$) and (b) the linear growth rate ($\gamma_{lin}$) for $\hat{s}=-0.1$ as functions of $q$ of the most unstable ballooning parity mode and tearing parity mode.  }
\label{fig:GKV_qscan_shear-0.1_omega}
\end{figure}
\begin{figure}[htb!]
\centering
\subfigure[]{ \includegraphics[width=0.24\textwidth ]{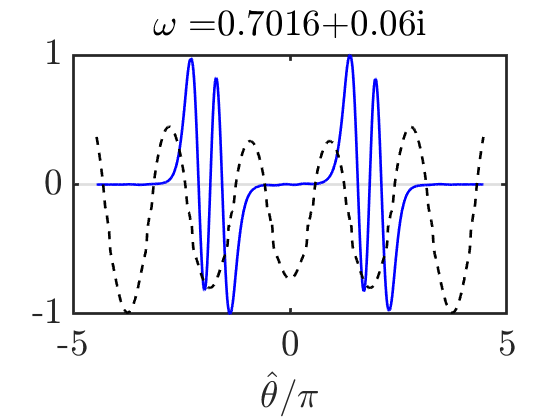}\label{fig:q6_shear-0.1_phiz_and_Q_T_E1}}
\subfigure[]{ \includegraphics[width=0.24\textwidth ]{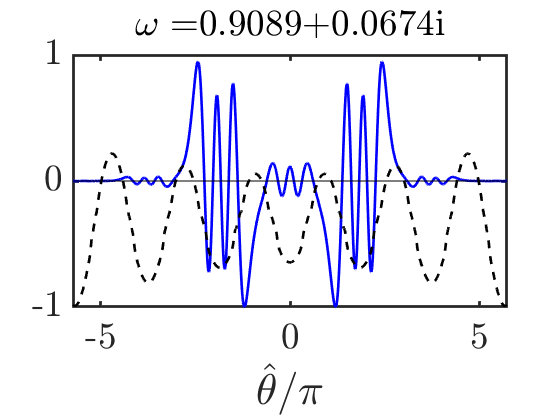}\label{fig:q6_shear-0.1_phiz_and_Q_B_E2}}
\subfigure[]{ \includegraphics[width=0.24\textwidth ]{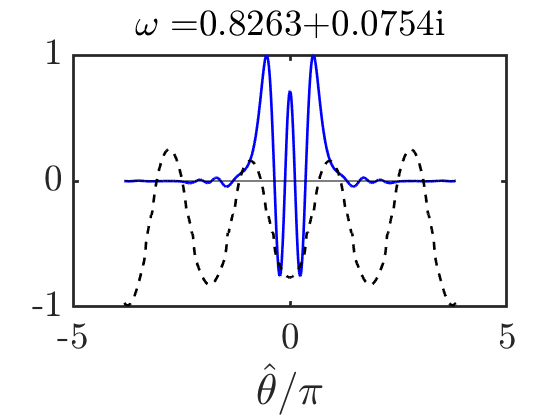}\label{fig:q6_shear-0.1_phiz_and_Q_B_E1}}
\caption{ Real part of the eigenfunction (blue solid line) and the effective potential (black dotted line) of various eigenstates for $\hat{s}=-0.1$ and $q=6.0$. }
\label{fig:q6_shear-0.1_phiz_and_Q_conv}
\end{figure}
\begin{figure}[htb!]
\centering
\subfigure[]{ \includegraphics[width=0.24\textwidth ]{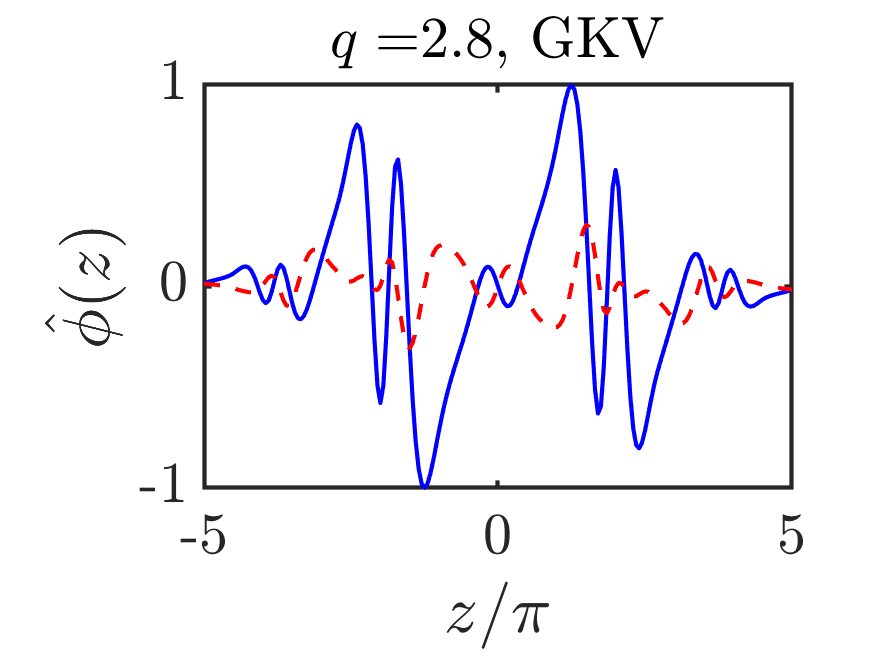}\label{fig:q2.8_shear-0.1_phi_z2_GKV_odd}}
\subfigure[]{ \includegraphics[width=0.24\textwidth ]{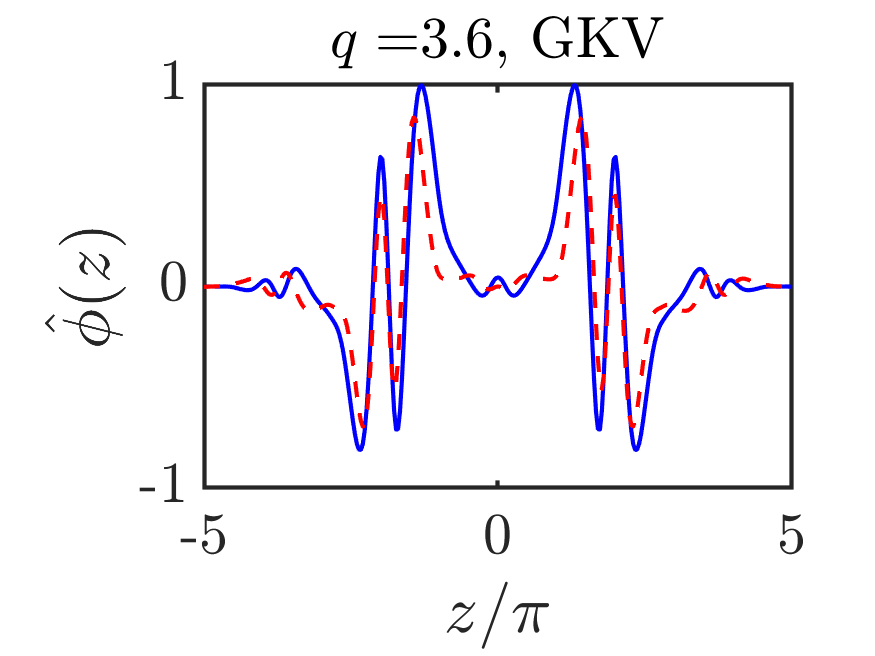}\label{fig:q3.6_shear-0.1_phi_z1_GKV_B}}
\subfigure[]{ \includegraphics[width=0.24\textwidth ]{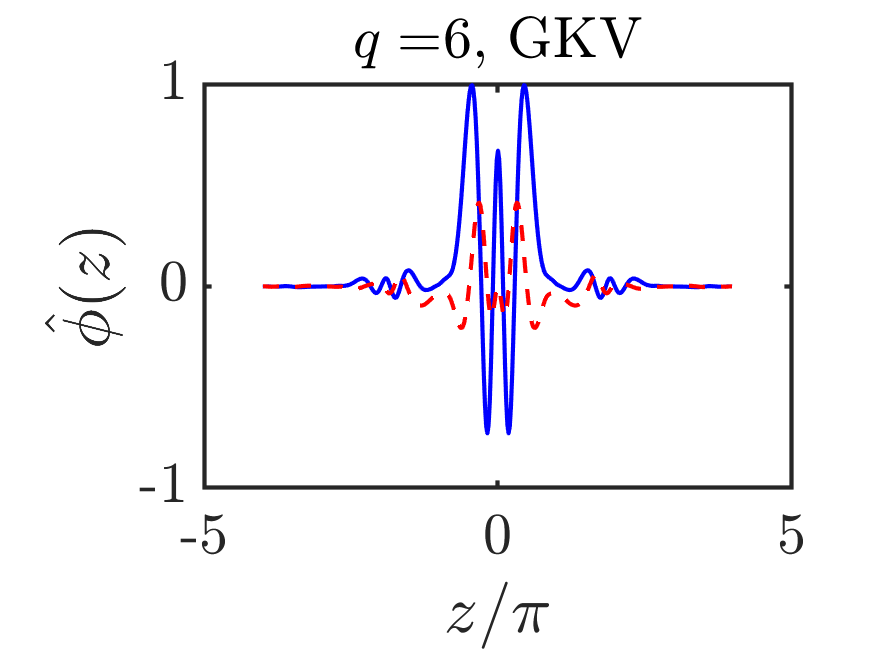}\label{fig:q6.0_shear-0.1_phi_z1_GKV_B}}
\caption{ Eigenmode structures of the most unstable tearing parity mode for (a) $q=2.8$, and most unstable ballooning parity mode for (b) $q=3.6$, and (c) $q=6.0$, calculated by the GKV code. }
\label{fig:GKV_phi_B_qscan_shear-0.1}
\end{figure}

\indent Figure \ref{fig:GKV_qscan_shear-0.1_omega} shows $\omega_r$ and $\gamma_{lin}$ as functions of the safety factor $q$ for $\hat{s}=-0.1$. The general behavior of $\omega_r$ and $\gamma_{lin}$ is similar to that observed in the $\hat{s}=-1.0$ case. The linear growth rate of the most unstable mode rapidly increases with $q$ before reaching a saturation regime with minor fluctuations. The linear growth rates calculated by both methods show significant differences in the case of $\hat{s}=-0.1$, which are larger than those of $\hat{s}=-1.0$ case. Since the resonant contribution of passing fast ions, mainly related to $k_\parallel$, becomes dominant for $\hat{s}=-0.1$, this result shows the limitation of local treatment of $k_\parallel$ for the shooting method, as discussed before. \\
\indent $\omega_r$ of both parity modes gradually decreases as $q$ increases, exhibiting several significant shifts related to eigenstate transitions. The mode structure undergoes more frequent eigenstate transitions compared to the $\hat{s}=-1.0$ case.  In contrast to the $\hat{s}=-1.0$ case, it is difficult to categorize the eigenstates because the oscillating toroidicity-induced potential is dominant. Not all eigenmodes are localized in a single local potential well. Figure \ref{fig:q6_shear-0.1_phiz_and_Q_conv} shows the eigenmode and effective potential structures of various eigenstates calculated by the shooting method for $q=6.0$. This result suggests that unstable modes not localized around $\hat{\theta}=0$ can appear, and their linear growth rate is comparable to that of the most unstable eigenstate. GKV simulation results further show that the eigenstates not localized around $\hat{\theta}=0$ can be the most unstable mode, as shown in figure \ref{fig:GKV_phi_B_qscan_shear-0.1}. These results represent the complex behavior of the fast ion-driven-drift instability and the complex interaction between the plasma parameters and the stability of the modes in a weak magnetic shear plasma as various eigenstates become dominant in different $q$ values.
%%%%%%%%%%%%%%%%%%%%%%%%%%%%%%%%%%%%%%%%%%%%%%%%%%%%%%%%%%%%%%%%%%%%%%%%%%%%%%%%%%%%%%%%%%%%%%%%%%%%%%%%%%%%%%%%%%%%%%%%%%%%%%%%%%%%%%%%%%%%%%%%%%%%%%%%%%%%%%%%%%%%%%%%%%%%
%%%%%%%%%%%%%%%%%%%%%%%%%%%%%%%%%%%%%%%%%%%%%%%%%%%%%%%%%%%%%%%%%%%%%%%%%%%%%%%%%%%%%%%%%%%%%%%%%%%%%%%%%%%%%%%%%%%%%%%%%%%%%%%%%%%%%%%%%%%%%%%%%%%%%%%%%%%%%%%%%%%%%%%%%%%%
%%%%%%%%%%%%%%%%%%%%%%%%%%%%%%%%%%%%%%%%%%%%%%%%%%%%%%%%%%%%%%%%%%%%%%%%%%%%%%%%%%%%%%%%%%%%%%%%%%%%%%%%%%%%%%%%%%%%%%%%%%%%%%%%%%%%%%%%%%%%%%%%%%%%%%%%%%%%%%%%%%%%%%%%%%%%
\subsection{$k_\theta$ variation}\label{Subsec: k_theta variation}
\begin{figure}[htb!]
\centering
\subfigure[]{ \includegraphics[width=0.3\textwidth ]{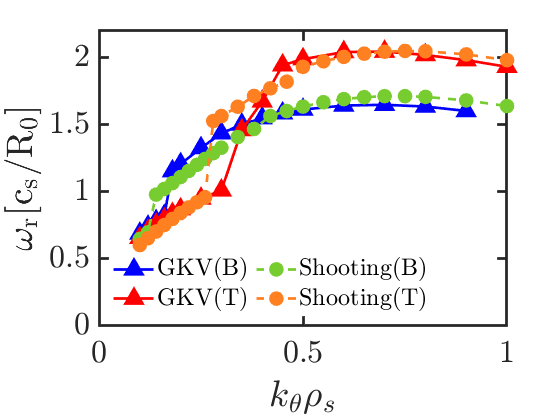}\label{fig:GKV_kyscan_shear0.1_omega_r2}}
\subfigure[]{ \includegraphics[width=0.3\textwidth ]{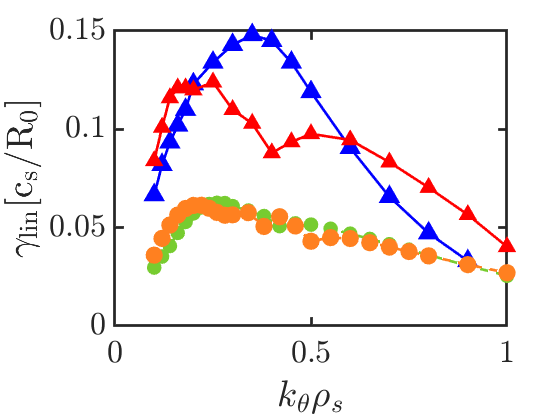}\label{fig:GKV_kyscan_shear0.1_omega_i2}}
\caption{ (a) Real frequency ($\omega_r$) and  (b) linear growth rate ($\gamma_{lin}$) as functions of $k_\theta \rho_s$ of the most unstable ballooning parity mode and tearing parity mode. }
\label{fig:GKV_kyscan_shear0.1_omega}
\end{figure}
\begin{figure}[htb!]
\centering
\subfigure[]{ \includegraphics[width=0.24\textwidth ]{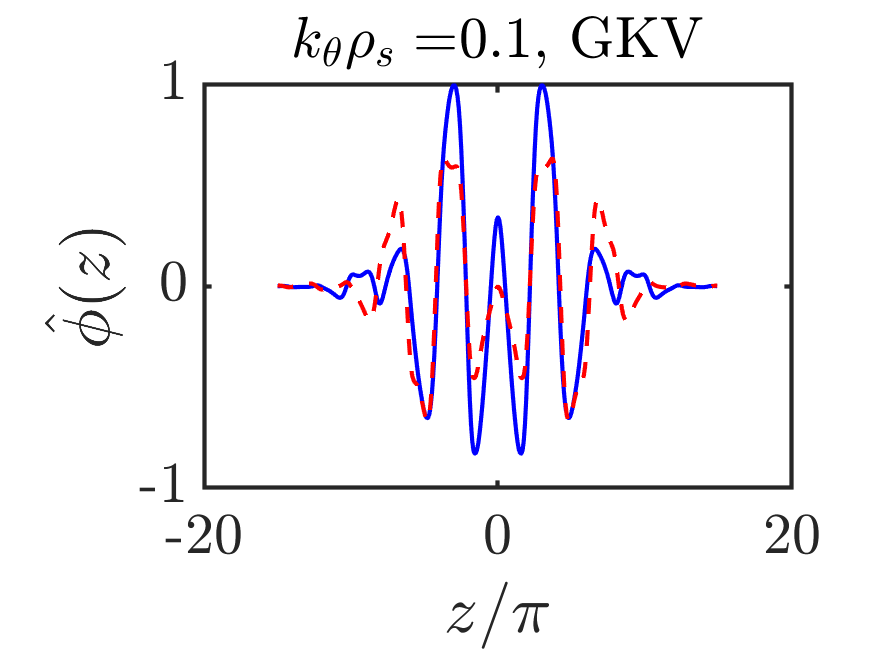}\label{fig:ky0.1_phi_z1_GKV_even}}
\subfigure[]{ \includegraphics[width=0.24\textwidth ]{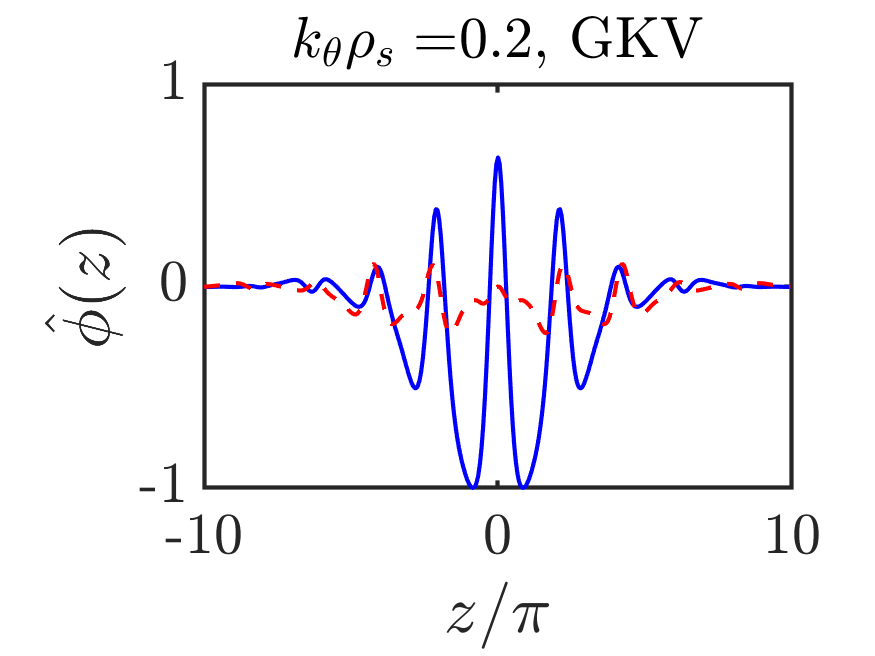}\label{fig:ky0.2_phi_z1_GKV_even}}
\subfigure[]{ \includegraphics[width=0.24\textwidth ]{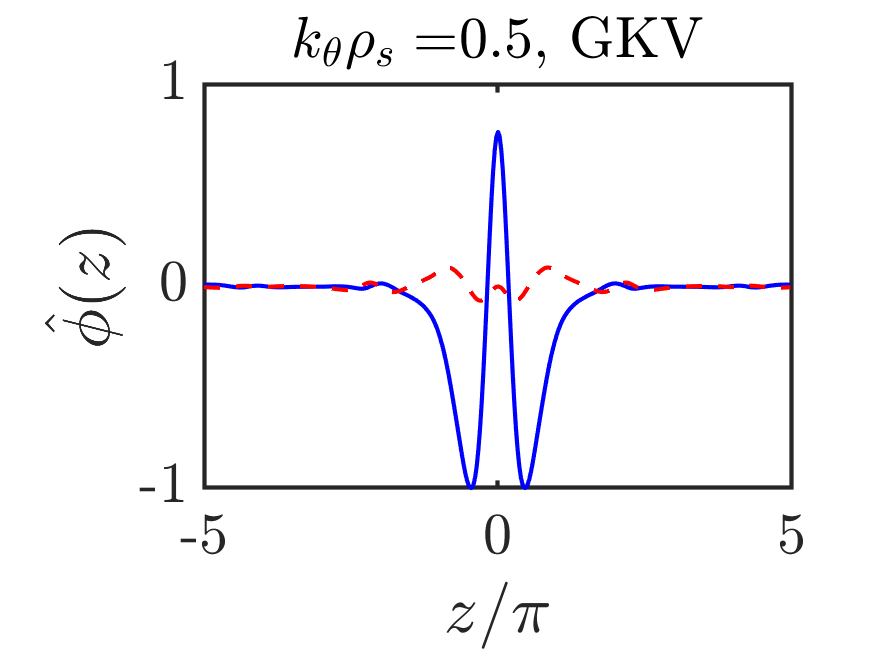}\label{fig:ky0.5_phi_z1_GKV_even}}
\caption{ Eigenmode structures of the most unstable ballooning parity mode for (a) $k_\theta \rho_s =0.1$, (b) $k_\theta \rho_s =0.2$, and (c) $k_\theta \rho_s =0.5$, calculated by the GKV code. }
\label{fig:GKV_phi_B_kyscan_shear0.1}
\end{figure}
\begin{figure}[htb!]
\centering
\subfigure[]{ \includegraphics[width=0.24\textwidth ]{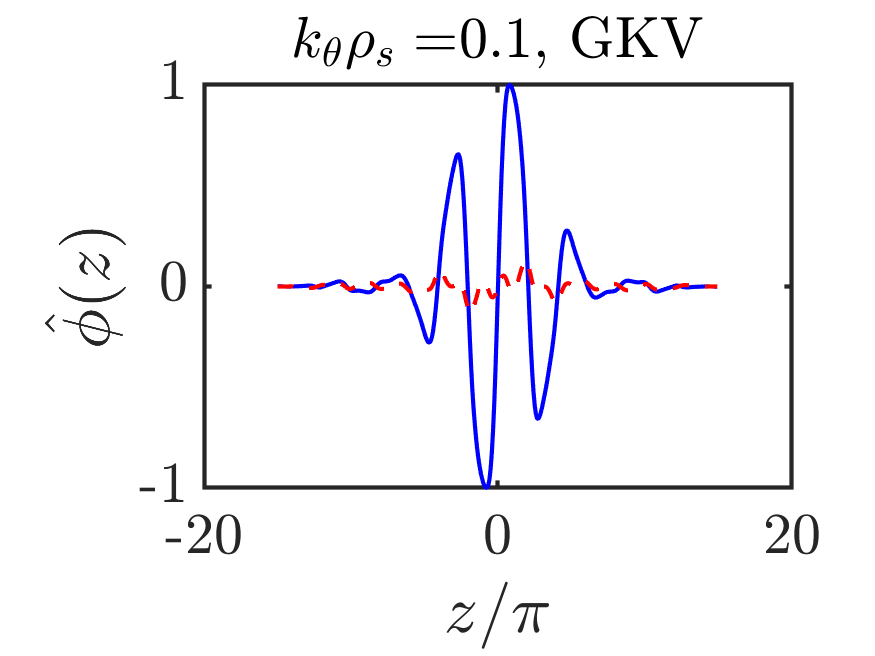}\label{fig:ky0.1_phi_z2_GKV_odd}}
\subfigure[]{ \includegraphics[width=0.24\textwidth ]{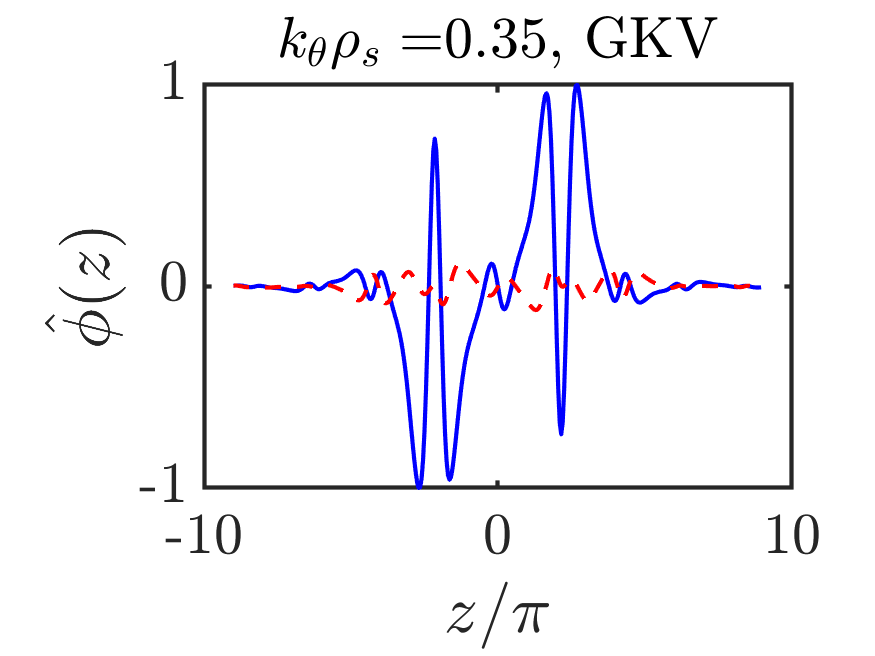}\label{fig:ky0.35_phi_z2_GKV_odd}}
\subfigure[]{ \includegraphics[width=0.24\textwidth ]{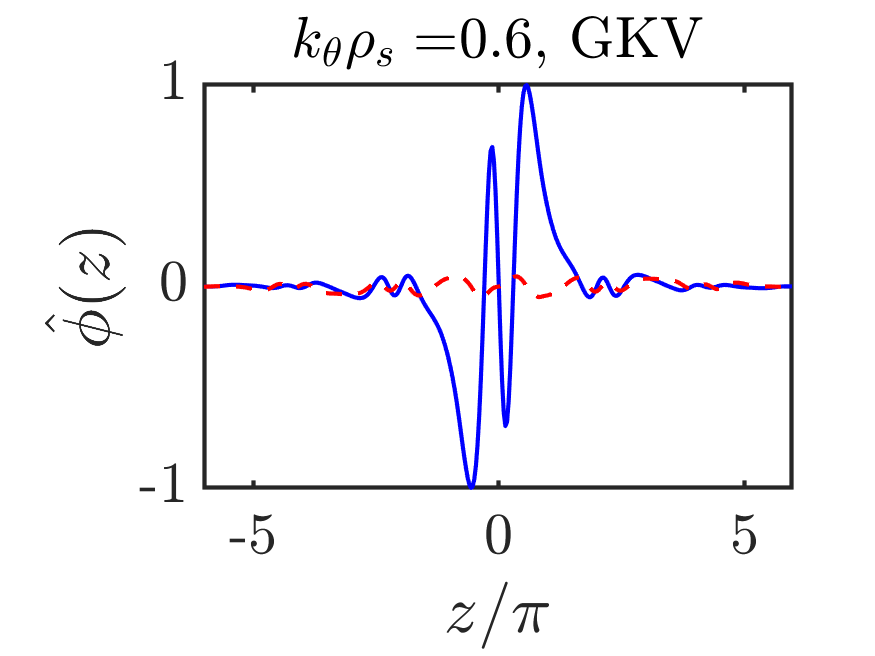}\label{fig:ky0.6_phi_z2_GKV_odd}}
\caption{ Eigenmode structures of the most unstable tearing parity mode for (a) $k_\theta \rho_s =0.1$, (b) $k_\theta \rho_s =0.35$, and (c) $k_\theta \rho_s =0.6$, calculated by the GKV code. }
\label{fig:GKV_phi_T_kyscan_shear0.1}
\end{figure}
\begin{figure}[htb!]
\centering
\subfigure[]{ \includegraphics[width=0.24\textwidth ]{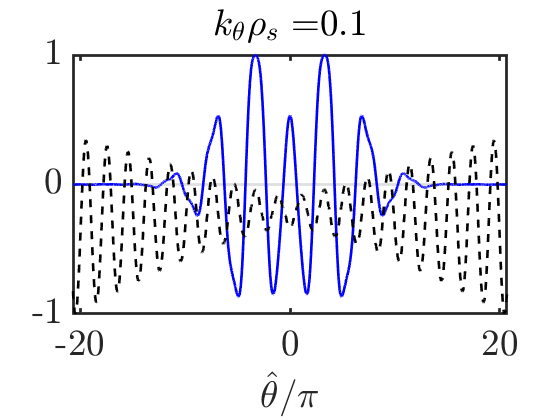}\label{fig:ky0.1_shear0.1_phiz_and_Q_B_MU.png}}
\subfigure[]{ \includegraphics[width=0.24\textwidth ]{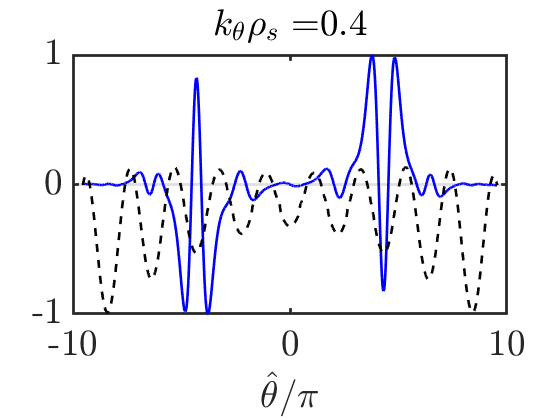}\label{fig:ky0.4_shear0.1_phiz_and_Q_T_MU.png}}
\subfigure[]{ \includegraphics[width=0.24\textwidth ]{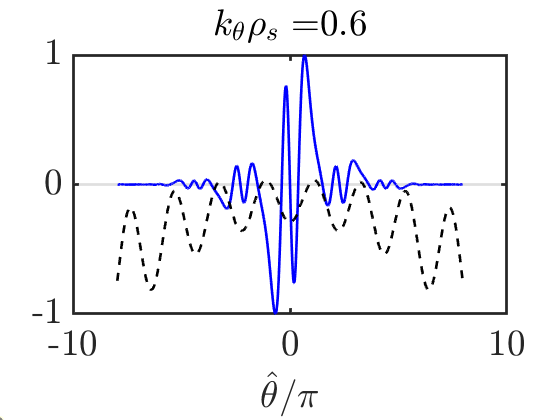}\label{fig:ky0.6_shear0.1_phiz_and_Q_T_MU.png}}
\caption{ Real part of the eigenfunction (blue solid line) and the effective potential (black dotted line) of the various eigenstates, calculated by the shooting method. }
\label{fig:kyscan_shear0.1_phiz_and_Q_B}
\end{figure}

The real frequency and linear growth rate as functions of $k_\theta\rho_s$ are shown in figure \ref{fig:GKV_kyscan_shear0.1_omega} for $r/a = 0.5$, $R_0/a=3$, $q=1.4$, $\hat{s}=0.1$, $T_{i0}/T_{e0}=0.1$, $T_{f0}/T_{e0}=50$, $R_0/{L_{Te}}=R_0/{L_{Ti}}=0$, $R_0/L_{ne}=R_0/L_{ni}=R_0/L_{nf}=3$, $R_0/L_{Tf} =30$, $n_{f0}/n_{e0}=0.1$. The linear growth rate of the ballooning parity mode initially increases with $k_\theta \rho_s$ until it peaks and then starts to decrease. On the other hand, the linear growth rate of the tearing parity mode follows a similar trend but exhibits several distinct local peaks. For the intermediate range of $k_\theta \rho_s$, the ballooning parity modes are more unstable compared to tearing parity modes, while the opposite is true for the other range of $k_\theta \rho_s$. These results indicate that the optimal range for $k_\theta\rho_s$ lies within $k_\theta \rho_s \sim 0.3$, where the ballooning parity modes become dominant. Although both methods show a similar trend, the linear growth rate calculated by the GKV code is almost twice as large as that calculated by the shooting method, associated with the local treatment of $k_\parallel$ for the shooting method.  Each parity mode undergoes eigenstate transitions as $k_\theta$ varies, with significant shifts in $\omega_r$. The overall trends of the eigenstate transition calculated by the GKV code are illustrated in figures \ref{fig:GKV_phi_B_kyscan_shear0.1} and \ref{fig:GKV_phi_T_kyscan_shear0.1}. The parallel mode structure becomes more localized in the inner region as $k_\theta$ increases. The tearing parity mode shows various eigenstate transitions, displaying an eigenstate not localized around $z=0$ (figure \ref{fig:ky0.35_phi_z2_GKV_odd}). As before, the effective potential structure is associated with the behavior of the eigenmode structure. Figure \ref{fig:kyscan_shear0.1_phiz_and_Q_B} shows effective potential and eigenmode structures of various eigenstates calculated by the shooting method. If $k_\theta \rho_s$ is small, the toroidicity-induced potential becomes dominant in almost all ballooning angles, and the oscillatory nature of the toroidicity-induced potential allows various eigenstates to appear. In contrast, if $k_\theta \rho_s$ is sufficiently large, $Q_{tor}$ becomes less dominant, and the effective potential structure becomes approximately an anti-well structure with few local potential wells. As a result, the eigenmode becomes localized in a single potential well for sufficiently large $k_\theta \rho_s$ values. 
These results suggest that non-ground states showing complex mode structures become important for $k_\theta \rho_s$ values where the linear growth rate of the instability is maximized due to the oscillating toroidicity-induced potential structure.
\clearpage
%%%%%%%%%%%%%%%%%%%%%%%%%%%%%%%%%%%%%%%%%%%%%%%%%%%%%%%%%%%%%%%%%%%%%%%%%%%%%%%%%%%%%%%%%%%%%%%%%%%%%%%%%%%%%%%%%%%%%%%%%%%%%
\subsection{$T_f$ variation}\label{Subsec: T_f variation}
\begin{figure}[htb!]
\centering
\subfigure[]{ \includegraphics[width=0.3\textwidth ]{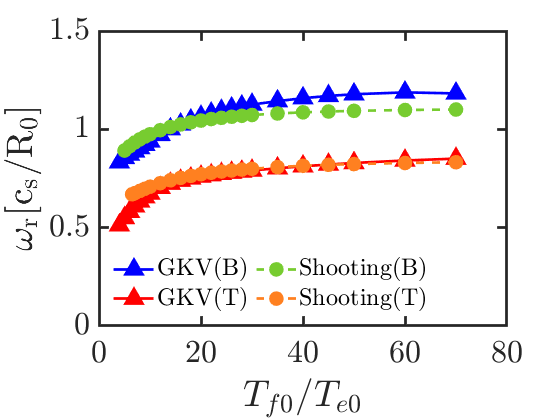}\label{fig:GKV_Tfscan_omega_r2}}
\subfigure[]{ \includegraphics[width=0.3\textwidth ]{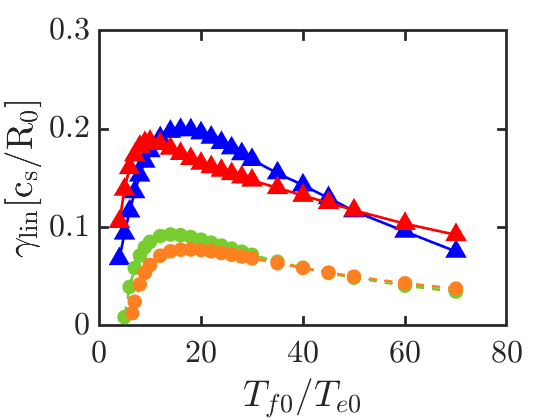}\label{fig:GKV_Tfscan_omega_i2}}
\caption{ (a) Real frequency ($\omega_r$) and  (b) linear growth rate ($\gamma_{lin}$) as functions of $T_{f0}/T_{e0}$ of the most unstable ballooning parity mode and tearing parity mode. }
\label{fig:GKV_Tfscan_omega}
\end{figure}
Figure \ref{fig:GKV_Tfscan_omega} illustrates the dependence of the real frequency ($\omega_r$) and linear growth rate ($\gamma_{lin}$) of the fast ion-driven drift instability on the fast ion temperature ratio ($T_{f0}/T_{e0}$). The calculations were performed at $r/a = 0.5$, $R_0/a=3$, $q=1.4$, $\hat{s}=0.2$, $k_\theta \rho_s = 0.2$, $T_{i0}/T_{e0}=0.1$, $R_0/{L_{Te}}=R_0/{L_{Ti}}=0$, $R_0/L_{ne}=R_0/L_{ni}=R_0/L_{nf}=3$, $R_0/L_{Tf} =30$, and $n_{f0}/n_{e0}=0.1$. Both GKV simulations and the shooting method generally show good agreement. As seen in figure \ref{fig:GKV_Tfscan_omega_r2}, the real frequency increases with $T_{f0}/T_{e0}$, saturating at higher fast ion temperature. 
Figure \ref{fig:GKV_Tfscan_omega_i2} reveals that the linear growth rate peaks at $T_{f0}/T_{e0}\approx 15$ before gradually decreasing. This behavior highlights the importance of resonance effects, which are strongest at intermediate fast ion temperatures. No eigenstate transitions were observed in this scan, suggesting that such transitions are more sensitive to other plasma parameters like the magnetic shear or safety factor.  
These trends align with the findings of \cite{KangPoP2020}, which reported similar behavior in reversed magnetic shear conditions ($\hat{s}=-1.0$). These results confirm that the fast ion temperature significantly impacts the linear growth rate of the fast ion-driven drift instability. 
\clearpage
\subsection{$n_f$ variation}\label{Subsec: n_f variation}
\begin{figure}[htb!]
\centering
\subfigure[]{ \includegraphics[width=0.3\textwidth ]{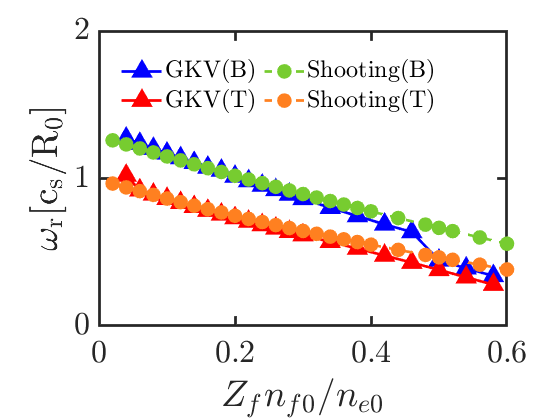}\label{fig:GKV_nfscan_omega_r2}}
\subfigure[]{ \includegraphics[width=0.3\textwidth ]{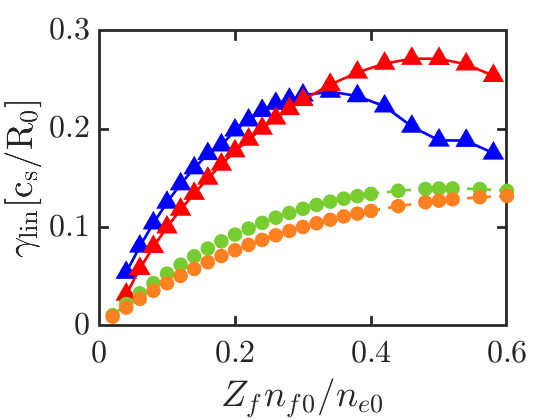}\label{fig:GKV_nfscan_omega_i2}}
\caption{ (a) Real frequency ($\omega_r$) and  (b) linear growth rate ($\gamma_{lin}$) as functions of $Z_f n_{f0}/n_{e0}$ of the most unstable ballooning parity mode and tearing parity mode. }
\label{fig:GKV_nfscan_omega}
\end{figure}
\begin{figure}[htb!]
\centering
\subfigure[]{ \includegraphics[width=0.24\textwidth ]{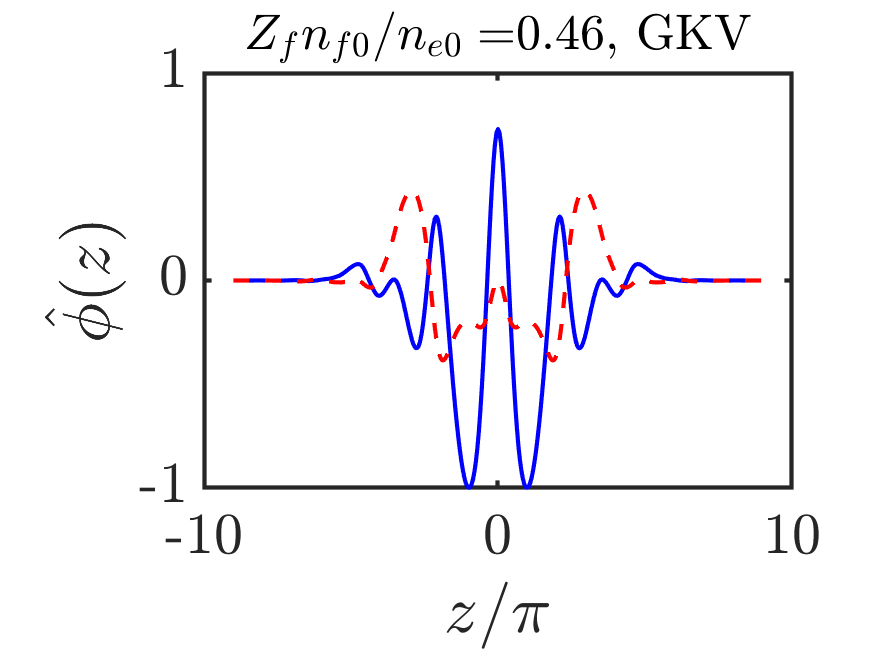}\label{fig:Zfnf0.46_phi_z1_GKV_B}}
\subfigure[]{ \includegraphics[width=0.24\textwidth ]{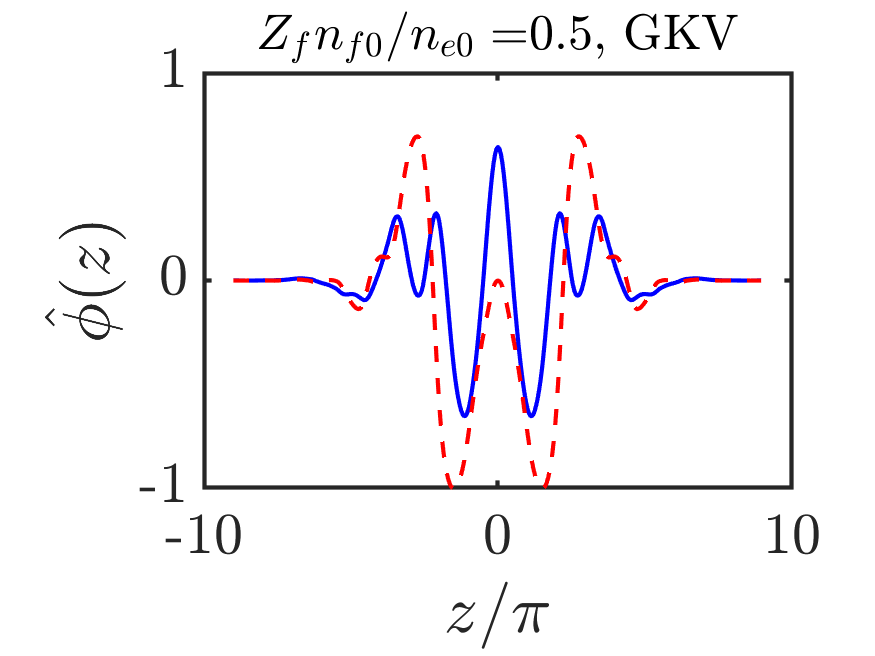}\label{fig:Zfnf0.5_phi_z1_GKV_B}}
\caption{ Eigenmode structures of the most unstable ballooning parity mode for (a) $Z_f n_{f0}/n_{e0}=0.46$, and (b) $Z_f n_{f0}/n_{e0}=0.5$, calculated by the GKV code. }
\label{fig:GKV_phi_B_Zfnfscan}
\end{figure}
Figure \ref{fig:GKV_nfscan_omega} shows the influence of the fast ions density ratio ($Z_f n_{f0}/n_{e0}$) on the instability. For this scan, parameters were set to $q=1.4$, $\hat{s}=0.2$, $k_\theta \rho_s = 0.2$, $T_{i0}/T_{e0}=0.1$, $R_0/{L_{Te}}=R_0/{L_{Ti}}=0$, $R_0/L_{ne}=R_0/L_{ni}=R_0/L_{nf}=3$, $R_0/L_{Tf} =30$. As shown in figure \ref{fig:GKV_nfscan_omega_r2}, the real frequency decreases consistently with increasing $Z_f n_{f0}/n_{e0}$ for both partiy modes. Interestingly, figure \ref{fig:GKV_nfscan_omega_i2} reveals non-monotonic behavior in the linear growth rate, which initially increases, and then decreases at higher fast ion density ffractions.  This trend suggests that the resonance effects driving the instability diminish when the bulk ion fraction becomes too small, reducing the wave-particle resonance. Recall that the fast ion-driven drift instability is sustained by the resonance between the fast ions and the electron drift wave, primarily caused by interactions between electrons and main bulk ions. Also, note that our theoretical model is valid for $Z_fn_{f0}/n_{e0} \ll 1$, where enough fraction of the main ion population is needed to sustain the electron drift wave. An eigenstate transition was observed in the GKV simulations for the ballooning parity mode at $Z_f n_{f0}/n_{e0} \approx 0.5$, indicating a subtle shift in the eigenmode structure (figure \ref{fig:GKV_phi_B_Zfnfscan}). These results underscore the importance of fast ion density in determining the behavior of the fast ion-driven drift instability.
\clearpage
\subsection{$L_{Tf}$ variation}\label{Subsec: L_Tf variation}
\begin{figure}[htb!]
\centering
\subfigure[]{ \includegraphics[width=0.3\textwidth ]{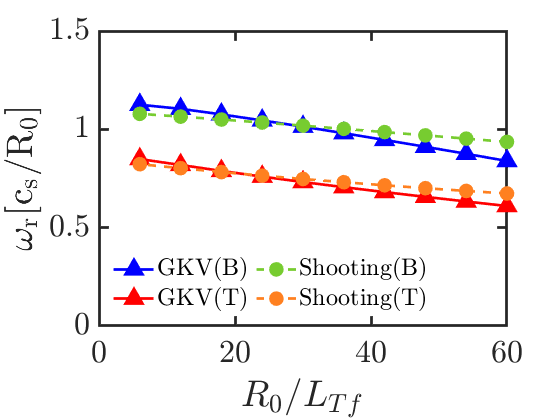}\label{fig:GKV_RLTfscan_omega_r2}}
\subfigure[]{ \includegraphics[width=0.3\textwidth ]{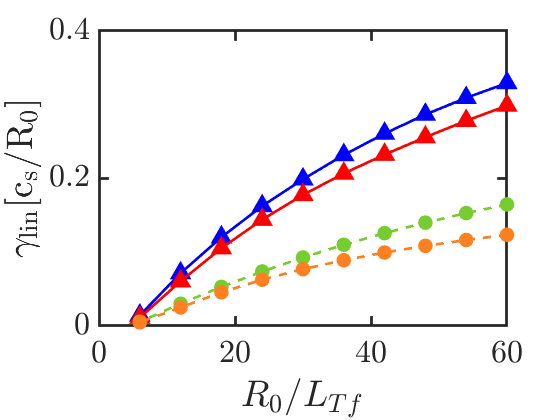}\label{fig:GKV_RLTfscan_omega_i2}}
\caption{ (a) Real frequency ($\omega_r$) and  (b) linear growth rate ($\gamma_{lin}$) as functions of $R_0/L_{Tf}$ of the most unstable ballooning parity mode and tearing parity mode. }
\label{fig:GKV_RLTfscan_omega}
\end{figure}
Figure \ref{fig:GKV_RLTfscan_omega} examines the dependence of  the real frequency ($\omega_r$) and linear growth rate ($\gamma_{lin}$) on the normalized fast ion temperature gradient ($R_0/L_{Tf})$. The scan was conducted with $q=1.4$, $\hat{s}=0.2$, $k_\theta \rho_s = 0.2$, $T_{i0}/T_{e0}=0.1$, $R_0/{L_{Te}}=R_0/{L_{Ti}}=0$, $R_0/L_{ne}=R_0/L_{ni}=R_0/L_{nf}=3$, $n_{f0}/n_{e0} =0.1$.
Figure \ref{fig:GKV_RLTfscan_omega_r2} shows that the real frequency decreases slightly with increasing $R_0/L_{Tf}$ for both parity modes. Meanwhile, the linear growth rate, as shown in figure \ref{fig:GKV_RLTfscan_omega_i2}, increases nearly linearly with $R_0/L_{Tf}$, indicating a strong destabilizing effect of the steep fast ion temperature gradient. These results confirm that the fast ion temperature gradient plays a critical role in enhancing the instability drive. Similar behavior was reported in reversed magnetic shear conditions ($\hat{s}=-1.0$) by \cite{KangPoP2020}, and these trends are consistent with those findings. No eigenstate transitions were observed for this parameter scan, indicating that the eigenmode structure remains consistent across variations in $R_0/L_{Tf}$.
\clearpage
\subsection{$L_{nf}$ variation}\label{Subsec: L_nf variation}
\begin{figure}[htb!]
\centering
\subfigure[]{ \includegraphics[width=0.3\textwidth ]{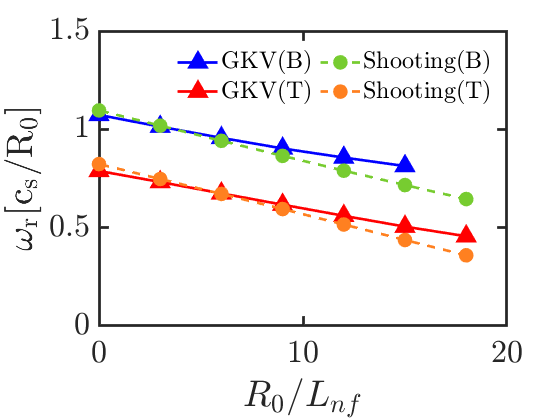}\label{fig:GKV_RLnfscan_omega_r2}}
\subfigure[]{ \includegraphics[width=0.3\textwidth ]{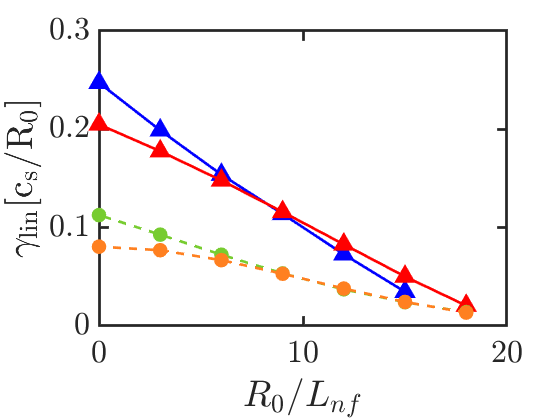}\label{fig:GKV_RLnfscan_omega_i2}}
\caption{ (a) Real frequency ($\omega_r$) and  (b) linear growth rate ($\gamma_{lin}$) as functions of $R_0/L_{nf}$ of the most unstable ballooning parity mode and tearing parity mode. }
\label{fig:GKV_RLnfscan_omega}
\end{figure}
Figure \ref{fig:GKV_RLnfscan_omega} shows the impact of the normalized fast ion density gradient ($R_0/L_{nf}$) on the instability for $q=1.4$, $\hat{s}=0.2$, $k_\theta \rho_s = 0.2$, $T_{i0}/T_{e0}=0.1$, $R_0/{L_{Te}}=R_0/{L_{Ti}}=0$, $R_0/L_{ne}=3$,  $R_0/L_{Tf}=30$, $n_{f0}/n_{e0} =0.1$.  In this case, $R_0/L_{ne}$ was held fixed, while $R_0/L_{ni}$ varied according to the quasi-neutrality condition. As illustrated in figure \ref{fig:GKV_RLnfscan_omega_r2}, the real frequency decreases steadily with increasing $R_0/L_{nf}$, while figure \ref{fig:GKV_RLnfscan_omega_i2} shows a decrease in the linear growth rate. This stabilizing effect of a steeper fast ion density gradient agrees with the theoretical prediction that the fast ion-driven drift instability is most pronounced when the temperature gradient dominates over the density gradient. No eigenstate transitions were observed in this parameter scan, suggesting that the dominant eigenstate remains unaffected as $R_0/L_{nf}$ varies.
\clearpage
%%%%%%%%%%%%%%%%%%%%%%%%%%%%%%%%%%%%%%%%%%%%%%%%%%%%%%%%%%%%%%%%%%%%%%%%%%%%%%%%%%%%%%%%%%%%%%%%%%%%%%%%%%%%%%%%%%%%%%%%%%%%%
\section{QUASILINEAR TRANSPORT}\label{Sec: Quasilinear transport}
\indent To analyze the effect of the fast ion-driven drift instability on plasma transport, we calculate the turbulence-driven fluxes with the quasilinear model. The quasilinear particle flux of the fast ion is calculated as:
\begin{equation}\label{QLPF_f}
\begin{split}
\Gamma_{f}^{turb} &= \sum_{\bold{k_\perp}} \Gamma_{f,\bold{k_\perp}}^{turb}\\
&=\sum_{\bold{k_\perp}} \Re \left\langle\left\langle i \frac{k_\theta}{B}\int d^3 \mathrm{v} \delta \hat{g}_{f\bold{k_\perp}} \delta \hat{\psi}_{\sigma \bold{k_\perp}}^* \right\rangle\right\rangle \\
&= \sum_{\bold{k_\perp}} \Re \left\langle\left\langle i\frac{k_\theta}{B} \left( C_T\alpha_T \hat{D}^T_{f\bold{k_\perp}} + \alpha_P \hat{D}^P_{f\bold{k_\perp}} \right) \frac{Z_f e|\delta \hat{\phi}_\bold{k_\perp}|^2}{T_{f0}} \right\rangle\right\rangle \\
& \sim -\sum_{\bold{k_\perp}}\left\langle\left\langle\frac{\gamma_{lin\bold{k_\perp}}}{\omega_{r\bold{k_\perp}}} k_\theta \rho_s c_s n_{e0} |\hat{\phi}_{\bold{k_\perp}}|^2\right\rangle\right\rangle,
\end{split}
\end{equation}
where $\left\langle\left\langle \cdots \right\rangle\right\rangle$ denotes a double average over the ensemble and the flux surface, $\delta \psi_{\sigma \bold{k_\perp}}=J_0(k_\perp \rho_{\perp\sigma})\delta \hat{\phi}_{\bold{k_\perp}}$ is the gyrophase-averaged perturbed potential, superscript $*$ represents the complex conjugate, and $\hat{\phi}_{\bold{k_\perp}}=\frac{e\delta \hat{\phi}_\bold{k_\perp}}{T_{e0}}$. Since we assumed the adiabatic electron response, the quasilinear particle flux of the background main ion can be easily calculated through the quasi-neutrality condition as 
\begin{equation}\label{QLPF_i}
\Gamma_i = -Z_f\Gamma_f.
\end{equation}
The turbulence-driven quasilinear energy flux $Q^{turb}_{\sigma}$ can be expressed as
\begin{equation}\label{QLEF}
\begin{split}
Q_{\sigma}^{turb} &=\frac{T_{\sigma 0}}{2} \sum_{\bold{k_\perp}} \Re \left\langle\left\langle i \frac{k_\theta}{B}\int d^3 \mathrm{v} \frac{\mathrm{v}^2}{\mathrm{v}_{t\sigma}^2}\delta \hat{g}_{\sigma\bold{k_\perp}} \delta \hat{\psi}_{\sigma \bold{k_\perp}}^* \right\rangle\right\rangle \\
&= {q_{\sigma}^{turb}} + \frac{5}{2}T_{\sigma 0}\Gamma_\sigma^{turb}
\end{split}
\end{equation}
where $q_\sigma^{turb}$ and $\frac{5}{2}T_{\sigma 0}\Gamma_{\sigma}^{turb}$ correspond to the conductive and convective parts of the energy flux, respectively. Note that the energy flux is the sum of the convective and conductive flux. Here, we consider the energy flux $Q^{turb}_\sigma$ rather than the purely conductive heat flux $q^{turb}_\sigma$, which would be more appropriate for the energy confinement. 
For a sufficiently steep fast ion temperature gradient, i.e., $\eta_f\gg 1$, the turbulent fast ion energy flux is primarily driven by the resonant contributions. Since relatively low-energy particles resonate with the electron drift wave \cite{KangPoP2019, KangPoP2020}, we estimate the energy flux of fast ions as:
\begin{equation}\label{QLEF_f}
\begin{split}
Q_{f}^{turb} &=\frac{T_{f 0}}{2} \sum_{\bold{k_\perp}} \Re \left\langle\left\langle i \frac{k_\theta}{B}\int d^3 \mathrm{v} \frac{\mathrm{v}^2}{\mathrm{v}_{tf}^2}\delta \hat{g}_{f\bold{k_\perp}} \delta \hat{\psi}_{f \bold{k_\perp}}^* \right\rangle\right\rangle\\
&\simeq  \hat{E}_{f0}T_{f0}\Gamma_f^{turb},
\end{split}
\end{equation}
where $\hat{E}_{f0}$ is the average normalized energy of resonant fast ions.  The energies of resonant trapped and passing fast ions are assumed to be in the same order as $\hat{E}_{f0} \sim \frac{T_{e0}}{T_{f0}\epsilon_{ne}}$ where $\epsilon_{ne}=L_{ne}/R_0$. \\
\indent To calculate the main ion energy flux, we restrict our analysis to long wavelength perturbations since $k_\theta \rho_s \sim 0.3$ for the fastest-growing mode. We further neglect the effect of the main ion temperature gradient ($\eta_i$) in $\delta \hat{g}_i$, and keep only the linear terms in $\gamma_{lin}/\omega_r \ll 1$. Then $Q_i^{turb}$ is calculated as
\begin{equation}\label{QLEF_i}
\begin{split}
&{Q_{i}^{turb}} = \frac{T_{i0}}{2} \sum_{\bold{k_\perp}} \Re \left\langle\left\langle i \frac{k_\theta}{B}\int d^3 \mathrm{v}  \frac{\mathrm{v}^2}{\mathrm{v}_{ti}^2} \delta \hat{g}_{i\bold{k_\perp}} \delta \hat{\psi}_{i\bold{k_\perp}}^* \right\rangle\right\rangle \\
&\simeq \frac{T_{i0}}{2} \sum_{\bold{k_\perp}} \left\langle\left\langle -\frac{k_\theta}{B}\int d^3 \mathrm{v} \frac{\mathrm{v}^2}{\mathrm{v}_{ti}^2} \frac{\gamma_{lin\bold{k_\perp}}}{\omega_{r\bold{k_\perp}}}\left[ \frac{\omega_{*i}}{\omega_r} + \frac{\omega_{di\mathrm{v}}}{\omega_r} \left( 2\frac{\omega_{*i}}{\omega_r} - 1 \right) + \frac{\mathrm{v}_{\parallel}^2k_\parallel^2}{\omega_r^2} \left( 3\frac{\omega_{*i}}{\omega_r} - 2\right)\right] f_{Mi} \frac{e|\hat{\delta\phi}_{\bold{k_\perp}}|^2}{T_{i0}}\right\rangle\right\rangle \\
&= \frac{n_{i0}T_{i0}}{2} \sum_{\bold{k_\perp}} \left\langle\left\langle \frac{k_\theta}{B}\frac{\gamma_{lin\bold{k_\perp}}}{\omega_{r\bold{k_\perp}}}\left[ 3\frac{\omega_{*i}}{\omega_r} + 10\frac{\omega_{di}}{\omega_r} \left( 2\frac{\omega_{*i}}{\omega_r} - 1 \right) + 5\frac{\mathrm{v}_{\parallel}^2k_\parallel^2}{\omega_r^2} \left( 3\frac{\omega_{*i}}{\omega_r} - 2\right)\right] \frac{e|\hat{\delta \phi}_{\bold{k_\perp}}|^2}{T_{i0}} \right\rangle\right\rangle \\
&\simeq \frac{3}{2} T_{i0} \Gamma_{i}^{turb}.
\end{split}
\end{equation}
When the instability occur, i.e., $\gamma_{lin}>0$, we obtain $\Gamma_f^{turb}<0$, $Q_f^{turb}<0$, $\Gamma_i^{turb}>0$, $Q_i^{turb}>0$ and $|Q_f^{turb}/Q_i^{turb}| \sim T_{e0}/(T_{i0}\epsilon_{ne}) \gg 1$. In other words, particle and energy fluxes of fast ions are inward, while those of main ions are outward, but the net energy flux is inward since the fast ions dominate the energy flux. \\
\indent Despite the steep temperature gradient of fast ions, the inward energy flux of fast ions may seem to contradict the second law of thermodynamics. However, it can be shown that the total turbulent fluxes do not contradict the second law of thermodynamics. To show this, we employ an entropy balance equation derived by Kato \cite{KatoPoP2024}, based on the quasilinear modeling of turbulent energy exchange by wavenumber spectral analysis of entropy balance in microturbulence \cite{SugamaPoP1996, SugamaPoP2009}. The entropy balance equation is written as 
\begin{equation}\label{Entropy}
\begin{split}
&\frac{\partial}{\partial t} \left\langle\left\langle  \int d^3 \mathrm{v} \frac{|\delta f_{\sigma\bold{k_\perp}}|^2}{2f_{M\sigma}} + \delta n_{\sigma \bold{k_\perp}} \frac{Z_\sigma e\delta\hat{\phi}^*_{\bold{k_\perp}}}{2 T_{\sigma 0}} 
 \right\rangle\right\rangle - N_{\sigma \bold{k}_\perp} - C_{\sigma \bold{k_\perp}}= \frac{\Gamma^{turb}_{\sigma \bold{k_\perp}}}{L_{p\sigma}} + \frac{q^{turb}_{\sigma\bold{k_\perp}}}{T_{\sigma 0}L_{T\sigma}} + \frac{Y_{\sigma\bold{k_\perp}}}{T_{\sigma 0}},
\end{split}
\end{equation}
where the electromagnetic perturbation is neglected. \\
In Eq. (\ref{Entropy}), $N_{\sigma \bold{k_\perp}}$ represents the entropy which the mode with the wavenumber vector $\bold{k}_\perp$ gains through nonlinear interaction with other modes. $C_{\sigma \bold{k_\perp}}$ represents collisional dissipation and $L_{p\sigma}^{-1}=L_{n\sigma}^{-1} + L_{T\sigma}^{-1}$ is pressure gradient scale length. $Y_\sigma$ indicates the turbulent energy exchange introduced by Candy \cite{CandyPoP2013}, expressed as
\begin{equation}\label{Ysigma}
\begin{split}
Y_\sigma = \sum_\bold{k_\perp} Y_{\sigma\bold{k_\perp}} = \frac{Z_\sigma e}{2} \sum_\bold{k_\perp} \Re \left\langle\left\langle \int d^3 \mathrm{v} \left( \delta \hat{g}_{\sigma\bold{k_\perp}}\frac{\partial \delta \hat{\psi}^*_{\sigma \bold{k_\perp}}}{\partial t} - \frac{\partial \delta\hat{g}_{\sigma\bold{k_\perp}} }{\partial t}  \delta\hat{\psi}^*_{\sigma \bold{k_\perp}} \right) \right\rangle\right\rangle.
\end{split}
\end{equation}
Note that this expression of $Y_\sigma$ rigorously satisfies $\sum_\sigma Y_\sigma = 0$.\\
Since we are interested in collisionless plasmas, the collisional dissipation $C_\sigma$ can be neglected.  We further restrict the discussion to the linear growing state and can neglect the nonlinear interaction $N_{\sigma}$. 
Multiplying Eq. (\ref{Entropy}) by $T_{\sigma 0}$ and adding them all together for fast ions and main ions, we obtain the following equation:
\begin{equation}\label{Entropy2}
\begin{split}
&\frac{\partial}{\partial t} \left\langle\left\langle  \int d^3 \mathrm{v} \left( T_{f0} \frac{|\delta f_{f\bold{k_\perp}}|^2}{2f_{Mf}} + T_{i0} \frac{|\delta f_{i\bold{k_\perp}}|^2}{2f_{Mi}}\right) + \delta n_{f \bold{k_\perp}} \frac{Z_f e\delta\hat{\phi}^*_{\bold{k_\perp}}}{2} + \delta n_{i \bold{k_\perp}} \frac{ e\delta\hat{\phi}^*_{\bold{k_\perp}}}{2} \right\rangle\right\rangle\\
&= \frac{T_{f0} \Gamma^{turb}_{f \bold{k_\perp}}}{L_{pf}} + \frac{T_{i0} \Gamma^{turb}_{i \bold{k_\perp}}}{L_{pi}} + \frac{q^{turb}_{f\bold{k_\perp}}}{L_{Tf}} + \frac{q^{turb}_{i\bold{k_\perp}}}{L_{Ti}} +Y_{f\bold{k_\perp}} + Y_{i\bold{k_\perp}}.
\end{split}
\end{equation}
In the right-hand side of Eq. (\ref{Entropy2}), $Y_{f\bold{k_\perp}} + Y_{i\bold{k_\perp}}=-Y_{e\bold{k_\perp}}=0$ because of the adiabatic electron response. Substituting Eqs. (\ref{QLPF_f}) - (\ref{QLEF_i}) into Eq. (\ref{Entropy2}) and assuming $\eta_f \gg 1$, we obtain
\begin{equation}\label{Entropy3}
\begin{split}
&\frac{\partial}{\partial t} \left\langle\left\langle  \int d^3 \mathrm{v} \left( T_{f0} \frac{|\delta f_{f\bold{k_\perp}}|^2}{2f_{Mf}} + T_{i0} \frac{|\delta f_{i\bold{k_\perp}}|^2}{2f_{Mi}}\right) + \frac{1}{2}n_{e0}T_{e0} |\hat{\phi}_{\bold{k_\perp}}|^2 \right\rangle\right\rangle\\
&= \left[ \left( \hat{E}_{f0} - \frac{3}{2} \right)\frac{1}{L_{Tf}} - Z_f \frac{T_{i0}}{T_{f0}L_{ni}} \right] T_{f0}\Gamma_{f\bold{k_\perp}}^{turb}
\end{split}
\end{equation}
The left-hand side of Eq. (\ref{Entropy3}) is the rate of change over time of the perturbed quantities and is proportional to the linear growth rate. Since $T_{f0}\gg T_{i0}$, dominant term of the right-hand side of Eq. (\ref{Entropy3}) is $\left(\hat{E}_{f0}-\frac{3}{2}\right)T_{f0}\Gamma_{f\bold{k_\perp}}^{turb}/L_{Tf}$, which is also proportional to the linear growth rate. Note that the condition $\hat{E}_{f0} - 3/2<0$ should hold because of Eq. (34) in \cite{KangPoP2019}. Therefore, quasilinear energy fluxes $Q_i^{turb}$ and $Q_f^{turb}$do not contradict the second law of thermodynamics because the entropy balance is still maintained by the entropy production of the main ions. These turbulent fluxes imply that fast ions are effectively confined within the core plasma region and can help maintain a high central temperature, enhancing the heating efficiency. However, recall that these estimations have been derived with simplified models. Therefore, extended studies, including the non-adiabatic electron response, are necessary to address the turbulent transport of instability more precisely. Although the estimation of turbulent fluxes in the linear growing state is consistent with the nonlinear gyrokinetic simulation results in \cite{KangPLA2021}, it was observed that the direction of fast ion energy flux in a nonlinear state varies in time without reaching a clear saturation state. Note that the term "heat flux" used in \cite{KangPLA2021} corresponds to the meaning of energy flux $Q^{turb}_\sigma$, defined in this section. It has been recently found that the ITG turbulence transfers energy from ions to electrons regardless of the relative temperature between electrons and ions, which is in marked contrast to the energy transfer by Coulomb collisions \cite{KatoPoP2024}. It would be practical interest to investigate the nonlinear interaction and turbulent energy exchange to address the temporal variation of fast ion energy flux. Understanding these properties is important to estimate the potential impact of turbulence transport driven by the fast ion-driven drift instability.
\clearpage
%%%%%%%%%%%%%%%%%%%%%%%%%%%%%%%%%%%%%%%%%%%%%%%%%%%%%%%%%%%%%%%%%%%%%%%%%%%%%%%%%%%%%%%%%%%%%%%%%%%%%%%%%%%%%%%%%%%%%%%%%%%%%%%%%%%%%%%%%%%%%%%%%%%%%%%%%%%%%%%%%%%%%%%%%%%%
%%%%%%%%%%%%%%%%%%%%%%%%%%%%%%%%%%%%%%%%%%%%%%%%%%%%%%%%%%%%%%%%%%%%%%%%%%%%%%%%%%%%%%%%%%%%%%%%%%%%%%%%%%%%%%%%%%%%%%%%%%%%%%%%%%%%%%%%%%%%%%%%%%%%%%%%%%%%%%%%%%%%%%%%%%%%
%%%%%%%%%%%%%%%%%%%%%%%%%%%%%%%%%%%%%%%%%%%%%%%%%%%%%%%%%%%%%%%%%%%%%%%%%%%%%%%%%%%%%%%%%%%%%%%%%%%%%%%%%%%%%%%%%%%%%%%%%%%%%%%%%%%%%%%%%%%%%%%%%%%%%%%%%%%%%%%%%%%%%%%%%%%%
\section{SUMMARY AND DISCUSSION}\label{Sec: Conclusion}
In this work, we have comprehensively studied the fast ion-driven drift instability using both analytical and numerical methods, extending previous models by incorporating the nonadiabatic response of passing fast ions. Through linear stability analyses and gyrokinetic simulations performed with the GKV code, we have demonstrated the significant impact of passing fast ions on the eigenstate transitions of the instability, particularly in plasmas with weak negative shear and moderate positive shear. We have identified that plasma parameters $\hat{s}$, $q$, and $k_\theta$ play critical roles in shaping the effective potential and determining the eigenmode structure. Our results show that eigenstate transitions from the ground to the non-ground state can widely occur in fast ion-driven drift instabilities under steep temperature gradients of fast ions. Furthermore, the oscillatory nature of the toroidicity-induced potential leads to complex eigenmode behaviors, with non-ground states often being the most unstable, especially in regimes with weak magnetic shear, high safety factor, and long wavelength perturbations. The quasilinear transport caused by the fast ion-driven drift instability has been calculated so that the net energy flux can be inward, consistent with the second law of thermodynamics.\\
\indent In addition to the electrostatic nature of the fast ion-driven drift instability, the role of eigenstate transitions has broader implications. It has been reported that transitions to non-ground states can significantly reduce transport coefficients and even reverse transport trends in electrostatic drift instabilities \cite{XiePRL2017}. Our study has shown that eigenstate transitions become particularly important at plasma parameters where the linear growth rate of the fast ion-driven drift instability is maximized. Moreover, parity transitions in the eigenmode structure could impact electromagnetic turbulent transport, particularly in finite-$\beta$ plasmas. It has been demonstrated that the parity of the mode plays a crucial role in the formation of magnetic islands, with the tearing parity mode and ballooning parity mode producing opposite directions of electromagnetic heat flux \cite{HatchPRL2012}. Thus, even though the fast ion-driven drift instability is primarily electrostatic, its influence on electromagnetic heat flux through eigenstate transitions and parity changes should be considered. Future nonlinear gyrokinetic simulations should investigate these effects with relevant parameters to estimate their impact on plasma confinement in fusion devices. \\
\indent Steep fast ion temperature profiles can be generated by the fast ion tails with the ICRH. In such conditions, fast ions can suppress ITG instability through the electrostatic wave-particle resonance \cite{SienaNF2018, SienaPoP2019}. In contrast, our results demonstrate that steep fast ion temperature gradients can lead to destabilization when fast ions resonantly interact with electron drift waves, mainly in reversed shear plasmas. Reversed shear plasmas are widely recognized as one of the key operational scenarios for steady-state tokamak reactors, offering good confinement, high stability, and a high fraction of self-sustaining bootstrap current \cite{KesselPRL1994, TurnbullPRL1995}. Their stabilizing influence on CTEMs, attributed to the precession reversal of trapped electrons \cite{BeerPoP1997, LiPPCF2002} is well known. Furthermore, reversed magnetic shear is known to amplify the stabilizing effect of the electrostatic wave-particle resonance on ITG modes \cite{SienaPoP2019}. Therefore, ICRH in reversed shear plasmas may create favorable conditions for fast ion-driven drift instability while suppressing microinstabilities such as ITG mode and CTEM. Investigating the fast ion-driven drift instability under experimental ICRH scenarios in reversed shear plasmas could provide valuable insights into the interplay between stabilization and destabilization mechanisms of fast ions in fusion plasmas.\\
\indent Even though some plasma parameters used in this study, such as $n_{f0}/n_{e0}$ and $R_0/L_{Tf}$, are idealized and far from typical experimental conditions, they were intentionally chosen to amplify the effect of fast ions and provide deeper insights into the underlying physics. Additionally, the choice of $T_{f0}/T_{e0}=50$ in section \ref{Subsec: Magnetic shear variation} - \ref{Subsec: k_theta variation} was made to explore the eigenstate transitions observed in \cite{KangPoP2020}. The parametric analysis in section \ref{Subsec: T_f variation} demonstrated that the fast ion temperature has a significant impact on the linear growth rates of the instability, with the maximum growth rate occurring near $T_{f0}/T_{e0} \approx 15$. This suggests that our results may be applicable to more realistic scenarios. \\
\indent Additionally, the Fast Ion Regulated Enhancement (FIRE) mode in the KSTAR tokamak recently achieved high-performance, steady-state conditions by maintaining a high fraction of fast ions in the core \cite{HanNature2022}. This suggests that maintaining a high fast ion fraction could be key to achieving steady-state fusion operations with high energy performance.
In the core region, the fast ion density fraction ($n_{f}/n_e$) of FIRE mode reached nearly 0.4, which is significantly higher than the value considered in our study. In such scenarios, electrostatic modes driven by resonances between fast ions and electrostatic waves could become significant in specific parameter regimes. The FIRE mode experiments further provided evidence that fast ions can trigger localized instabilities within the central region of the plasma, and gyrokinetic simulations have identified fast ion-relevant modes under FIRE mode parameters \cite{KimNF2024}. 
Furthermore, recent studies such as \cite{SienaPRL2021} have shown that the electrostatic wave-particle resonance effects can trigger transport barriers, leading to substantial confinement improvements in ASDEX Upgrade discharge. These findings support the potential importance of the fast ion-driven drift instability in future experiments, and it should be carefully investigated using more realistic parameters.\\
\indent Finally, although this study has focused on electrostatic modes, electromagnetic effects such as the interaction between fast ions and Alfv\'enic instabilities can play a critical role in plasma stability and transport. Recent work by Chen et al. \cite{ChenNF2022} showed that drift wave turbulence can scatter Toroidal Alfv\'en Eigenmodes (TAEs), generating short-wavelength kinetic Alfv\'en waves (KAWs) that are subsequently Landau damped by electrons. These complex interactions highlight the necessity of considering electromagnetic effects in future studies. Including these effects in future models will be essential for a more comprehensive understanding of fast ion-driven instabilities. Future research should consider these electromagnetic effects, especially in regimes where the stabilizing role of fast ions competes with or complements their destabilizing effects on drift instabilities. Such studies could provide deeper insights into how complex interactions between fast ions, drift wave turbulence, and TAEs influence overall confinement in fusion plasmas, which will be particularly relevant for optimizing the performance of next-generation fusion devices like ITER.
\clearpage
\section*{CRediT authorship contribution statement}
\noindent \textbf{B. J. Kang}: Conceptualization, Data curation, Formal analysis, Investigation, Methodology, Writing - original draft, Writing - review \& editing.  \\
\textbf{H. Sugama}: Conceptualization, Formal analysis, Methodology, Writing - original draft, Writing - review \& editing, \\
\textbf{T. -H. Watanabe}: Data curation, Investigation, Methodology, Software, Writing - review \& editing, \\
\textbf{M. Nunami}: Data curation, Investigation, Software.
\section*{Declaration of competing interest}
The authors declare that they have no known competing financial interests or personal relationships that could have appeared to influence the work reported in this paper.
\section*{Data availability}
Data will be made available on request.
\section*{Acknowledgements}
This study was supported in part by the JSPS Grants-in-Aid for Scientific Research (Grant Nos. 19H01879 and 24K07000) and in part by the NINS program of Promoting Research by Networking among Institutions (Grant No. 01422301). Simulations in this work were performed on “Plasma Simulator” (NEC SX-Aurora TSUBASA) of NIFS with the support and under the auspices of the NIFS Collaboration Research program (Grant No. NIFS23KIPT009).
%%%%%%%%%%%%%%%%%%%%%%%%%%%%%%%%%%%%%%%%%%%%%%%%%%%%%%%%%%%%%%%%%%%%%%%%%%%%%%%%%%%%%%%%%%%%%%%%%%%%%%%%%%%%%%%%%%%%%%%%%%%%%%%%%%%%%%%%%%%%%%%%%%%%%%%%%%%%%%%%%%%%%%%%%%%%
%%%%%%%%%%%%%%%%%%%%%%%%%%%%%%%%%%%%%%%%%%%%%%%%%%%%%%%%%%%%%%%%%%%%%%%%%%%%%%%%%%%%%%%%%%%%%%%%%%%%%%%%%%%%%%%%%%%%%%%%%%%%%%%%%%%%%%%%%%%%%%%%%%%%%%%%%%%%%%%%%%%%%%%%%%%%
%%%%%%%%%%%%%%%%%%%%%%%%%%%%%%%%%%%%%%%%%%%%%%%%%%%%%%%%%%%%%%%%%%%%%%%%%%%%%%%%%%%%%%%%%%%%%%%%%%%%%%%%%%%%%%%%%%%%%%%%%%%%%%%%%%%%%%%%%%%%%%%%%%%%%%%%%%%%%%%%%%%%%%%%%%%%


\begin{thebibliography}{99} 
\bibitem{RoesnbluthPRL1975} M. N. Rosenbluth and P. H. Rutherford, Phys. Rev. Lett. 34, 1428 (1975).
\bibitem{HasegawaPF1976} A. Hasegawa and L. Chen, Phys. Fluids 19, 1924 (1976).
\bibitem{ChenRMP2016} L. Chen and F. Zonca, Rev. Mod. Phs. 88, 015008 (2016). %https://journals.aps.org/rmp/abstract/10.1103/RevModPhys.88.015008
\bibitem{AngioniPoP2008} C. Angioni and A. G. Peeters, Phys. Plasmas 15, 052307 (2008).
\bibitem{ZhangPRL2008} W. Zhang, Z. Lin, and L. Chen, Phys. Rev. Lett. 101, 095001 (2008).
\bibitem{YangPoP2018} S. M. Yang, C. Angioni, T. S. Hahm, D. H. Na, and Y. S. Na, Phys. Plasmas 25, 122305 (2018).
\bibitem{CitrinPPCF2023} J. Citrin and P. Mantica, Plasma Phys. Control. Fusion 65, 033001 (2023).
\bibitem{KazakovNF2015} Y. O. Kazakov, D. Van Eester, R. Dumont and J. Ongena, Nucl. Fusion 55, 032001 (2015).
\bibitem{SienaNF2018} A. Di Siena, T. G' orler, H. Doerk, E. Poli and R. Bilato, Nucl. Fusion 58, 054002 (2018).
\bibitem{SienaPoP2019} A. Di Siena, T. Göler, E. Poli, R. Bilato, H. Doerk, and A. Zocco, Phys. Plasmas 26, 052504 (2019).
\bibitem{KesselPRL1994} C. Kessel, J. Manickam, G. Rewoldt, and W. M. Tang, Phys. Rev. Lett. 72, 1212 (1994).
\bibitem{TurnbullPRL1995} A. D. Turnbull, T. S. Taylor, Y. R. Lin-Liu, and H. S. John, Phys. Rev. Lett. 74, 718 (1995).
\bibitem{LevintonPRL1995} F. M. Levinton, M. C. Zarnstorff, S. H. Batha, M. Bell, R. E. Bell, R. V. Budny, C. Bush, Z. Chang, E. Fredrickson, A. Janos, J. Manickam, A. Ramsey, S. A. Sabbagh, and G. L. Schimidt, Phys. Rev. Lett. 75, 4417 (1995).
\bibitem{StraitPRL1995} E. J. Strait, L. L. Lao, M. E. Mauel, B. W. Rice, T. S. Taylor, K. H. Burrell, M. S. Chu, E. A. Lazarus, T. H. Osborne, S. J. Thompson, and A. D. Turnbull, Phys. Rev. Lett. 75, 4421 (1995).
\bibitem{NazikianPRL2005} R. Nazikian, K. Shinohara, G. J. Kramer, E. Valeo, K. Hill, T. S. Hahm, G. Rewoldt, S. Ide, Y. Koide, Y. Oyama, H. Shirai, and W. Tang, Phys. Rev. Lett. 94, 135002 (2005).
\bibitem{BeerPoP1997} M. A. Beer, G. W. Hammett, G. Rewoldt, E. J. Synakowski, M. C. Zarnstorff, and W. Dorland, Phys. Plasmas 4, 1792-1799 (1997).
\bibitem{LiPPCF2002} J. Li and Y. Kishimoto, Plasma Phys. Controlled Fusion 44, A479 (2002).
\bibitem{KangPoP2019} B.J. Kang and T.S. Hahm, Phys. Plasmas 26, 042501 (2019). %https://doi.org/10.1063/1.5086935
\bibitem{KangPoP2020} B.J. Kang, Y.J. Kim, C. Angioni and T.S. Hahm, Phys. Plasmas 27, 072510 (2020). %https://doi.org/10.1063/5.0010098
\bibitem{KangPLA2021} B. J. Kang, C. Angioni and T. S. Hahm, Phys. Lett. A 414, 127632 (2021).  %https://doi.org/10.1016/j.physleta.2021.127632
\bibitem{ChenPF1980} L. Chen and C.Z. Cheng, Phys. Fluids 23, 2242 (1980). %http://dx.doi.org/10.1063/1.862907
\bibitem{HortonRMP1999} W. Horton, Rev. Mod. Phys. 71, 735 (1999). %https://journals.aps.org/rmp/abstract/10.1103/RevModPhys.71.735
\bibitem{WangNF2012} E. Wang, X. Xu, J. Candy, R. Groebner, P. Snyder, Y. Chen, S. Parker, W. Wan, G. Lu, and J. Dong, Nucl. Fusion 52, 103015 (2012). %http://dx.doi.org/10.1088/0029-5515/52/10/103015
\bibitem{XiePoP2015} H. S. Xie and Y. Xiao, Phys. Plasmas 22, 090703 (2015). %http://dx.doi.org/10.1063/1.4931072
\bibitem{XiePoP2016} H. S. Xie and B. Li, Phys. Plasmas 23, 082513 (2016). %https://doi.org/10.1063/1.4931072
\bibitem{XiePRL2017} H. S. Xie, Y. Xiao, and Z. Lin, Phys. Rev. Lett. 118, 095001 (2017). %https://doi.org/10.1103/PhysRevLett.118.095001
\bibitem{ConnorPRS1979} J.W. Connor, R.J. Hastie, and J.B. Taylor, Proc. R. Soc. 365, 1 (1979). %https://doi.org/10.1098/rspa.1979.0001
\bibitem{WatanabeNF2006} T.-H. Watanabe and H. Sugama, Nucl. Fusion 46, 24 (2006).
\bibitem{FriemanPF1982} E.A. Frieman, Liu Chen, Phys. Fluids 25, 502 (1982). %http://dx.doi.org/10.1063/1.863762
\bibitem{HahmPF1988} T.S. Hahm, Phys. Fluids 31, 2670 (1988). %https://doi.org/10.1063/1.866544
\bibitem{GangPFB1990} F.Y. Gang and P.H. Diamond, Phys. Fluids, B Plasma Phys. 2, 2976 (1990). %https://doi.org/10.1063/1.859363
\bibitem{FongPoP1999} B.H. Fong and T.S. Hahm, Phys. Plasmas 6, 188 (1999). %https://doi.org/10.1063/1.873272
\bibitem{RomanelliPFB1989} F. Romanelli, Phys. Fluids B 1, 1018 (1989). %https://doi.org/10.1063/1.859023
\bibitem{SugamaPoP2006} H. Sugama and T. -H. Watanabe, Phys. Plasmas 13, 012501 (2006).
\bibitem{SugamaJPP2006} H. Sugama and T. -H. Watanabe, J. Plasma Phys. 72(6), 825 (2006).
\bibitem{SugamaPPCF2011} H. Sugama, T.-H. Watanabe, M. Nunami, and S. Nishimura, Plasma Phys. Contrl. Fusion 53, 024004 (2011). %doi:10.1088/0741-3335/53/2/024004
\bibitem{IshizawaJPP2015} A. Ishizawa, S. Maeyama, T.-H. Watanabe, H. Sugama, and N. Nakajima, J. Plasma Phys. 81, 435810203 (2015).
\bibitem{KadomtsevSPJ1967} B. B. Kadomtsev and O. P. Pogutse, Sov. Phys. JETP 24, 1172 (1967).
\bibitem{BeerPoP1996} M.A. Beer and G.W. Hammett, Phys. Plasmas 3, 4046 (1996). %https://doi.org/10.1063/1.871574
\bibitem{Fried1961} B.D. Fried and S.D. Conte, The Plasma Dispersion Function. Academic Press, New York (1961). % https://doi.org/10.11316/butsuri1946.17.6.472_1
\bibitem{WhiteJCP1979} R.B. White, J. Compy. Phys. 31, 409 (1979).
\bibitem{Haeseleer1991} W. D. D'Haeseleer, W. N. G. Hitchon, J. D. Callen, and J. L. Shohet, Flux Coordinates and Magnetic Field Structure (Springer, Berlin, 1991).
\bibitem{LapillonnePoP2009} X. Lapillonne, S. Brunner, T. Dannert, S. Jolliet, A. Marinoni, L. Villard, T. G\"orler, F. Jenko, and F.Merz, Phys. Plasmas 16, 032308 (2009).
\bibitem{PeetersCPC2009} A.G. Peeters, Y. Camenen, F.J. Casson, W.A. Hornsby, A.P. Snodin, D. Strintzi, G.Szepesi, Comput. Phys. Commun. 180(12) (2009) 2650.
\bibitem{KatoPoP2024} T. Kato, H. Sugama, T. -H Watanabe, and M. Nunami, Phys. Plasmas 31, 062510 (2024).
\bibitem{SugamaPoP1996} H. Sugama, M. Okamoto, W. Horton, and M. Wakatani, Phys. Plasmas 3, 2379 (1996).
\bibitem{SugamaPoP2009} H. Sugama, T.-H. Watanabe, and M. Nunami, Phys. Plasmas 16, 112503 (2009).
\bibitem{CandyPoP2013} J. Candy, Phys. Plasmas 20, 082503 (2013).
\bibitem{HatchPRL2012} D. R. Hatch, M. J. Pueschel, F. Jenko, W. M. Nevins, P. W. Terry, and H. Doerk, Phys. Rev. Lett. 108, 235002 (2012).
\bibitem{HanNature2022} H. Han, S. J. Park, C. Sung, J. Kang, Y. H. Lee, J. Chung, T. S. Hahm, B. Kim, J.-K. Park, J. G. Bak, M. S. Cha, G. J. Choi, M. J. Choi, J. Gwak, S. H. Hahn, J. Jang, K. C. Lee, J. H. Kim, S. K. Kim, W. C. Kim, J. Ko, W. H. Ko, C. Y. Lee, J. H. Lee, J. H. Lee, J. K. Lee, J. P. Lee, K. D. Lee, Y. S. Park, J. Seo, S. M. Yang, S. W. Yoon, and Y.-S. Na, Nature 609, 269-275 (2022).
\bibitem{KimNF2024} D. Kim, S. J. Park, G. J. Choi, Y. W. Cho, J. Kang, H. Han, J. Candy, E. A. Belli, Y. -S. Na, T. S. Hahm and C. Sung, Nucl. Fusion 64, 066013 (2024).
\bibitem{SienaPRL2021} A. Di Siena, R. Bilato, T. Görler,  A. Bañón Navarro, E. Poli, V. Bobkov, D. Jarema, E. Fable, C. Angioni, Ye. O. Kazakov, R. Ochoukov, P. Schneider, M. Weiland, F. Jenko, and the ASDEX Upgrade Team, Phys. Rev. Lett. 127, 025002 (2021).
\bibitem{ChenNF2022} Liu Chen, Zhiyong Qiu, and Fulvio Zonca, Nucl. Fusion 62, 094001 (2022).



\end{thebibliography}
\end{document}